\documentclass[final,1p,times]{elsarticle}
\usepackage{amsmath,mathptmx,amssymb,amsthm}
\newtheorem{assumption}{Assumption}[section]
\newtheorem{problem}{Problem}[section]
\newtheorem{lemma}{Lemma}[section]

\newtheorem{definition}{Definition}[section]
\newtheorem{theorem}{Theorem}[section]
\newtheorem{example}{Example}[section]
\usepackage{graphicx} 
\usepackage {epstopdf}
\graphicspath{{Figures/}}
\usepackage{caption}
\usepackage{subcaption}
\usepackage{xcolor}
\renewcommand{\qed}{\hfill\blacksquare}
\newcommand{\qedwhite}{\hfill \ensuremath{\Box}}
\makeatletter
\def\ps@pprintTitle{%
	\let\@oddhead\@empty
	\let\@evenhead\@empty
	\let\@oddfoot\@empty
	\let\@evenfoot\@oddfoot
}
\makeatother
\begin{document}

\begin{frontmatter}

\title{Realization of multi-input/multi-output switched linear systems from Markov parameters}

\author[inst1]{Fethi~Bencherki \fnref{label1}}
\ead{fethi.bencherki@control.lth.se}
\fntext[label1]{Fethi Bencherki was with the Department of Electrical and Electronics Engineering, Eskisehir 
	Technical University.}
\affiliation[inst1]{organization={Department of Automatic Control},
            addressline={Lund University, Box 117}, 
            postcode={SE-221 00}, 
            country={Sweden}
        }

\author[inst2]{Semiha~T\"urkay}
\ead{semihaturkay@eskisehir.edu.tr}
\affiliation[inst2]{organization={Department of Electrical and Electronics Engineering},%Department and Organization
	addressline={Eskisehir Technical University}, 
    city={Eskisehir},
	postcode={26555}, 
	country={Turkey}
}

\author[inst2]{H\"useyin Ak\c{c}ay \corref{label3}}
\ead{huakcay@eskisehir.edu.tr}
\cortext[label3]{H\"useyin Ak\c{c}ay is the corresponding author. Tel: +90 222 321 3550-X 6459. Fax: +90 222 323 9501.}

\begin{abstract}
This paper presents a four-stage algorithm for the realization of multi-input/multi-output (MIMO) switched linear systems (SLSs) 
from Markov parameters. In the first stage, a linear time-varying (LTV) realization that is topologically equivalent to the true 
SLS on intervals of interest is derived from the Markov parameters assuming that the discrete states have a common MacMillan 
degree and a mild condition on their dwell times holds. In the second stage, stationary point set of a Hankel matrix with fixed 
block column and row dimensions and built from the Markov parameters is examined. Splitting of this set into a union of disjoint 
intervals and intersection of their complements reveals linear time-invariant dynamics prevailing on these intervals. 
Absolute sum of modal eigenvalues is used as a feature for clustering to extract the discrete states, up to arbitrary similarity
transformations. Recovery of the discrete states is complete if a mild unimodality assumption holds and each discrete 
state visits at least one segment and stays there long enough. In the third stage, the switching sequence is estimated by 
three schemes. The first scheme is based on correcting the state-space matrices estimated by the LTV realization algorithm 
from the Markov parameters. It starts from a discrete state estimate, proceeds along the positive and/or the negative real axes, 
and continues until a switch or a switch pair is detected. This process is repeated until all switches compatible with the 
dwell time requirements are detected. The second scheme is based on matching the estimated and the true Markov parameters of 
the SLS system over segments of the switching sequence. The third scheme is also based on matching the Markov parameters, but 
it is a discrete optimization/hypothesis testing algorithm. The three schemes operate on different dwell time and model 
structure requirements, but their dwell time requirements are weaker than the one needed to recover the discrete states. In 
the fourth stage, the discrete states estimated in the second stage are brought to a common basis by applying a novel basis
transformation method which is necessary before performing output predictions to prescribed inputs. Robustness of this algorithm 
to amplitude bounded noise is studied and it is shown that small perturbations may only produce small deviations in the 
estimates that vanish as noise amplitude diminishes. Time complexity of the four-stage algorithm is also studied. 
A numerical example illustrates the derived results. A key role in this algorithm is played by a time-dependent switching 
sequence that partitions the state-space according to time, unlike many other works in the literature in which partitioning 
is state and/or input dependent.

\end{abstract}


\end{frontmatter}

%\linenumbers

\section{Introduction}

Hybrid systems  have attracted a vast interest lately due to their universal modeling capabilities of nonlinear dynamical 
systems. Switching among a finite set of linear submodels facilitated the difficult nonlinear identification problem into an
identification of several affine submodels \cite{Ruietal:2016,Paolettietal:2007} and has found its place in many versatile
applications in video and texture segmentation \cite{Vidal:2008,Ozayetal:2011,Ozayetal:2015}, air traffic systems 
\cite{Jin&Huang:2010}, human control behavior \cite{Smith:1998}, and  evolution of HIV infection \cite{Mestre:2010}. 

Most works in hybrid systems literature abide to input-output representations, namely, piecewise auto-regressive exogenous 
(PWARX) and switched auto-regressive exogenous (SARX) modeling techniques  \cite{Bemporadetal:2005,Ferrarietal:2003,
Juloskietal:2005,Lassoued&Abderrahim:2014,Vidaletal:2003,Ohlsson&Ljung:2013,Hartmannetal:2015} while especially in MIMO systems,
state-space models are most preferred \cite{Lopesetal:2013,Sefidmazgietal:2016}. Controllability, observability, fault detection, 
and observer design concepts are also well understood for state-space models. Equivalence between input-output and state-space 
models in hybrid systems was introduced in \cite{Weilandetal:2006} by assuming pathwise observability. An SLS is inspired by 
classical state-space representation based on linear time-invariant (LTI) models and is characterized by its hard switching 
points between its possible operating subdynamics, unlike a linear parameter-varying (LPV) system whose model parameters most 
often change smoothly over time \cite{Toth:2010}. Actually, an SLS is nothing but an LPV system with non-smooth and abrupt 
changes \cite{Lauer&Bloch:2019}. Equivalence between SLSs and the LPVs was  formally addressed in
\cite{Petreczky:2011,Petreczkyetal:2013} and a power series based algorithm was proposed for the realization of the latter 
which is actually the Ho-Kalman algorithm \cite{Ho&Kalman:1965} adapted to these particular model structures. Markov parameters 
are useful to estimate state-space matrices describing best the underlying system dynamics when the system's pulse response is 
known \cite{Phanetal:1998}.

State-space SLS models are an important subclass of hybrid systems and their estimation is known to be a difficult problem 
since neither the knowledge of the data partitioning nor the submodels (discrete states) are available {\em a priori}. The 
continuous state is also unknown, which further obstructs  treating  it as a regression problem.  Additionally, estimated 
state-space submodels reside in a different state basis, which hinders their use for output prediction to prescribed input 
signals \cite{Bakoetal:2009a}. A state-space realization algorithm for LPV input-output models was presented in 
\cite{Tothetal:2011}. Relationships between state-space models and auto-regressive exgoneous (ARX) and auto-regressive 
moving-average exogenous (ARMAX) input-output models were explored in \cite{Phan&Longman:1996} by introducing observer 
Markov parameters initially in LTI setting and later in LTV setting \cite{Majjietal:2010}. Complexity of the SLS 
identification process requires a set of assumptions on dwell times of the discrete states \cite{Huangetal:2004}, 
observability \cite{Vidal&Chiuso&Soatto:2002}, or knowledge of the switching sequence \cite{Verdult&Verhaegen:2004} 
which can be vital for the success of a proposed algorithm.

\subsection{Related work}

Methods that utilize subspace techniques to detect switches, identify active submodels between 
switches, and merge similar submodels to estimate a switching sequence were reported in 
\cite{Borgesetal:2005,Pekpeetal:2004}. In \cite{Bakoetal:2013}, input-to-state-to-output measurements 
were assumed to be available and discrete state parameters were estimated by a sparsity-inducing optimization 
approach. A novel approach in \cite{Bencherki&Turkay&Akcay:2020} transformed identification problem into 
an SARX estimation problem by modulated output injection with observer deadbeat gains and discrete state 
parameters were estimated by minimizing a sparsity-promoting norm.  An online structured subspace 
identification method equipped with a switch detection strategy was reported in \cite{Bakoetal:2009b}. 
An approximate expectation-maximization method for learning maximum-likelihood parameters of a linear switching 
system was proposed in \cite{Blackmoreetal:2007}, which might get stuck to a local minimum. This method
was improved in \cite{Gil&Williams:2009} by adjusting search direction. Online approach proposed in 
\cite{Pekpe&Lecoeuche:2008} alternates between Markov parameter estimation and system classification to 
obtain local models.

The eigensystem realization algorithm (ERA) proposed in \cite{Juang&Pappa:1985} is a noisy version of the famous 
Ho-Kalman algorithm \cite{Ho&Kalman:66}. It uses numerically robust, but expensive singular value decomposition (SVD) 
and returns a minimal balanced realization of the LTI system. A modified  so-called ERA/DC algorithm was 
presented in \cite{Juangetal:1988} where data correlations rather than actual response values were 
used. The methodology here serves to average-out the effect of noise on the estimated state-space 
realization under  the observation that output covariance matrices are identical to Markov 
parameter covariance matrices when the system is excited with white noise inputs. Online version of 
the ERA was reported in \cite{Cooper:1997}. Extented version  of the ERA dealing with 
arbitrary-variations was presented in \cite{Majjietal:2010b} as a time-varying ERA 
(LTV-ERA). 

\subsection{Motivation for the SLS realization}
This paper is about the realization of SLSs described by a discrete convolution equation
\begin{equation}\label{convolution}
y(k)=h(k,l)\ast u(l)+e(k) \qquad k=1,\cdots,N
\end{equation}
where $u(l)$ is the input, $h(k,l)$, $-\infty \leq l\leq k$ are the doubly indexed Markov parameters, 
$y(k)$ and $e(k)$ are respectively the output and noise. The collection $h(k,l)$, $l\leq k \leq N$ grows as $N$ 
increases. The state-space description in Section~\ref{probform} provides a compact representation of 
(\ref{convolution}) where the state-space matrices are assumed to change at switches.

The goal of system identification is to learn a model for (\ref{convolution}) from the input-output data
and provide a convergence guarantee, which is often asymptotic in $N$. Finite sample complexity issues
have mostly been ignored in system identification literature. There is a surge of interest from the 
machine learning community in data-driven control and non-asymptotic analysis. In \cite{Oymak&Ozay:2019}, 
sample complexity results in learning the Markov parameters from single trajectories of LTI systems were 
derived. On the same theme, but with an emphasis on the realization issue, further results were obtained
in \cite{Sarkaretal:2021}. In contrast to LTI systems, relatively few results are available for 
the identification and the realization of SLSs in the state-space framework. Sparse optimization approach 
applied to SLS parameter estimation in \cite{Bakoetal:2013,Bencherki&Turkay&Akcay:2020} has been
a promising alternative to least-squares methods. 

Realization theory for LPV systems is not complete despite many advances \cite{Petreczky:2011,Tothetal:2011b,
Petreczkyetal:2013,Petreczkyetal:2017}. The realization problem studied in this paper is an extension of a 
result derived in \cite{Bencherki&Turkay&Akcay:2020} to MIMO setting and draws on the early LTV realization results
\cite{Shokoohi&Silverman:87}. A key role in the realization is played by a time-dependent switching sequence that 
partitions the state-space according to time, unlike many other works in the literature in which partitioning is 
state and/or input dependent. In jump Markov linear systems, for example, switching sequence evolves according to a
finite state Markov chain \cite{Costa&Fragoso&Marques:2000}. In a recent study \cite{Singh&Sznaier&Ljung:2018},
time-varying parameters were modeled  as an output vector of a finite-dimensional linear system driven by the 
input-output data (\ref{convolution}).

\subsection{Organization of the paper}

In Section~\ref{probform}, we formulate the SLS realization problem from finite Markov parameter sequences. 
Section~\ref{LTVhankel} builds on the realization problem for MIMO-LTV systems from input-output data studied in
\cite{Shokoohi&Silverman:87}. It is demonstrated that under mild assumptions on the dwell times of the discrete states one 
can extract an LTV realization that is topologically equivalent to the true SLS on every time interval of sufficient length. 

Section~\ref{discstate} starts by examining {\em stationary point set} of a Hankel matrix with fixed block and column dimensions 
and generated by the Markov parameters of the SLS. This is a special case of the sum-of-norms regularization method applied to 
the segmentation problem in SARX models \cite{Ohlsson&Ljung&Boyd:2010} in which a parameter vector replaces the Hankel matrix. 
The switches of the SARX model are estimated in \cite{Ohlsson&Ljung&Boyd:2010} by minimizing a quadratic norm of mismatch error 
regularized by a sum of parameter changes. On interval subsets of a stationary point set, an SLS behaves like an LTI system. 
It is demonstrated that every sufficiently long segment of a switching sequence contains an interval from the stationary point 
set if the discrete states satisfy a unimodality condition. This result paves the road for the proposal of a discrete state 
estimation algorithm based on clustering. Features used for clustering and clustering algorithms are briefly discussed. A feature
based on absolute sum of modal eigenvalues used for clustering extracts all discrete states up to arbitrary similarity 
transformations if each discrete state visits at least one segment and dwells long enough there.

Sections~\ref{switch-estI} and \ref{switch-estII} are devoted to estimation of the switching sequence. 
We present three schemes. The first scheme is based on correcting the state-space matrices estimated by the LTV
realization algorithm from the Markov parameters. It starts from a discrete state estimate and proceeds along the positive
and/or the negative real axes and continues until a switch or a switch pair is detected. This process is repeated until 
all switches, compatible with the dwell time requirements, are detected. The second scheme is based on matching the estimated 
and the true Markov parameters of the SLS system over a segment. This scheme too similarly starts from a discrete state estimate 
and proceeds along the positive and/or negative real axes until a switch or a switch pair is detected. The third scheme 
is also based on matching the Markov parameters, but it is a discrete optimization/hypothesis testing algorithm. The three
switch detection schemes operate on different dwell time and model structure requirements, but their dwell time 
requirements are weaker than that one needed to recover the discrete states.

The discrete state estimates in Section~\ref{discstate} are only similar to the true ones. If they will be 
used for predicting outputs to prescribed inputs, it is necessary to bring them into a common basis with 
only one similarity transformation chosen freely. By exploiting minimality of the discrete states and 
putting a mild restriction on the minimum dwell time, we present an elegant solution to this problem
in Section~\ref{basistransform}. In Section~\ref{hyperalg}, we put all derivations together to form a 
meta-algorithm. Sensitivity analysis of this meta-algorithm to amplitude bounded noise 
is performed, more specifically it is shown that small perturbations in the Markov parameters may only lead to small 
deviations in the estimates that vanish as noise amplitude diminishes. Time complexity of each stage in the meta-algorithm 
is studied in detail. A numerical example in Section~\ref{example} 
illustrates the derived results. Section~\ref{conclus} concludes the paper.

\section{Problem formulation}\label{probform}
In this paper, we consider a class of MIMO--LTV systems represented by the state-space equations
\begin{eqnarray}
	x(k+1) &=& A(k) x(k) +B(k) u(k), \label{ssx} \\
	y(k) &=& C(k) x(k) + D(k) u(k) \label{ssy}
\end{eqnarray}
where $u(k) \in \mathbb{R}^{m}$, $y(k) \in \mathbb{R}^{p}$, and $x(k) \in \mathbb{R}^{n}$, 
$k=1,\cdots,N$ are respectively the input, the output, and the state sequences. The state 
dimension $n$ is constant and assumed to be known.

Let $\mathbb{N}$ denote the set of positive integers and $\varphi$ be a switching sequence, that is, 
a map from $\mathbb{N}$ onto a finite set $\mathbb{S} =\left\{1,\cdots,\sigma \right\}$ with a fixed and known 
$\sigma \in \mathbb{N}$. Let ${\mathcal P}(k) = (A(k),  B(k),C(k),D(k))$ for $k=1,\cdots,N$. Since 
$\varphi(k) \in \mathbb{S}$ for all $k$, we define the collection of the discrete states by 
${\mathcal P}=\left\{{\mathcal P}_1,\cdots,{\mathcal P}_\sigma \right\}.$ The MIMO--LTV system 
(\ref{ssx})--(\ref{ssy}) with the state-space matrices changed 
by $\varphi$ is an SLS. The switching sequence $\varphi(k)$ partitions $[1\;\;N]$ to a collection of 
disjoint intervals $[1\;\;k_1)$, $[k_i\;\;k_{i+1}),\,\cdots,\,[k_{i^*}\;\;N]$ such that   
\begin{equation}\label{varphit}
	\varphi(k)  = \left\{ \begin{array}{l} \varphi(1), \qquad 1\leq k <k_1 \\ \varphi(k_i), 
		\qquad k_i \leq  k < k_{i+1} \\ 
		\varphi(k_{i^*}), \qquad k_{i^*} \leq k < N. \end{array} \right.
\end{equation}
Denoting this segmentation by $\chi$, we define the dwell times $\delta_i(\chi)$ and the minimum dwell time 
$\delta_*(\chi)$ by
\begin{equation}\label{mindwell}
	\delta_i(\chi) = k_{i+1}-k_{i} \;\;\;\mbox{and}\;\;\;  
	\delta_*(\chi) = \min_{1\leq i < i^*} \delta_i(\chi). 
\end{equation}
Thus, $\delta_i(\chi)$ is the waiting time of the discrete state active in the segment $[k_i,k_{i+1})$ 
and $\delta_*(\chi)$ is the smallest waiting time of all discrete states in the interval $[1\;\;k_{i^*})$. Set 
$\delta_0(\chi)=k_1-1$ and $\delta_{i^*}(\chi)=N-k_{i^*}$.

By introducing the {\em Markov parameters} and the {\em state transition matrix} associated with the 
{\em homogeneous part} of (\ref{ssx}) defined respectively by
\begin{equation}\label{defMarkov}
	h(k,\ell) = \left\{ \begin{array}{ll} C(k) \Phi(k,\ell+1) B(\ell), & k > \ell \\ D(k), 
		& k=\ell \\ 0, & k<\ell, \end{array} \right. 
\end{equation}
and
\begin{equation}\label{defsttran}
	\Phi(k,\ell) = \left\{ \begin{array}{ll} A(k-1) \, \cdots \, A(\ell), & k > \ell \\ I_n, & k=\ell
	\end{array} \right. 
\end{equation}
where $I_n \in \mathbb{R}^{n \times n}$ is the identity matrix, from (\ref{ssx})--(\ref{ssy}), we can
calculate the response of the system to any initial state $x(\ell)$, $\ell \geq 1$ and a prescribed input 
sequence $u(k)$, $k \geq 1$ as follows
\begin{equation}\label{yresp}
	y(k)=C(k) \Phi(k,\ell) x(\ell) + \sum_{j=\ell}^{k} h(k,j) u(j), \;\;\; \ell \leq k \leq N.
\end{equation}

Since the state-space matrices are changed by the switching sequence, the input-output trajectory of 
(\ref{ssx})--(\ref{ssy}) strongly depends on $\varphi$. In most works on SLSs, $\varphi$ 
is a typically state or/and input dependent signal whereas in the current work it is an unknown signal. 
The realization problem we study in this paper is described as follows. 

\begin{problem}\label{problem}
	Given a set of doubly indexed Markov parameters $h(k,\ell)$ of the MIMO--SLS model  
	(\ref{ssx})--(\ref{varphit}) for $1\leq k \leq N$ and $1 \leq \ell \leq k$ , estimate each 
	${\mathcal P}_j \in {\mathcal P}$, $j=1,\cdots,\sigma$ up to a similarity transformation 
	and $\varphi(k)$.
\end{problem}

In the course of developing a realization algorithm that solves Problem~\ref{problem}, 
we will impose some conditions on the system structure and the switching sequence.
We will also discuss the freedom in choosing similarity transformation.

\section{LTV realization from Markov parameters}\label{LTVhankel}

This section prepares the stage for a discrete-state estimation algorithm and three switch detection 
schemes presented in Sections~\ref{discstate}, \ref{switch-estI}, \ref{switch-estII}, and \ref{switch-estIII}. 
In this section, we will review the realization problem for MIMO-LTV systems from input-output data. Recall that 
the description of an LTV system in terms of state-space matrices is not unique: Two given realizations 
$(A,B,C,D)$ and $(\check{A},\check{B},\check{C},\check{D})$ have the same Markov parameters if they are 
{\em topologically equivalent}, i.e., there exists a bounded $T(k) \in \mathbb{R}^{n \times n}$ with bounded 
inverse such that for all 
$k \in \mathbb{N}$,
\begin{eqnarray*}
	\check{A}(k) = T(k+1) A(k) T^{-1}(k), \qquad \check{B}(k) &=& T(k+1) B(k), \\
	\check{C}(k) = C(k)T^{-1}(k), \qquad \check{D}(k) &=& D(k).
\end{eqnarray*}
We call $T(k)$ a {\em Lyapunov transformation}. As far as input-output behavior is 
concerned, it suffices to estimate the Markov parameters from input-output data. 

Let $q,r \in \mathbb{N}$ be two numbers to be fixed later. For all $k > r$, we define
the {\em Hankel matrices} by
\begin{equation}\label{gHankel}
	{\mathcal H}_{q,r}(k) = \left[ \begin{array}{ccc} h(k,k-1) & \cdots & h(k,k-r) \\  
		& \ddots & \vdots \\  & \cdots & h(k+q-1,k-r)   \end{array} \right]
\end{equation}
which can be factorized from \cite{Shokoohi&Silverman:87} as follows
\begin{equation}\label{defHqr}
	{\mathcal H}_{q,r}(k) = {\mathcal O}_q(k) \, {\mathcal R}_r(k-1)
\end{equation}
with ${\mathcal O}_q(k)$ and ${\mathcal R}_r(k-1)$ denoting the extended observability and the controllability 
matrices defined by
\begin{equation}\label{defobser}
	{\mathcal O}_q(k)=\left[\begin{array}{c} C(k) \\ C(k+1) \Phi(k+1,k) \\ \vdots \\  
		C(k+q-1) \Phi(k+q-1,k) \end{array}\right] \in \mathbb{R}^{qp \times n} 
\end{equation}
and
\begin{equation}\label{defcont}
	{\mathcal R}_r(k-1) = \left[B(k-1) \;\;\cdots \;\; \Phi(k,k-r+1) B(k-r)\right].
\end{equation}

The observability and the controllability Gramians for the LTV system (\ref{ssx})--(\ref{ssy}) are defined by
\begin{eqnarray}
     {\mathcal G}_{\rm o}(k;q) &=& {\mathcal O}_q(k)^T {\mathcal O}_q(k) \\
     {\mathcal G}_{\rm c}(k;r) &=& {\mathcal R}_r(k-1) {\mathcal R}_r^T(k-1)
\end{eqnarray}
If $A(k),B(k),C(k)$ are bounded matrices for all $k$, $(A(k),B(k),C(k))$ is said a {\em bounded realization}. 
A bounded realization $(A(k),B(k),C(k))$ is {\em uniformly observable} iff  $A(k)$ is
nonsingular and $\exists \, q \in \mathbb{N}$ such that ${\mathcal G}_{\rm o}(k;q) \geq \alpha_{\rm o} I_n$ for some 
$\alpha_{\rm o}>0$ and $\forall k$. This fact is a restatement of Lemma~2 in \cite{Shokoohi&Silverman:87} with changes
in the notation. Similarly, a bounded realization $(A(k),B(k),C(k))$ is {\em uniformly controllable} iff  
$A(k)$ is nonsingular and $\exists \, r \in \mathbb{N}$ such that ${\mathcal G}_{\rm c}(k;r) \geq \alpha_{\rm c} I_n$ 
for some $\alpha_{\rm c}>0$ and $\forall k$. This fact restates Lemma~1 in \cite{Shokoohi&Silverman:87} with 
appropriate changes in the notation. The following definition is adapted from \cite{Shokoohi&Silverman:87}.

\begin{definition}
	A bounded realization $(A(k),B(k),C(k))$ is uniform if it is uniformly observable and controllable.
\end{definition}

For a bounded realization $(A(k),B(k),C(k))$, the definitions of uniformly observable, uniformly controllable, 
and uniform  applies to all $k \in \mathbb{N}$. However, our observation interval is limited to $[1\;\;N]$. 
Moreover from (\ref{defobser}) and (\ref{defcont}), $k$ must satisfy the inequalities $k+q-1 \leq N$ and $k-r\geq 1$.
Hence, the above definitions can only be checked for all $k \in [r+1\;\;N-q+1]$. It is our best interest to pick
$q$ and $r$ as small as possible while ${\mathcal G}_{\rm o}(k;q)$ and ${\mathcal G}_{\rm c}(k;r)$ are positive-definite.  
For fixed $k$, ${\mathcal G}_{\rm o}(k;q)$ and ${\mathcal G}_{\rm c}(k;r)$ are non-decreasing functions of $q$ 
and $r$, i.e., ${\mathcal G}_{\rm o}(k;q_1) \geq {\mathcal G}_{\rm o}(k;q_2)$ iff  $q_1\geq q_2$ and 
${\mathcal G}_{\rm c}(k;r_1) \geq {\mathcal G}_{\rm c}(k;r_2)$ iff $r_1 \geq r_2$. 

Observability is a critical aspect in observer design and identification algorithms for SLSs. Several observability 
definitions for SLSs (mode observability, strong mode observability, observability, strong observability 
{\em etc.}) have appeared in hybrid systems literature. Checking observability for SLSs is formidable. For 
example, checking the observability of (\ref{ssx})--(\ref{varphit}) for all possible discrete state sequences of 
length $N$ can involve $O(\sigma^N)$ rank tests. To deal with this exponential complexity, 
in \cite{Elhamifiar&Petreczky&Vidal:2009} switching times were assumed to be separated by a minimum dwell time
that depends on $n$. Under this assumption, computational complexity was shown to reduce from 
$O(\sigma^N)$ to $O(\sigma^2)$. The minimum dwell time is not very restrictive since
rapidly changing modes may cause instability problems \cite{Bakoetal:2009a}.

Bounded realization assumption follows from the following.

\begin{assumption}\label{sysasmp}
	Every discrete state ${\mathcal P}_j \in {\mathcal P}$, $1 \leq j \leq \sigma$ is 
	bounded-input/bounded-output (BIBO) stable and has a Macmillan degree $n$. $\qed$
\end{assumption}

We also impose a mild restriction on the minimum dwell time as follows.

\begin{assumption}\label{varphiassmp}
	The switching sequence $\varphi$ satisfies $\delta_*(\chi) \geq n$. $\qed$
\end{assumption}

The following result was derived in \cite{Bencherki&Turkay&Akcay:2020}.

\begin{lemma}\label{lemreal}
	Suppose Assumptions~\ref{sysasmp}--\ref{varphiassmp} hold and $q,r\geq 2n$. Then, the MIMO-SLS model 
	(\ref{ssx})--(\ref{varphit}) is uniform on $[2n+1 \;\;N-2n+1]$. $\qed$
\end{lemma}

This lemma enables one to extract an LTV realization which is topologically equivalent to (\ref{ssx})--(\ref{varphit}) 
from the pairs $\{{\mathcal H}_{q,r}(k),{\mathcal H}_{q,r}(k+1)\}$, $k>r$. In fact, uniform realization 
property implies that ${\mathcal O}_q(k)$ and ${\mathcal R}_r(k-1)$ are full-rank matrices and the relationship 
${\rm rank}({\mathcal H}_{q,r}(k))=n$ holds from an application of the Sylvester inequality \cite{Kailath:80} to the 
product in (\ref{defHqr}). The extended observability and controllability matrices are then determined up to similarity
transformations from the SVDs of ${\mathcal H}_{q,r}(k)$ and ${\mathcal H}_{q,r}(k+1)$. In fact, let $J_{\uparrow}$ 
and $J_{\downarrow}$ denote the shift matrices of block-row up 
and down and $J_{\leftarrow}$ and $J_{\rightarrow}$ denote the block-column left and right shift matrices, respectively, defined by
\begin{eqnarray}
	J_{\uparrow} = \left[0_{(q-1)p\times p} \;\;I_{(q-1)p}\right], \;\;\;
	J_{\downarrow} &=& \left[I_{(q-1)p} \;\;0_{(q-1)p\times p}\right], \label{fac1} \\
	J_{\leftarrow} = \left[\begin{array}{l} 0_{m \times (r-1)m} \\ I_{(r-1)m} \end{array} \right],\;\;\;
	J_{\rightarrow} &=& \left[\begin{array}{l} I_{(r-1)m}\\ 0_{m \times(r-1)m} \end{array} \right] \label{fac2}
\end{eqnarray}
where $0_{m \times n}$ denotes the $m$ by $n$ matrix of zeros. Then, from the factorization formula 
(\ref{defHqr}) with $k$ and $k+1$ plugged in and (\ref{fac1})--(\ref{fac2}) we derive the following relations
\begin{eqnarray}
	J_{\uparrow} {\mathcal O}_q(k) &=& \left(J_{\downarrow} {\mathcal O}_q(k+1) \right) A(k) 
	= {\mathcal O}_{q-1}(k+1) A(k) \label{adet1} \\
	{\mathcal R}_r(k) J_{\leftarrow} &=& A(k)\left({\mathcal R}_r(k-1) J_{\rightarrow} \right)
	=  A(k) {\mathcal R}_{r-1}(k-1). \label{adet2} 
\end{eqnarray}
Hence, from (\ref{adet1}) and/or (\ref{adet2}) depending on if $q>2n$ and/or $r>2n$, we retrieve $A(k)$ from 
either/both of the formulas
\begin{eqnarray}
	A(k) &=& {\mathcal O}_{q-1}^{\dag} (k+1)J_{\uparrow} {\mathcal O}_q(k)
	= {\mathcal R}_r(k) J_{\leftarrow}{\mathcal R}_{r-1}^{\dag}(k-1) \label{Aeq}
\end{eqnarray}
where $X^\dag$ denotes the unique left or right-inverse of a given full-rank matrix $X$ defined as $(X^TX)^{-1}X^T$ 
or $X^T(XX^T)^{-1}$ depending on the context. We fix $q$ and $r$ as $q=2n+1$ and $r=2n$. Then, $k>2n$. 
For an upper bound on $k$, the maximum block-row index of ${\mathcal H}_{qr}(k+1)$ must satisfy $k+q \leq  N-2n+1$ 
or $k \leq N-4n$. Hence, $k^\prime \leq k \leq k^{\prime\prime}$ where $k^\prime=2n+1$ and $k^{\prime\prime}=N-4n$.

It remains to estimate ${\mathcal O}_q(k)$ and ${\mathcal O}_q(k+1)$. We apply SVD 
to ${\mathcal H}_{q,r}(k)$ and ${\mathcal H}_{q,r}(k+1)$
\begin{eqnarray}
	{\mathcal H}_{q,r}(k) &=& U_q(k) \Sigma(k) V_r^T(k) , \nonumber 
	\\[-1ex] \label{svdHqr} \\[-1ex]
	{\mathcal H}_{q,r}(k+1) &=& U_q(k+1) \Sigma(k+1) V_r^T(k+1) \nonumber
\end{eqnarray}
and let
\begin{eqnarray}
	\hat{\mathcal O}_q(k) =  U_q(k) \Sigma^{1/2}(k), \;\;\;
	\hat{\mathcal R}_r(k-1) &=&  \Sigma^{1/2}(k) V_r^T(k),  \label{Oqr} \\
	\hat{\mathcal O}_q(k+1)=  U_q(k+1) \Sigma^{1/2}(k+1), \;\;\; 
	\hat{\mathcal R}_r(k) &=& \Sigma^{1/2}(k+1) V_r^T(k+1). \label{Obqrp}
\end{eqnarray}
Up to similarity transformations $T(k)$ and $T(k+1)$, $\hat{\mathcal O}_q(k)$ and $\hat{\mathcal O}_q(k+1)$
provide estimates of ${\mathcal O}_q(k)$ and ${\mathcal O}_q(k+1)$,  i.e., 
${\mathcal O}_q(k)=\hat{\mathcal O}_q(k)T(k)$ and ${\mathcal O}_q(k+1)=\hat{\mathcal O}_q(k+1)T(k+1)$. Let
\begin{equation}\label{Aes}
	\hat{A}(k)=\left(J_{\downarrow} \hat{\mathcal O}_q(k+1) \right)^{\dag} J_{\uparrow} \hat{\mathcal O}_q(k)
\end{equation}
so that
\begin{equation}\label{topA}
	\hat{A}(k) = T(k+1) A(k) T^{-1}(k).
\end{equation}

The estimation of $B(k)$ and $C(k)$ is in order. To this end, let
\begin{equation}\label{CBdef}
	J_{\rm C} = \left[I_p \;\;O_{p \times (q-1)p} \right], \;\;\; J_{\rm B} = \left[\begin{array}{l} I_m \\ 
		0_{(r-1)m \times m} \end{array} \right]. 
\end{equation}
Then, from (\ref{defobser}) and (\ref{defcont})
\begin{equation}
C(k) = J_{\rm C} {\mathcal O}_q(k), \;\;\; B(k) = {\mathcal R}_r(k) J_{\rm B}.
\end{equation}
As the estimates of $C(k)$ and $B(k)$, we set
\begin{equation}\label{CBes} 
	\hat{C}(k) = J_{\rm C} \hat{\mathcal O}_q(k), \;\;\; \hat{B}(k)=\hat{\mathcal R}_r(k) J_{\rm B}. 
\end{equation}
Then, from $ \hat{\mathcal R}_r(k)=T(k+1) {\mathcal R}_r(k)$,
\begin{equation}\label{topCB}
	\hat{C}(k) = C(k) T^{-1}(k), \;\;\; \hat{B}(k) = T(k+1) B(k).
\end{equation}
Set $\hat{D}(k)=h(k,k)$. Then, from (\ref{topA}) and (\ref{topCB}) it follows that  
$\hat{\mathcal P}(k)=(\hat{A}(k),\hat{B}(k),\hat{C}(k),\hat{D}(k))$ is topologically equivalent to 
${\mathcal P}(k)$ on the interval $[k^\prime\;\;k^{\prime\prime}]$. 

If $T(k) =T(k+1)$, $\hat{A}(k)$ is similar to $A(k)$ denoted by the notation $\hat{A}(k) \sim A(k)$. It is
the case when $H_{q,r}(k)$ equals $H_{q,r}(k+1)$. For an LTI system, i.e., $\sigma=1$ in Assumption~\ref{sysasmp}, 
$\hat{A}(k) \sim A(k)$. The derivations in this section are outlined in Appendix as a subspace-based realization 
algorithm. The realization returned by Algorithm~1 is topologically equivalent to (\ref{ssx})--(\ref{varphit}). The 
quadruples returned are not similar to ${\mathcal P}(k)$, yet match the Markov parameters. Matching the 
discrete states will be achieved by modifying the state-space matrices as described in Section~\ref{switch-estI}. 

The LTV realization problem was studied in \cite{Majjietal:2010} from a rigid body dynamics perspective. 
The scheme presented above is based on single input-output data batches. If an ensemble of input-output 
data are available, state-space models can be identified without constraints on time-variations of 
the Markov parameters \cite{Verhaegen&Yu:95,Liu:1997}. The derivations above are summarized in the following
result.

\begin{theorem}\label{realprop}
	Consider Algorithm~1 with the noiseless Markov parameters of (\ref{ssx})--(\ref{varphit}) 
	satisfying Assumptions~\ref{sysasmp}--\ref{varphiassmp}. Then, Algorithm~1 returns a realization 
	topologically equivalent to (\ref{ssx})--(\ref{varphit}) on $[k^\prime\;\;k^{\prime\prime}]$. $\qed$
\end{theorem}

Notice that ${\mathcal H}_{q,r}(k+1)$ is formed from the $4n$ Markov parameters and a submatrix of 
${\mathcal H}_{q,r}(k)$. In fact, partition $\mathcal{H}_{q,r}(k)$ as follows
\begin{equation}\label{gHankel12}
	{{\cal H}_{q,r}}(k) = \left[ {\begin{array}{*{20}{c}}{h(k,k - 1)\;}& \cdots &{h(k,k - r)}\\
			{{\cal H}_{q,r}^\prime (k)}&{}& \vdots \\ {}&{}&{h(k + q - 1,k - r)}
	\end{array}} \right]
\end{equation}
and observe that
\begin{eqnarray}\label{gHankel2}
	{\mathcal H}_{q,r}(k+1)= \left[ \begin{array}{ccc} h(k+1,k) &  &  \\  
		\vdots	&  & {\mathcal H}_{qr}^\prime(k) \\ h(k+q,k) & \cdots & h(k+q,k-r+1)   
	\end{array} \right].
\end{eqnarray}
Algorithm~1 uses $qr+(q+r-1)(k^{\prime\prime}-k^{\prime}+1)$ Markov parameters if it computes $\hat{\mathcal P}(k)$ 
for all $k \in [k^\prime\;\;k^{\prime\prime}]$ and  if it computes for one $k$, $qr+q+r-1$ Markov parameters. 
Plug $q=2n+1$ and $r=2n$ in: $(4N-20n+2)n \approx 4Nn$ and $4n^2+6n \approx 4n^2$. If $\hat{\mathcal P}(k)$ is 
computed at every $2n+1$th sample,  $(4n^2+6n)(k^{\prime\prime}-k^\prime+1)/(2n+1) \approx 2Nn$ Markov parameters 
roughly will be used since ${\cal H}_{2n+1,2n}(k)$ and ${\cal H}_{2n+1,2n}(k+2n)$ share no common entries. Hence, 
more than 50\% of the Markov parameters need to be stored if calculations are done more frequently than $O(n)$. 
The algorithms developed in the sequel will need all Markov parameters to be stored. Time complexity of
Algorithm~1, however, depends on the frequency of calculations as the following analysis demonstrates.

\subsection{Time complexity of Algorithm~1}\label{complexity1}
In Algorithm~1, calculation of the SVDs in (\ref{svdHqr}) is computationally the most expensive step without 
exploiting the block matrix structure and the nesting property in (\ref{gHankel12})--(\ref{gHankel2}). According 
to \cite{Golub&VanLoan:1989}, the best algorithms for SVD computation of an $M \times L$ matrix have time complexity 
$O(c_1 M^2 L + c_2  L^3)$ where $c_1,c_2>0$  are absolute constants which are $4$ and $22$ for an algorithm called R-SVD. 
Recall that $q=2n+1$ and $r=2n$, from $M=qp$ and $L=rm$. Each SVD in (\ref{svdHqr}) has 
then a time complexity $O(poly(n))$ where $poly(n)$ is a third order polynomial in $n$. Pseudo-inverse of an $M \times L$ 
($M>L$) matrix calculated by SVD has complexity $O(L M^2)$. Hence, with $M=(q-1)p$ and $L=n$, 
$\left(J_{\downarrow} \hat{\mathcal O}_q(k+1) \right)^{\dag}$ in (\ref{Aes}) has complexity $O(n^3)$. Multiplication 
of two matrices $X \in \mathbb{R}^{M \times K}$ and $Y \in \mathbb{R}^{K \times L}$ has complexity $O(MKL)$ when 
calculated directly. Hence, with $M=n$, $K=(q-1)p$, $L=n$ we get $O(poly(n))$ for the complexity of multiplication 
$\left(J_{\downarrow} \hat{\mathcal O}_q(k+1) \right)^{\dag}$ times $J_{\uparrow} \hat{\mathcal O}_q(k)$. Adding up 
we get $O(poly(n))$ for the complexity of $\hat{A}(k)$. Since $J_{\rm B},\hat{\mathcal R}_r(k),J_{\rm C},
\hat{\mathcal O}_q(k)$ are $n \times rm$, $rm \times m$, $p \times qp$, $qp \times n$ matrices respectively,  
applying the complexity result to the matrix multiplications in (\ref{CBes}) we derive $O(poly(n))$ for the 
complexities of $\hat{B}$ and $\hat{C}$ where $poly(n)$ is a second order polynomial in $n$. It follows that 
$\hat{\mathcal P}(k)$ has complexity $O(poly(n))$ where $poly(n)$ is a third order polynomial. Denote the number 
of the elements in $\kappa$ by $|\kappa|$. Then, Algorithm~1 has time complexity $O(|\kappa|poly(n))$. Hence, 
Algorithm~1 subject to Assumption~\ref{varphiassmp} has a worst-case time complexity $O(Npoly(n))$ where $poly(n)$ 
is a second order polynomial since $|{\mathcal K}|=O(Nn^{-1})$ for evenly distributed switches. If ${\mathcal K}$ 
is a singleton, then time complexity of Algorithm~1 will be $O(poly(n))$ with $poly(n)$ a third order polynomial 
in $n$. These calculations took into consideration only $n$ and $N$ since $p$ and $m$ are fixed small numbers. There 
is also a similar concept called space complexity which is usually weaker than time complexity. In the setting of 
this paper, for example, the SVD has space complexity $O(n^2)$. 

\subsection{Robustness of Algorithm~1}

Suppose that the Markov parameters in (\ref{defMarkov}) are corrupted by `unknown-but-bounded' type noise
\begin{equation}\label{corruptMarkov}
\tilde{h}(k,\ell)=h(k,\ell)+e(k,\ell), \qquad \ell \leq k\;\;\;\mbox{and} \;\;\;k^\prime \leq k \leq k^{\prime\prime}
\end{equation}
where $\|e(k,\ell)\|_F \leq \varepsilon$ with $\|\cdot\|_F$ denoting the Frobenius norm of a given matrix defined as 
the square root of the sum of its squared elements. Our goal is to study how this noise affects  the estimates 
$\hat{\mathcal P}(k)$, $k \in [k^{\prime}\;\;k^{\prime\prime}]$ for small $\varepsilon>0$. 

Let $\tilde{\mathcal H}_{2n+1,2n}(k)$ be the Hankel matrix obtained by populating ${\mathcal H}_{2n+1,2n}(k)$ 
in (\ref{gHankel}) with $\tilde{h}(k,\ell)$. Split $\tilde{\mathcal H}_{2n+1,2n}(k)$
\begin{equation}\label{corruptHankel}
\tilde{\mathcal H}_{2n+1,2n}(k)={\mathcal H}_{2n+1,2n}(k)+E(k).
\end{equation}
Robustness of Algorithm~1 to unknown-but-bounded type noise follows from the following result.

\begin{theorem}\label{Alg1robustness}
Consider (\ref{ssx})--(\ref{varphit}) with the noisy Markov parameters in (\ref{corruptMarkov}).
Suppose that Assumptions~\ref{sysasmp}--\ref{varphiassmp} hold. Let 
$\tilde{\mathcal P}(k) =(\tilde{A}(k),\tilde{B}(k),\tilde{C}(k),\tilde{D}(k))$
and $\hat{\mathcal P}(k)=(\hat{A}(k),\hat{B}(k),\hat{C}(k),\hat{D}(k))$ denote the models estimated by Algorithm~1 
in the noisy and noiseless cases. Then, there exists a realization ${\mathcal P}^\varepsilon(k) =(A^\varepsilon(k),B^\varepsilon(k),C^\varepsilon(k),D^\varepsilon(k))$ that is topologically equivalent to
$\tilde{\mathcal P}(k)$ satisfying $\|A^\varepsilon(k)-\hat{A}(k)\|_F \leq c \varepsilon$, 
$\|B^\varepsilon(k)-\hat{B}(k)\|_F \leq c \varepsilon$, $\|C^\varepsilon(k)-\hat{C}(k)\|_F \leq c \varepsilon$,
$\|D^\varepsilon(k)-\hat{D}(k)\|_F \leq c \varepsilon$ for all $\varepsilon \leq \varepsilon_*$ and some 
$c,\varepsilon_*>0$. The Lyapunov transformation $S(k)$ mapping $\tilde{\mathcal P}(k)$ to ${\mathcal P}^\varepsilon(k)$ 
is asymptotically orthogonal, {\em that is}, $S^T(k)S(k) \rightarrow I_n$ as $\varepsilon \rightarrow 0$. $\qed$
\end{theorem}

{\em Proof.} Our proof is based on perturbation techniques employed in subspace identification, but far more difficult 
due to the fact Lyapunov transformations replace time-invariant similarity transformations in the current setup. 
See, Lemma~4 in Section~III.A of \cite{McKelvey&Akcay&Ljung:96}. Due to noise not only the singular values of 
${\mathcal H}_{2n+1,2n}(k)$, but also its singular subspaces will be perturbed which will cause changes in the 
state-space matrices. The perturbation term in (\ref{corruptHankel}) is bounded in the Frobenius norm as follows 
$$
\|E(k)\|_F \leq \sqrt{2n(2n+1)} \, \varepsilon < 4n \varepsilon. 
$$
In our perturbation analysis, we assume that $\varepsilon$ can be made as small as we are pleased. Write the first SVD in
(\ref{svdHqr}) 
\begin{equation}\label{svdHqr2}
 	{\mathcal H}_{2n+1,2n}(k) = \left[U_{2n+1}(k)\;\;U_e(k)\right] 
 	\left[\begin{array}{cc} \Sigma(k) & 0 \\ 0 & 0 \end{array}\right] 
 	\left[\begin{array}{c} V_{2n}^T(k) \\ V_e^T(k)\end{array}\right]
\end{equation} 
by completing the singular subspaces with $U_e(k)$ and $V_e(k)$. Let 
\begin{equation}\label{F1}
\left[\begin{array}{c} U_{2n+1}^T(k) \\ U_e^T(k)\end{array}\right]E(k) \left[V_{2n}(k)\;\;V_e(k)\right]=
\left[\begin{array}{cc} X_{11}(k) & X_{12}(k) \\ X_{21}(k) & X_{22}(k) \end{array}\right]=X(k).	
\end{equation}
From (\ref{F1}), notice that $\|X(k)\|_F =\|E(k)\|_F <  4n \varepsilon$. Moreover, $\|X_{st}(k)\|_2 
\leq \|X_{st}(k)\|_F$, $1 \leq s,t \leq 2$ since the Euclidean norm is dominated by the Frobenius norm. 
Let ${\sigma}_j(k)$ denote the $j$th largest singular value of ${\mathcal H}_{2n+1,2n}(k)$. Then, 
\begin{equation}\label{F2}
\delta \stackrel{\Delta}{=} {\sigma}_n(k)-\|X_{11}(k)\|_2-\|X_{22}(k)\|_2 >  {\sigma}_n(k) -8n \varepsilon.
\end{equation}
Choose $\varepsilon_0>0$ such that ${\sigma}_n(k)  > 16n \varepsilon_0$ for all $k$. This is possible since there are
only $\sigma$ discrete states. Then, from (\ref{F2}) we get $\delta > \frac{1}{2} \, {\sigma}_n(k)$ for all 
$\varepsilon\leq \varepsilon_0$. We also have
\begin{equation}\label{karpuzmu}
\frac{\|X_{21}(k) \;\;X^T_{12}(k)\|_F}{\delta} < \frac{4n \varepsilon}{\frac{1}{2} \, 
	{\sigma}_n(k)}  < \frac{1}{2}, \qquad \forall \, \varepsilon \leq \varepsilon_0.
\end{equation}
Running Algorithm~1 with the corrupted Markov parameters (\ref{corruptMarkov}) accounts for replacing the first SVD 
in (\ref{svdHqr}) by
\begin{equation}\label{corruptsvdHqr}
	\tilde{\mathcal H}_{2n+1,2n}(k) = \left[\tilde{U}_{2n+1}(k)\;\;\tilde{U}_e(k)\right] \left[\begin{array}{cc} 
		\tilde{\Sigma}(k) & 0 \\ 0  & \tilde{\Sigma}_e(k) \end{array}\right] \left[\begin{array}{c} \tilde{V}_{2n}^T(k) \\ \tilde{V}_e^T(k)\end{array}\right].
\end{equation} 
Simply plug $k+1$ in (\ref{corruptsvdHqr}) for the second SVD in (\ref{svdHqr}). Now, the chain of inequalities 
(\ref{karpuzmu}) implies that there exist matrices $P_e(k)$ and $Q_e(k)$ satisfying
\begin{equation}\label{dondurmaci}
\left\| \left[\begin{array}{c} Q_e(k) \\ P_e(k) \end{array}\right] \right\|_F \leq 2 \frac{\varepsilon}{\delta}
\end{equation}
such that $range\left(V_{2n}(k)+V_e(k) Q_e(k)\right)$ and $range\left(U_{2n+1}(k)+U_e(k) P_e(k)\right)$ are a pair of 
singular subspaces for the perturbed Hankel matrix $\tilde{\mathcal H}_{2n+1,2n}(k)$. See, for example Theorem~8.3.5 
in \cite{Golub&VanLoan:1989}. Since the range spaces are equal and $U_{2n+1}(k)$ is of full rank, there exists a unique
nonsingular matrix $\tilde{T}(k)$ such that
\begin{equation}\label{konik}
\tilde{U}_{2n+1}(k) = \left(U_{2n+1}(k)+U_e(k) P_e(k)\right) \tilde{T}(k).
\end{equation}
A similar expression holds for $\tilde{U}_{2n+1}(k+1)$ by plugging $k+1$ in (\ref{konik}). Let 
$\tilde{\mathcal O}_{2n+1}(k)=\tilde{U}_{2n+1}(k) \tilde{\Sigma}^{1/2}(k)$. Then, with 
$\tilde{\Delta}_e(k)=U_e(k) P_e(k) \Sigma^{1/2}(k)$ and $S(k)=\Sigma^{-1/2}(k)\tilde{T}(k)\tilde{\Sigma}^{1/2}(k)$,
from $\hat{\mathcal O}_{2n+1}(k)=U_{2n+1}(k) \Sigma^{1/2}(k)$ and (\ref{konik}) we get 
$\tilde{\mathcal O}_{2n+1}(k)=\left(\hat{\mathcal O}_{2n+1}(k)+\tilde{\Delta}_e(k)\right)S(k)$. Thus, 
\begin{eqnarray*}\label{Aes2}
	\tilde{A}(k) &\stackrel{\Delta}{=}& \left(J_{\downarrow} \tilde{\mathcal O}_{2n+1}(k+1) \right)^{\dag} 
	J_{\uparrow} \tilde{\mathcal O}_{2n+1}(k) \\
	&=& S^{-1}(k+1) \left(J_{\downarrow} \left(\hat{\mathcal O}_{2n+1}(k+1)+\tilde{\Delta}_e(k+1)\right) \right)^{\dag} J_{\uparrow} \left(\hat{\mathcal O}_{2n+1}(k)+\tilde{\Delta}_e(k)\right)S(k)
\end{eqnarray*}
and therefore
\begin{equation}\label{LSrobust}
S(k+1)\tilde{A}(k) S^{-1}(k) = \left(J_{\downarrow} \left(\hat{\mathcal O}_{2n+1}(k+1)
+\tilde{\Delta}_e(k+1)\right) \right)^{\dag} J_{\uparrow}\left(\hat{\mathcal O}_{2n+1}(k)+\tilde{\Delta}_e(k)\right). 
\end{equation}
The right-hand side of (\ref{LSrobust}) is the least-squares solution of the inconsistent equations
\begin{equation}\label{lambacimi}
\left(J_{\downarrow}\hat{\mathcal O}_{2n+1}(k+1)+J_{\downarrow}\tilde{\Delta}_e(k+1)\right) A^\sharp(k) = 
J_{\uparrow}\hat{\mathcal O}_{2n+1}(k)+J_{\uparrow}\tilde{\Delta}_e(k)
\end{equation}
The perturbations on the left and right-hand sides of (\ref{lambacimi}) are bounded from (\ref{dondurmaci}) as follows
\begin{equation*}
\|J_{\downarrow}\tilde{\Delta}_e(k+1)\|_F \leq \|\tilde{\Delta}_e(k+1)\|_F =\| P_e(k) \Sigma^{1/2}(k)\|_F
\leq {\sigma}_1(k) \|P_e(k)\|_F \leq 2 \frac{{\sigma}_1(k)}{\delta} \varepsilon 
< 4 \mu_{\Sigma^{1/2}(k)} \varepsilon 
\end{equation*}
and $\| J_{\uparrow}\tilde{\Delta}_e(k)\|_F \leq 4 \mu_{\Sigma^{1/2}(k)} \varepsilon$ where $\mu_{\Sigma^{1/2}(k)}
\stackrel{\Delta}{=} {\sigma}_1(k)/{\sigma}_n(k)$ is the condition number of ${\Sigma}^{1/2}(k)$. 
If $\varepsilon=0$, then $A^\sharp(k)$ equals to $\hat{A}(k)$ in (\ref{Aes}) and if $\varepsilon$ is small, say 
$\varepsilon \leq \varepsilon_1$, $A^\sharp(k)$ satisfies $\|A^\sharp(k)-\hat{A}(k) \|_F \leq c_1 \varepsilon$ for 
some $c_1>0$. See, for example Theorem~5.3.1 in \cite{Golub&VanLoan:1989}. Let 
$\varepsilon_2=\min\{\varepsilon_0,\varepsilon_1\}$. 
Then, we have shown that
\begin{equation}
\|S(k+1)\tilde{A}(k) S^{-1}(k)-\hat{A}(k)\|_F \leq c_1 \varepsilon, \qquad \forall \, \varepsilon \leq \varepsilon_2. 
\end{equation}
The rest of the perturbed state-space matrices are calculated similarly starting with $\tilde{C}(k)$ as follows
\begin{equation*}
\tilde{C}(k) \stackrel{\Delta}{=} J_{\rm C} \tilde{\mathcal O}_{2n+1}(k)=J_{\rm C} 
\left(\hat{\mathcal O}_{2n+1}(k)+\tilde{\Delta}_e(k)\right) S(k).
\end{equation*}
Thus,
\begin{equation}
\|\tilde{C}(k)S^{-1}(k)-\hat{C}(k)\|_F \leq c_2 \varepsilon, \qquad \forall \, \varepsilon \leq \varepsilon_2.  
\end{equation}
Let $\tilde{D}(k) \stackrel{\Delta}{=}\tilde{h}(k,k)$. Since $\hat{D}(k)=h(k,k)$, 
\begin{equation}
\|\tilde{D}(k)-\hat{D}(k)\|_F=\|e(k,k)\|_F \leq \varepsilon.
\end{equation}
Let $\tilde{B}(k-1) \stackrel{\Delta}{=} \tilde{\mathcal R}(k-1) J_B$. From (\ref{corruptsvdHqr}), we first derive 
an expression for $\tilde{V}_{2n}^T(k)$ as follows
\begin{equation}
\tilde{V}_{2n}^T(k)=\tilde{\Sigma}^{-1}(k)\tilde{U}_{2n+1}^T(k) \tilde{\mathcal H}_{2n+1,2n}(k).
\end{equation}
Then, from the first equation in (\ref{Obqrp})
\begin{eqnarray}
\tilde{B}(k-1) &=& \tilde{\Sigma}^{1/2} (k) \tilde{V}_{2n}^T(k) J_{\rm B}=\tilde{\Sigma}^{-1/2}(k)
\tilde{U}_{2n+1}^T(k) \tilde{\mathcal H}_{2n+1,2n}(k)J_{\rm B} \nonumber 
\\[-1ex] \label{inek} \\[-1ex] 
&=& \tilde{\Sigma}^{-1/2}(k)\tilde{U}_{2n+1}^T(k) {\mathcal H}_{2n+1,2n}(k)J_{\rm B} +
 \tilde{\Sigma}^{-1/2}(k)\tilde{U}_{2n+1}^T(k) E(k) J_{\rm B} \nonumber
\end{eqnarray}
where we used (\ref{corruptHankel}). Let $\tilde{\sigma}_j(k)$ denote the $j$th largest singular value of 
$\tilde{\mathcal H}_{2n+1,2n}(k)$. We bound the second term in (\ref{inek})
\begin{eqnarray}
\|\tilde{\Sigma}^{-1/2}(k)\tilde{U}_{2n+1}^T(k) E(k) J_{\rm B}\|_F &\leq& \|\tilde{\Sigma}^{-1/2}(k)\tilde{U}_{2n+1}^T(k) 
E(k)\|_F \leq \|\tilde{\Sigma}^{-1/2}(k)\| \|\tilde{U}_{2n+1}^T E(k) \|_F \nonumber
\\[-1ex] \label{maymun} \\[-1ex]
&\leq& \tilde{\sigma}_n^{-1}(k)\|E(k)\|_F \leq 4n \tilde{\sigma}_n^{-1}(k) \varepsilon \nonumber
\end{eqnarray}
where $\|\cdot\|$ denotes the spectral norm. From Corollary~8.3.2 in \cite{Golub&VanLoan:1989}, note that
\begin{equation}\label{biktimbe0}
|\tilde{\sigma}_n(k)-{\sigma}_n(k)| \leq \|E(k)\| < 4n \varepsilon.
\end{equation}
Hence, if $\varepsilon \leq \varepsilon_2$ from ${\sigma}_n(k) > 16n \varepsilon_0$
\begin{equation}\label{biktimbe}
\tilde{\sigma}_n(k) \geq {\sigma}_n(k) -4n \varepsilon \geq \frac{3}{4}{\sigma}_n(k). 
\end{equation}
Thus, from (\ref{maymun}) and (\ref{biktimbe}) if $\varepsilon \leq \varepsilon_2$
\begin{equation}\label{B1x}
\|\tilde{\Sigma}^{-1/2}(k)\tilde{U}_{2n+1}^T E(k) J_{\rm B}\|_F < 6n {\sigma}_n^{-1}(k) \varepsilon
\end{equation}
Split the first term in the second equation of (\ref{inek}) as 
\begin{eqnarray*}
\tilde{\Sigma}^{-1/2}(k)\tilde{U}_{2n+1}^T(k) {\mathcal H}_{2n+1,2n}(k)J_{\rm B} &=&\tilde{\Sigma}^{-1/2}(k)\tilde{T}^T(k)
U_{2n+1}^T(k){\mathcal H}_{2n+1,2n}(k)J_{\rm B} \\
&{}& + \tilde{\Sigma}^{-1/2}(k)\tilde{T}^T(k){\Sigma}^{-1/2}(k)\tilde{\Delta}_e^T(k){\mathcal H}_{2n+1,2n}(k+1)J_{\rm B} \\
&=& \tilde{\Sigma}^{-1/2}(k)\tilde{T}^T(k) \Sigma^{1/2}(k) \hat{B}(k-1)+
\tilde{\Sigma}^{-1}(k)S^T(k)\tilde{\Delta}_e^T(k){\mathcal H}_{2n+1,2n}(k+1)J_{\rm B}.
\end{eqnarray*}
If $\varepsilon \leq \varepsilon_2$, the second term above may be bounded from (\ref{biktimbe}) and (\ref{dondurmaci}) 
as follows
\begin{eqnarray}
\|\tilde{\Sigma}^{-1}(k)S^T(k)\tilde{\Delta}_e^T(k){\mathcal H}_{2n+1,2n}(k+1)J_{\rm B}\|_F &\leq& 
\tilde{\sigma}_n^{-1}(k) \|{\mathcal H}_{2n+1,2n}(k)\| \|J_{\rm B}\| \|\tilde{\Delta}_e(k) S(k)\|_F \nonumber \\
&\leq& \frac{\sigma_1(k)}{\tilde{\sigma}_n(k)} \|\tilde{\Delta}_e(k)\|_F \|S(k)\| \leq \frac{4}{3} \mu_{\Sigma^{1/2}(k)} 
\sigma_1(k) \|P_e(k)\|_F \|S(k)\| \nonumber
\\[-1.5ex] \label{sapsal}  \\[-1.5ex]
&\leq & \frac{4}{3} \mu_{\Sigma^{1/2}(k)} \sigma_1(k) \frac{4\varepsilon}{\sigma_n(k)} \|S(k)\|< 
6 \mu_{\Sigma^{1/2}(k)}^2 \|S(k)\| \varepsilon. \nonumber
\end{eqnarray}
It remains to bound $\|S(k)\|$. If $\varepsilon \leq \varepsilon_2$, from the definition of $S(k)$ and (\ref{biktimbe0})
\begin{equation}\label{sapsalak}
\|S(k)\| \leq \sigma_n^{-1}(k) \tilde{\sigma}_1(k) \|\tilde{T}(k)\| \leq \frac{\sigma_1(k)+4n\varepsilon}{\sigma_n(k)}
\|\tilde{T}(k)\| < \frac{\sigma_1(k)+\frac{1}{4}\sigma_n(k)}{\sigma_n(k)} \|\tilde{T}(k)\| <
2 \mu_{\Sigma^{1/2}(k)} \|\tilde{T}(k)\|.
\end{equation}
Next, we derive a bound on $\|\tilde{T}(k)\|$. To this end, we multiply the left and right-hand sides of (\ref{konik})
from left with $\tilde{U}^T_{2n+1}(k)$ and recall that $\tilde{U}_{2n+1}(k)$, $U_{2n+1}(k)$, $U_e(k)$ are unitary
matrices and $U_{2n+1}(k) \perp U_e(k)$. Thus, 
\begin{equation}\label{tyr}
I_n-\tilde{T}^T(k)\tilde{T}(k)=\tilde{T}^T(k)P_e^T(k) P_e(k) \tilde{T}(k).
\end{equation}
Since the right-hand side of (\ref{tyr}) is positive semi-definite, $\|\tilde{T}(k)\| \leq 1$. Then, from (\ref{sapsal}) 
and (\ref{sapsalak})
\begin{equation}\label{B2x}
\|\tilde{\Sigma}^{-1}(k)S^T(k)\tilde{\Delta}_e^T(k){\mathcal H}_{2n+1,2n}(k+1)J_{\rm B}\|_F < 12  
\mu_{\Sigma^{1/2}(k)}^3 \varepsilon.
\end{equation}
In the last step, we perform a splitting again as follows
\begin{eqnarray}
\tilde{\Sigma}^{-1/2}(k)\tilde{T}^T(k) \Sigma^{1/2}(k) \hat{B}(k-1) &=& \tilde{\Sigma}^{-1/2}(k)\tilde{T}^{-1}(k) 
\Sigma^{1/2}(k) \hat{B}(k-1) \nonumber
\\[-1.5ex] \label{vuvuzela}\\[-1.5ex]
&{}& +\tilde{\Sigma}^{-1/2}(k)\left(\tilde{T}^T(k)\tilde{T}(k)-I_n\right)\tilde{T}^{-1}(k) \Sigma^{1/2}(k) 
\hat{B}(k-1)\nonumber
\end{eqnarray}
where we recognize the relation $\tilde{\Sigma}^{-1/2}(k)\tilde{T}^{-1}(k) \Sigma^{1/2}(k) 
\hat{B}(k-1)=S^{-1}(k)\hat{B}(k-1)$.
We will bound now the second term in (\ref{vuvuzela}). To this end, multiplying (\ref{tyr}) from left by 
$\tilde{T}^{-T}(k)$ and from right by $\tilde{T}^{-1}(k)$ we get
\begin{equation*}
\tilde{T}^{-T}(k)\tilde{T}^{-1}(k)=I_n+P_e^T(k) P_e(k).
\end{equation*}
Then, if $\varepsilon \leq \varepsilon_2$ from (\ref{dondurmaci})
\begin{equation}\label{jungle1}
\|\tilde{T}^{-1}(k)\|^2 \leq \|I_n+P_e^T(k) P_e(k)\| \leq \|I_n\|+\|P_e(k)\|^2 < 1+4 \sigma_n^{-1}(k) \varepsilon
<1+\frac{\varepsilon}{4n\varepsilon_0} \leq \frac{5}{4}.
\end{equation}
Next, from (\ref{tyr})
\begin{equation}\label{jungle2}
\|\tilde{T}^T(k)\tilde{T}(k)-I_n\| \leq \|\tilde{T}(k)\|^2 \|P_e(k)\|_2^2 < \frac{16\varepsilon^2}{\sigma_n^2(k)}.
\end{equation}
Hence, from (\ref{biktimbe}), (\ref{jungle1}), and (\ref{jungle2}) if $\varepsilon \leq \varepsilon_2$,
\begin{eqnarray}
\|\tilde{\Sigma}^{-1/2}(k)\left(\tilde{T}^T(k)\tilde{T}(k)-I_n\right)\tilde{T}^{-1}(k) \Sigma^{1/2}(k) 
\hat{B}(k-1)\|_F \leq \frac{\sigma_1(k)}{\tilde{\sigma}_n(k)} \|\tilde{T}^T(k)\tilde{T}^{-1}(k)-I_n\|\|\tilde{T}^{-1}(k)\|
\|\hat{B}(k-1)\|_F \nonumber 
\\[-1.5ex]  \label{domuz}  \\[-1.5ex] 
\leq \frac{4}{3} \mu_{\Sigma^{1/2}(k)}\frac{16\varepsilon^2}{\sigma_n^2(k)} \sqrt{\frac{5}{4}} 
\max_k \|\hat{B}(k-1)\|_F < \frac{24}{\sigma_n^2(k)} (\varepsilon \max_k \|\hat{B}(k-1)\|_F) \, \varepsilon. \nonumber
\end{eqnarray}
Since $\hat{\mathcal P}(k)$ is topologically equivalent to ${\mathcal P}(k)$, the Lyapunov transformation mapping 
${\mathcal P}(k)$ to $\hat{\mathcal P}(k)$ is uniformly bounded on compact time intervals. Since $B(k)$ is bounded for 
all $k$, $\|\hat{B}(k-1)\|_F$ will be bounded above by some finite $\gamma$ on $[k^\prime\;\;k^{\prime}]$. 
Choose $0<\varepsilon_3 \leq \varepsilon_2$ satisfying $\varepsilon_3 \gamma < 1$. Then, from  (\ref{domuz}) 
if $\varepsilon \leq \varepsilon_3$
\begin{equation}\label{Bx3}
	\|\tilde{\Sigma}^{-1/2}(k)\left(\tilde{T}^T(k)\tilde{T}^{-1}(k)-I_n\right)\tilde{T}^{-1}(k) \Sigma^{1/2}(k) 
	\hat{B}(k-1)\|_F < \frac{24}{\sigma_n^2(k)} \varepsilon. 
\end{equation}
Let $c_3^\prime=6n\sigma_n^{-1}(k)+12 \mu_{\Sigma^{1/2}(k)}^3+24 \sigma_n^{-2}(k)$. Then, from (\ref{inek}), (\ref{B1x}), 
(\ref{B2x}), and (\ref{Bx3}) if $\varepsilon \leq \varepsilon_3$
\begin{equation}\label{oplk}
\|\tilde{B}(k-1)-S^{-1}(k)\hat{B}(k-1)\|_F \leq c_3^\prime \varepsilon.
\end{equation}
Since the derived bounds are uniform in $k$, we may substitute $k+1$ in place of $k$. Furthermore, as 
$\varepsilon \rightarrow 0$
\begin{equation*}
S^T(k) S(k) =\tilde{\Sigma}^{1/2}(k) \tilde{T}^T(k) \tilde{T}(k)\Sigma^{-1/2}(k) \rightarrow I_n.
\end{equation*}
Thus, $\|S^{-1}(k)\| \rightarrow 1$ as $\varepsilon \rightarrow 0$ and if necessary reducing $\varepsilon_3$ to 
$\varepsilon_*>0$ and increasing $c_3^\prime$ to $c_3<\infty$ we derive 
\begin{equation}\label{cambaz}
	\|S(k+1)\tilde{B}(k)-\hat{B}(k)\|_F \leq c_3 \varepsilon
\end{equation}
if $\varepsilon \leq \varepsilon_*$ by reorganizing (\ref{oplk}) and plugging $k+1$ in.  Let $c=\max\{c_1,c_2,c_3\}$
and note that $S(k)$ is the sought Lyapunov transformation. We took a long path to derive (\ref{cambaz}). If we had
taken a shorter path by considering the perturbed range space of $V_{2n}(k)$, this would introduce 
another matrix $\tilde{T}^\prime(k)$ without leading to a Lyapunov transformation. 
$\qedwhite$

A close examination of the proof shows that $S(k)$ has some uniqueness properties up to $n$ sign changes, 
but this property will not be needed in the subsequent analyses.

\section{Discrete state estimation}\label{discstate}

Algorithm~1 delivers a time-varying model that is only topologically equivalent to (\ref{ssx})--(\ref{ssy}) on a
given subset of $[k^\prime\;\;k^{\prime\prime}]$. Further processing of this model is necessary to reveal the discrete 
states and the switching sequence. If $\varphi$ is an arbitrary signal, there is no hope to recover ${\mathcal P}$ 
and $\varphi$ from the LTV model since $T(k)$ may possibly be a completely arbitrary time-varying matrix. When the dwell 
times of (\ref{ssx})--(\ref{ssy}) are sufficiently large, it is possible to devise algorithms to estimate ${\mathcal P}$ 
and $\varphi$ from the LTV model as will be demonstrated in this section and Sections~\ref{switch-estI},
\ref{switch-estII}, \ref{switch-estIII}.

We start by taking the differences of (\ref{gHankel12}) and (\ref{gHankel2})
\begin{equation}\label{firstdiff}
	\delta_{\mathcal H}(k)={\mathcal H}_{q,r}(k+1)-{\mathcal H}_{q,r}(k), \qquad k \in [k^\prime\;\;k^{\prime\prime}]
\end{equation}
and define $\mathbb{Z}_{\mathcal H}$ as the set of the zeros of the first order difference function in (\ref{firstdiff})
as follows

\begin{equation} \label{zeroset}
	\mathbb{Z}_{\mathcal H}=\{k \in [k^\prime\;\;k^{\prime\prime}]: \, \delta_{\mathcal H}(k)=0\}.
\end{equation}
Recall that we fixed $q=2n+1$ and $r=2n$. We would like $\mathbb{Z}_{\mathcal H}$ to have a resemblance to $\chi$,
differing only by a few points around switches on every segment or at least on long segments. It is also desired that each sufficiently 
long interval in $\mathbb{Z}_{\mathcal H}$ must reside only in one segment. We will show that both aims are feasible if we
put some mild restrictions on the dwell times and the model structure.  

Let us choose the $\ell_1$-norm of eigenvalues of square matrices defined by
\begin{equation}\label{Meigdef}
	{\mathcal M}(X)=\sum_{i=1}^n |\lambda_i(X)|, \qquad  X \in \mathbb{R}^{n \times n}.
\end{equation}
as the feature used for clustering. We will elaborate on this feature and clustering later, but for the time being 
note that it is invariant to matrix similarity transformations as can be verified easily. We assume that ${\mathcal M}$ 
separates the discrete states in ${\mathcal P}$ as described in the following assumption.

\begin{assumption}\label{unimodality}
	The discrete states of the MIMO--SLS model (\ref{ssx})--(\ref{varphit}) satisfy
	\begin{equation}
		{\mathcal M}(A(k)) \neq {\mathcal M}(A(l)) \Longleftrightarrow \varphi(k) \neq \varphi(l). 
	\end{equation}
 $\qed$
\end{assumption}

We derive the following result when Assumption~\ref{unimodality} holds and the dwell times satisfy 
mild conditions.

\begin{lemma}\label{hoppala1}
	Consider the MIMO--SLS model (\ref{ssx})--(\ref{varphit}). Suppose that Assumption~\ref{unimodality} 
	holds and $\delta_*(\chi) \geq 4n+2$, $\delta_{i^*}(\chi)>8n$, $\delta _0(\chi) \geq 6n + 2$. 
	Let $\mathbb{Z}_{\mathcal H}$ be as in (\ref{zeroset}). Then,  $\{k_{i+1}-1,k_{i+1}\} \not \subset 
	\mathbb{Z}_{\mathcal H}$ and there exists a collection of closed intervals $S_i \subset 
	\mathbb{Z}_{\mathcal H}$, $0 \leq i \leq i^*$ such that
	\begin{equation}
		[k^\prime+2n \;\;k_1-2n-2] \subseteq S_0 \subset [k^\prime \;\; k_1), \label{Si0}
	\end{equation}
	\begin{equation}
		[k_i+2n \;\;k_{i+1}-2n-2] \subseteq S_i \subset [k_i\;\;k_{i+1}), \;\; 0<i<i^*, \label{Si1}
	\end{equation}
	\begin{equation}
		[k_{i^*}+2n \;\; k^{\prime\prime}-2n-1] \subseteq S_{i^*} \subset [k_{i^*} \;\; k^{\prime\prime}]. 
		\label{Si*}
	\end{equation}
	The closed intervals $S_i$, $0 \leq i \leq i^*$ are maximal, {\em i.e.,} any closed interval 
	$\tilde{S_i} \subset \mathbb{Z}_{\mathcal H}$ satisfying one chain of the inequalities 
	in (\ref{Si0})--(\ref{Si*}) is a subset of $S_i$. $\qed$
\end{lemma} 

{\em Proof.} The $(s,t)$ block entry of $\delta_{\mathcal H}(k)$ denoted by 
$\delta_{\mathcal H}^{st}(k)$ can be written as 
\begin{eqnarray*}
	\delta_{\mathcal H}^{st}(k) &=& h(k+s,k-t+1)-h(k+s-1,k-t) \\
	\hspace{5mm} &=& C(k+s)A(k+s-1) \, \cdots \, A(k-t+1) B(k-t+1) \\
	&{}& -C(k+s-1)A(k+s-2) \, \cdots \, A(k-t) B(k-t) 
\end{eqnarray*} 
for $1\leq s \leq 2n+1$ and $1 \leq t \leq 2n$. We first consider the case $0<i<i^*$. Then, 
$\delta_{\mathcal H}^{st}(k)=0$ if $k+2n+1 < k_{i+1}$ and $k-2n \geq k_i$ for some $i \in (0\;\;i^*)$. 
Thus, $\delta_{\mathcal H}(k)=0$ for all  $k$ in the interval $[k_i+2n \;\;k_{i+1}-2n-2]$. Let 
$S_i \subset \mathbb{Z}_{\mathcal H}$ be the largest closed interval with
$[k_i+2n \;\; k_{i+1}-2n-2] \subset S_i$.

Now, assume that $\delta_{\mathcal H}(k_{i+1}-1)=\delta_{\mathcal H}(k_{i+1})=0$. Then, 
\begin{equation}\label{j1}
	{\mathcal H}_{2n+1,2n}(k_{i+1}-1) = {\mathcal H}_{2n+1,2n}(k_{i+1})={\mathcal H}_{2n+1,2n}(k_{i+1}+1). 
\end{equation}
The first equality in (\ref{j1}) implies that $T(k_{i+1}-1) = T(k_{i+1})$ and $\hat{A}(k_{i+1}-1) \sim A(k_{i+1}-1)$
as a result.
Likewise, from the second equality, we get  $T(k_{i+1}) = T(k_{i+1}+1)$ and $\hat{A}(k_{i+1}) \sim A(k_{i+1})$.
But, $\hat{A}(k_{i+1}-1)=\hat{A}(k_{i+1})$. Thus, $A(k_{i+1}-1) \sim A(k_{i+1})$ or 
$A_{\varphi(k_{i+1}-1)} \sim A_{\varphi(k_{i+1})}$ and therefore 
${\mathcal M}(A_{\varphi(k_{i+1}-1)})={\mathcal M}(A_{\varphi(k_{i+1})})$. From Assumption~\ref{unimodality},
we then have $\varphi(k_{i+1}-1)=\varphi(k_{i+1})$. Since $\varphi(k_{i+1}-1)=\varphi(k_i)$ and 
$\varphi(k_i) \neq \varphi(k_{i+1})$, we reach a contradiction. Thus, both $k_{i+1}-1$ and $k_{i+1}$ 
can not be in $\mathbb{Z}_{\mathcal H}$ for all $0 < i < i^*$. Hence, if $\delta_{\mathcal H} (k_{i+1})=0$, 
then $k_{i+1}-1 \notin \mathbb{Z}_{\mathcal H}$ which implies that $S_i \subset (-\infty\;\;k_{i+1}-2]$.
If $\delta_{\mathcal H} (k_{i+1}) \neq 0$, then $S_i \subset (-\infty\;\;k_{i+1}-1]$. Combining these two results, 
we get  $S_i \subset (-\infty \;\;k_{i+1}-2] \cup (-\infty \;\;k_{i+1}-1]=(-\infty \;\;k_{i+1})$. The same 
arguments applied to $\delta_{\mathcal H}(k_i-1)$ and $\delta_{\mathcal H}(k_i)$ show that 
$\{k_i-1,k_i\} \not \subset \mathbb{Z}_{\mathcal H}$. Thus, if 
$k_i-1 \in \mathbb{Z}_{\mathcal H}$, then $k_i \notin \mathbb{Z}_{\mathcal H}$ and therefore, 
$S_i \subset (k_i \;\; \infty)$. If $k_i\in \mathbb{Z}_{\mathcal H}$, on the other hand, 
$k_i-1 \notin \mathbb{Z}_{\mathcal H}$. Therefore, $S_i \subset [k_i \;\;\infty)$. Combining
both cases we get $S_i \subset (k_i \;\;\infty)\cup [k_i \;\;\infty)=[k_i \;\;\infty)$. It follows that 
$S_i$ is a subset of $(-\infty \;\;k_{i+1}) \cap [k_i\;\;\infty)=[k_i\;\;k_{i+1})$. The case $i=0$ follows
from (\ref{Si1}) on plugging $k^\prime \Rightarrow k_i$ and $k_1\Rightarrow k_{i+1} $. The last case 
$i=i^*$ follows from (\ref{Si1}) on plugging $k_{i^*} \Rightarrow k_i$ and noting that $k^{\prime\prime}$
is not a switch.  $\qedwhite$

Each of the intervals $S_i$, $0\leq i \leq i^*$ in the lemma contain at least 
\begin{equation}
	N_S = \min \left\{\delta_0(\chi)-6n-1, \delta_*(\chi)-4n-1, \delta_{i*}(\chi)-8n\right\}
\end{equation}
points provided that $N_S>0$. When $N_S>0$, each segment of $\chi$ contains a closed interval of at least $N_S$ 
points that is a subset of $\mathbb{Z}_{\mathcal H}$. As a result, these intervals are disconnected.
This observation suggests an estimation algorithm to extract the discrete states by clustering.

Let $\mathbb{Z}_{\mathcal H,\varepsilon}=\left\{k \in [k^\prime\;\;k^{\prime\prime}]: 
\|\delta_{\mathcal H}(k)\|_F < \varepsilon \right\}$. We have $\mathbb{Z}_{\mathcal H}=\cap_{\varepsilon>0} 
\mathbb{Z}_{\mathcal H,\varepsilon}$. Since $N<\infty$ and the number of the switches is finite, 
$\mathbb{Z}_{\mathcal H} =\mathbb{Z}_{\mathcal H,\varepsilon_{\mathbb{Z}}}$ for some $\varepsilon_{\mathbb{Z}}>0$. 
Suppose $N_S \geq \nu n$ which is guaranteed if $\delta_0(\chi)\geq (6+\nu)n+1$, $\delta_*(\chi) \geq (4+\nu)n+1$, and $\delta_{i*}(\chi) \geq (8+\nu)n$. Consider two segments $[k_i\;\;k_{i+1})$ and $[k_{i+1}\;\;k_{i+2})$ for 
some $0\leq i<i^*$. By Lemma~\ref{hoppala1} and the remark just made, they contain two closed intervals 
$S_i=[\alpha_i\;\;\beta_i]$ and $S_{i+1}=[\alpha_{i+1}\;\;\beta_{i+1}]$ in 
$\mathbb{Z}_{\mathcal H,\varepsilon_{\mathbb{Z}}}$.
Let $\gamma_i$ be the midpoints of $S_i$, $0\leq i \leq i^*$, i.e., $\gamma_i=(\alpha_i+\beta_i)/2$. Since
we assumed $N_S\geq \nu n$, $[\alpha_i\;\;\beta_i]$ for all $0\leq i \leq i^*$ and from (\ref{Si0})--(\ref{Si*})
\begin{equation}
\alpha_{i+1}-\beta_i  \leq k_{i+1}+2n- (k_{i+1}-2n-2)=4n+2, \qquad 0<i<i^*.
\end{equation}
Thus, if $[k^\prime \;\;k^{\prime\prime}]$ is split to the sets  $\mathbb{Z}_{\mathcal H,\varepsilon_{\mathbb{Z}}}$ and 
$\mathbb{Z}_{\mathcal H,\varepsilon_{\mathbb{Z}}}-[k^\prime\;\;k^{\prime\prime}]$, the first set will contain disjoint 
intervals of length at least $\nu n$ and separated from each other by a distance at most $4n+2$. If $\nu >5$, $S_i$ 
is detectable by visual inspection since any transition band around it is not larger than $4n+2$. Now, use the feature 
(\ref{Meigdef}) for clustering with the estimators $\hat{A}(\gamma_i)$, $0\leq i \leq i^*$ satisfying 
$k^\prime \leq  k_i \leq k^{\prime\prime}$. Since $\hat{A}(\gamma_i) \sim A(\gamma_i)$, ${\mathcal M}(\hat{A}(\gamma_i))
={\mathcal M}(A(\gamma_i))$. Hence, there are at most $\sigma$ clusters. It remains to show that the number of the
clusters is $\sigma$, but this follows from the following assumption. 

\begin{assumption}\label{sysasmp2}
	Every discrete state in $\mathcal{P}$ is visited at least once in $[k^\prime\;\;k^{\prime\prime}]$ and 
	$N_S > 5n$. $\qed$
\end{assumption}

In Section~\ref{example}, we will use the density-based clustering algorithm, DBSCAN, proposed in \cite{Esteretal:96}
implemented by the {\tt dbscan} command in MATLAB. Another popular method is the $k$-means clustering algorithm 
\cite{Arthur&Vassilvitskii:2007}. It is implemented by the {\tt kmeans} command in MATLAB. Unlike the $k$-means 
clustering algorithm, the density-based clustering algorithm does not need the number of clusters to be specified 
{\em a priori}. In principle, any clustering algorithm can be used. See, for example, \cite{Hastie&Tibshirani&Friedman:2001} 
for further information on clustering algorithms. The time-varying ${\mathcal H}_2$ norm of $\hat{\mathcal P}(k)$ or the 
sum of the Frobenius norms of $\hat{A}(k),\hat{B}(k),\hat{C}(k),\hat{D}(k)$ were proposed in \cite{Bencherki&Turkay&Akcay:2020} 
as alternative features for clustering. When $D_k$, $1 \leq k \leq \sigma$ is a unimodal collection, $\hat{D}(k)$  may 
be used as a feature for clustering though it may be hard to enforce the unimodal assumption in practice since more than 
one discrete state could be strictly proper.

If $k \in S_i$ for some $0\leq i \leq i^*$, then $T(k)=T(k+1)$. Hence, $\hat{\mathcal P}(k) \sim {\mathcal P}(k)
= {\mathcal P}(\gamma_i)$ for all $k \in S_i$ and we set $\hat{\mathcal P}(k)={\mathcal P}(\gamma_i)$ there.
Therefore, we also determine the switching sequence on $S_i$ as $\varphi(k)=\varphi(\gamma_i)$. This leaves
$\varphi(k)$ undetermined only on $[k_i,k_{i+1})-S_i$. From Lemma~\ref{hoppala1}, this set has at most $4n+1$ 
elements if $0<i<i^*$ and even less if $i=0$ or $i=i^*$. This is the advantage of working with 
$\mathbb{Z}_{{\mathcal H},\varepsilon_{\mathbb{Z}}}$. As an example, consider the extreme case of one switch. 
Once two discrete states are identified, identifying a sequence $\varphi(\alpha_2-1),\varphi(\alpha_2-2),\cdots$ 
we find $k_1$ by no more than $2n$ trials, no matter how large $N$! Another advantage working with 
$\mathbb{Z}_{{\mathcal H},\varepsilon_{\mathbb{Z}}}$ is that $S_i$, $0\leq i \leq i^*$ are determined without
bothering the similarity or the Lyapunov transformations. 

The second part of Assumption~\ref{sysasmp2} is not really necessary. As long as the first part is satisfied by 
$\sigma$ sufficiently long segments of $\chi$, the conclusion drawn still holds. Picking a large lower bound for $N_S$ 
helps visualize $S_i$ intervals by overlooking transition bands between them. The number of the clusters 
may exceed $\sigma$ when the Markov parameters are corrupted by noise. A re-clustering will be necessary to reduce this 
number by eliminating clusters over short segments.  We summarize the results derived in this section as an estimation 
algorithm in Appendix. This algorithm estimates the discrete states in two stages. The first stage is divided into Steps~1-5
while the second into Steps~6--7. This division reduces computational burden of the clustering algorithm since there are 
at most $O(\delta_*^{-1}N)$ clusters. In \cite{Lopesetal:2013}, it was suggested to directly cluster 
the Markov parameters in ${\mathcal H}_{2n+1,2n}(k)$ after putting them into a vector form for each $k$. This is also 
a feasible approach; but, increases computational burden of the clustering algorithm.

The results derived in this section for the noiseless Markov parameters case are summarized in the following.

\begin{theorem}\label{main2}
	Consider Algorithm~2 with the noiseless Markov parameters of the MIMO-SLS model (\ref{ssx})--(\ref{varphit}). 
	Suppose that $\chi$ and $\mathcal{P}$ satisfy Assumptions~\ref{sysasmp}--\ref{varphiassmp} and \ref{unimodality}--\ref{sysasmp2}. Then, Algorithm~2 recovers every discrete state in ${\mathcal P}$ 
	up to similarity transformations.
\end{theorem}

\subsection{Time complexity of Algorithm~2}

In Steps~1--3 of Algorithm~2, $\mathbb{Z}_{{\mathcal H},\varepsilon_{\mathbb{Z}}}$ is found by calculating
$\|\delta_{\mathcal H}(k)\|_F$, $\forall k \in [k^\prime,k^{\prime\prime}]$ and a given $\varepsilon_{\mathbb{Z}}>0$.
Since $\|\delta_{\mathcal H}(k)\|_F^2={\rm trace}(\delta_{\mathcal H}^T(k)\delta_{\mathcal H}(k))$, time complexity 
of the matrix multiplication inside the trace operator is $O(n^2)$. Thus, calculation of 
$\mathbb{Z}_{{\mathcal H},\varepsilon_{\mathbb{Z}}}$ has time complexity $O(Nn^2)$ without considering the 
nesting property (\ref{gHankel2}). In Steps~4--6, Algorithm~1 is run over a subset 
$\kappa \subset [k^\prime\;\;k^{\prime\prime}]$. Recall that $O(1) \leq |\kappa| \leq O(Nn^{-1})$ where $O(1)$ is
for a few long intervals or switches in $\chi$ and $O(Nn^{-1})$ for evenly distributed switches. Hence, from 
Section~\ref{complexity1} we infer that Steps~4--6 have a worst-case time complexity $O(Npoly(n))$. Time complexity 
of clustering varies from one algorithm to another. For example, DBSCAN algorithm has complexity $O(|\kappa|\log |\kappa|)$ 
or $O(|\kappa|^2)$ depending on the hyper-parameter values selected and whether the data are degenerate or not. Space complexity,
i.e., memory requirement for recomputations in the DBSCAN, changes from $O(|\kappa|)$ to $O(|\kappa|^2)$.  The $k$-means algorithm 
has complexity $O(\tau |\kappa|\sigma)$ where $\tau$ is the number of iterations needed to complete clustering process.
It is observed that $\tau \approx O(|\kappa|)$. Hence, the k-means algorithm has time complexity $O(|\kappa|^2\sigma)$. Attempts
have been made to improve the iteration complexity and reduce the time-complexity of the k-means algorithm to 
$O(|\kappa|\sigma)$. Time or store complexity depends also on whether a supervised or unsupervised choice is made for 
clustering. An important factor is robustness to noise.

\subsection{Robustness of Algorithm~2} 
   
Recall that $\mathbb{Z}_{\mathcal H}=\mathbb{Z}_{{\mathcal H},\varepsilon_{\mathbb{Z}}}$ for some 
$\varepsilon_{\mathbb{Z}}$ when the Markov parameters are noise-free. Consider the bounded-but-unknown noise
model in (\ref{corruptMarkov}). Let $\tilde{\delta}_{\mathcal H}(k)=\tilde{\mathcal H}_{2n+1,2n}(k+1)-
\tilde{\mathcal H}_{2n+1,2n}(k)$. If ${\delta}_{\mathcal H}(k)=0$,  $\tilde{\delta}_{\mathcal H}(k)=E(k+1)-E(k)$ 
and $\|\tilde{\delta}_{\mathcal H}(k)\|_F \leq \sqrt{4n(2n+1)} < 4n \varepsilon$. Assume 
$4n\varepsilon < \varepsilon_{\mathbb{Z}}$. Then, no need to change Steps~1--4 in Algorithm~1. 
For a fixed $c>0$ and all sufficiently small $\varepsilon$,  under Assumptions~\ref{sysasmp}--\ref{varphiassmp}, 
Theorem~\ref{Alg1robustness} guarantees that a topological equivalent ${\mathcal P}^\varepsilon(k)$ of $\tilde{\mathcal P}(k)$ 
satisfies $\|A^\varepsilon(k)-\hat{A}(k)\|_F \leq c \varepsilon$. Since (\ref{Meigdef}) is invariant to similarity 
transformations and topological equivalence is transitive, Step~7 discloses all discrete-states under
Assumptions~\ref{unimodality}--\ref{sysasmp2} as $\varepsilon \rightarrow 0$ up to arbitrary similarity transformations.
The following extends Theorem~\ref{main2} to the noisy Markov parameters case.

\begin{theorem}\label{main2r}
	Consider Algorithm~2 with the noisy  Markov parameters in (\ref{corruptMarkov}). 
	Suppose that $\chi$ and $\mathcal{P}$ satisfy Assumptions~\ref{sysasmp}--\ref{varphiassmp} and 
	\ref{unimodality}--\ref{sysasmp2}. Then, Algorithm~2 recovers every discrete state in ${\mathcal P}$ 
	up to arbitrary similarity transformations as  $\varepsilon \rightarrow 0$.
\end{theorem}

\section{Switch detection from a modified LTV model}\label{switch-estI}

In this section,  we propose an estimation algorithm to find the complements of $S_i$ in the segments they
are lying for $0\leq i \leq i^*$, cf. Lemma~\ref{hoppala1}. This algorithm is iterative in time and starts from a given 
point in $S_i$. It is based on correcting the quadruples delivered 
by Algorithm~1 and proceeds in the positive and/or negative directions until a switch or a switch pair trapping $S_i$ 
is detected. The lower bound constraints on the dwell times in Lemma~\ref{hoppala1} apply. This process is repeated 
for all segments fulfilling the dwell time constraints in Lemma~\ref{hoppala1}.  

We first examine advancement in the positive direction recalling that the true/estimated extended observability matrices 
are related by
\begin{equation}\label{ekmek}
	\hat{\mathcal O}_{2n+1}(k+\ell)={\mathcal O}_{2n+1}(k+\ell) T^{-1}(k+\ell), 
	\qquad k \in [k^\prime\;\;k^{\prime\prime}];\;\;\ell=0,1.
\end{equation}
Let 
${\mathcal V}_{2n+1}(k) = {\mathcal O}_{2n+1}^\dag(k){\mathcal O}_{2n+1}(k+1)$ and
\begin{equation}\label{kahkaha1}
	\hat{\mathcal V}_{2n+1}(k) =\hat{\mathcal O}_{2n+1}^\dag(k) \hat{\mathcal O}_{2n+1}(k+1).
\end{equation} 
Then, from (\ref{ekmek}) with $\ell=0$ and $\ell=1$ we get what we call the foreward correction operator
\begin{equation}\label{kahkaha11}
	\hat{\mathcal V}_{2n+1}(k)=T(k) {\mathcal V}_{2n+1}(k) T^{-1}(k+1).
\end{equation}
Premultiplying $\hat{A}(k)$ and $\hat{B}(k)$ in (\ref{topA}) and (\ref{topCB}) with 
$\hat{\mathcal V}_{2n+1}(k)$, we get from (\ref{kahkaha11})
\begin{eqnarray}
	A_{\rm f}(k) &=& \hat{\mathcal V}_{2n+1}(k) \hat{A}(k) = T(k) {\mathcal V}_{2n+1}(k) A(k) T^{-1}(k), \nonumber 
	\\[-1.5ex] \label{backcorrect} \\[-1.5ex]
	B_{\rm f}(k) &=& \hat{\mathcal V}_{2n+1}(k)\hat{B}(k) = T(k) {\mathcal V}_{2n+1}(k) B(k). \nonumber
\end{eqnarray}
We leave $\hat{D}(k)$ and $\hat{C}(k)$ as they are, i.e., $\hat{D}(k)=D(k)$ and $\hat{C}(k)=C(k)T^{-1}(k)$. 
Thus, $T(k)$ maps the corrected quadruple $({\mathcal V}_{2n+1}(k) A(k),{\mathcal V}_{2n+1}(k) B(k), C(k),D(k))$ 
to ${\mathcal P}_{\rm f}(k)= (A_{\rm f}(k),B_{\rm f}(k),\hat{C}(k),\hat{D}(k))$. It is a time-varying similarity 
transformation satisfying ${\mathcal P}_{\rm f}(k) \sim {\mathcal P}(k)$ if ${\mathcal V}_{2n+1}(k)=I_n$.

Now, consider the block-rows of ${\mathcal O}_{2n+1}(k)$ given by $C(k)$ and 
\begin{equation}
	C(k+j) \Phi(k+j,k) =  C(k+j) A(k+j-1) \,\cdots \, A(k), \qquad 1 \leq j \leq 2n.	
\end{equation}	
We need $C(k+2n+1)$ and $C(k+j)$, $A(k+j)$, $0 \leq j \leq 2n$ to calculate ${\mathcal V}_{2n+1}(k)$. They are equal
to $C(k_i)$ or $A(k_i)$ for all $k \in [k_i\;\;k_{i+1}-2n-2]$ if $0<i< i^*$ and $\delta_i(\chi)\geq 2n+2$ and therefore, 
${\mathcal O}_{2n+1}(k)={\mathcal O}_{2n+1}(k+1)$. Consequently, ${\mathcal V}_{2n+1}(k)={\mathcal O}_{2n+1}(k)^\dag 
{\mathcal O}_{2n+1}(k+1)=I_n$. Hence, ${\mathcal P}_{\rm f}(k) \sim {\mathcal P}(k_i)$ provided that the switches $k_i$ 
and $k_{i+1}$ are known. 

Consider the case $0 < i <i^*$ and set $C_i=C(k_i)$, $A_i=A(k_i)$, and $\check{C} =C(k+2n+1)$. Assuming
 $k \in [k_i \;\;k_{i+1}-2n-1]$, we have
\begin{eqnarray}
	{\mathcal O}_{2n+1}(k+1) &=& \left[\begin{array}{c} C(k+1)\\ \vdots \\ C(k+2n) A(k+2n-1) \cdots A(k+1) \\ 
	C(k+2n+1) A(k+2n) \cdots A(k+1)\end{array} \right] = \left[\begin{array}{c} C_i  \\\vdots \\ C_i A_i^{2n-1}
		\\ \check{C} A_i^{2n} \end{array} \right], \nonumber
	\\[-0ex]  \label{alicengiz}   \\[-0ex] 
	{\mathcal O}_{2n+1}(k)   &=& \left[\begin{array}{c} C(k) \\ \vdots \\ C(k+2n-1)A(k+2n-2)\cdots A(k)\\ 
		C(k+2n)A(k+2n-1)\cdots A(k)\end{array}\right]=\left[\begin{array}{c} C_i \\ \vdots \\ C_i 
		A_i^{2n-1} \\ C_i A_i^{2n}\end{array} \right]. \nonumber
\end{eqnarray}
From (\ref{alicengiz}),
\begin{eqnarray}
	{\mathcal O}_{2n+1}^T(k){\mathcal O}_{2n+1}(k+1) &=&{\mathcal O}_{2n+1}^T(k)
	{\mathcal O}_{2n+1}(k)+(A_i^{2n})^T C_i^T (\check{C}-C_i) A_i^{2n} \nonumber 
	\\[-0.5ex]   \\[-0.5ex]
	&=& {\mathcal G}_{\rm o}(k;2n+1) \left(I_n +{\mathcal G}_{\rm o}^{-1}(k;2n+1)(A_i^{2n})^T C_i^T 
	(\check{C}-C_i) A_i^{2n}\right). \nonumber 
\end{eqnarray}
Hence,
\begin{eqnarray}
	\hat{\mathcal V}_{2n+1}(k) &=& T(k){\mathcal O}_{2n+1}^\dag(k){\mathcal O}_{2n+1}(k+1)T^{-1}(k+1) \nonumber 
	\\[-0.5ex] \label{haci} \\[-0.5ex] 
	&=& T(k)\left(I_n+{\mathcal G}_{\rm o}^{-1}(k;{2n+1})(A_i^{2n})^T C_i^T (\check{C}-C_i) 
	A_i^{2n}\right) T^{-1}(k+1). \nonumber
\end{eqnarray}
Since ${\mathcal M}$ is invariant to similarity transformations, we get 
\begin{equation}\label{portakalci}
	{\mathcal M}(\hat{\mathcal V}_{2n+1}(k)) = {\mathcal M}\left(I_n+{\mathcal G}_{\rm o}^{-1}(k;{2n+1})(A_i^{2n})^T 
	C_i^T (\check{C}-C_i) A_i^{2n}\right). 
\end{equation}
Since $k \leq k_{i+1}-2n-2 \Rightarrow \check{C}=C_i$, from (\ref{portakalci}) ${\mathcal M}(\hat{\mathcal V}_{2n+1}(k))=n$. 
Conversely, suppose ${\mathcal M}(\hat{\mathcal V}_{2n+1}(\hat{k})) \neq n$ implies that $\hat{k}=k_{i+1}-2n-1$. It is
the smallest $k$ satisfying ${\mathcal M}(\hat{\mathcal V}_{2n+2}(k)) \neq n$. Then, $\check{C}=C(\hat{k}+2n+1)=C(k_{i+1})$. 
It remains to pick an initial value $\check{k}$ for $k$. From Lemma~\ref{hoppala1}, the distance of $S_i=[\alpha_i\;\;\beta_i]$ 
to $k_{i+1}$ denoted by $d(S_i,k_{i+1})$ satisfies $1 \leq d(S_i,k_{i+1}) \leq 2n+2$. If we choose $\check{k}=\beta_i$,
then $\beta_i=k_{i+1}-1$ is a possible value, yet the assumption $\check{k} \leq k_{i+1}-2n-1$ is violated. Therefore,
we choose $\check{k}=\beta_i-2n$. Starting from $\check{k}$, we reach to $\hat{k}$ in $\hat{k}-\check{k}=
k_{i+1}-1-\beta_i$ steps. Tight upper and lower bounds on $\hat{k}-\check{k}$ are derived as follows
\begin{equation*}
0 \leq k_{i+1}-1-(k_{i+1}-1) \leq \hat{k}-\check{k}=k_{i+1}-1-\beta_i \leq k_{i+1}-1-(k_{i+1}-2n-2)=2n+1.
\end{equation*}
This case extends to $i=0$ by letting $k_0=1$. For $i=i^*$, we simply let $\varphi(k)=\varphi(\beta_{i^*})$ for all 
$k \in (\beta_{i^*},k^{\prime\prime}]$ since $k^{\prime\prime}$ is not a switch. Now, we examine the backward 
corrections case. 

Recall that the true extended controllability matrices and their estimates are related by
\begin{equation}\label{ekmek2}
	\hat{\mathcal R}_{2n}(k-\ell)=T(k-\ell+1) {\mathcal R}_{2n}(k-\ell) , \qquad 
	k \in [k^\prime\;\;k^{\prime\prime}];\;\;\ell=0,1.
\end{equation}
Let ${\mathcal W}_{2n}(k) = {\mathcal R}_{2n}(k-1) {\mathcal R}_{2n}^\dag(k)$ and
\begin{equation}\label{kahkaha2}
	\hat{\mathcal W}_{2n}(k) =\hat{\mathcal R}_{2n}(k-1) \hat{\mathcal R}_{2n}^\dag(k). 
\end{equation} 
Then, from (\ref{ekmek2}) with $\ell=0$ and $\ell=1$ we get what we call the backward correction operator
\begin{equation}\label{kahkaha22}
	\hat{\mathcal W}_{2n}(k)=T(k) {\mathcal W}_{2n}(k) T^{-1}(k+1)
\end{equation}
Post-multiplying $\hat{A}(k)$ in (\ref{topA}) and $\hat{C}(k)$ in (\ref{topCB}) with 
$\hat{\mathcal W}_{2n}(k)$, we get from (\ref{kahkaha22})
\begin{eqnarray}
	A_{\rm b}(k) &=& \hat{A}(k) \hat{\mathcal W}_{2n}(k) = T(k+1)A(k){\mathcal W}_{2n}(k) T^{-1}(k+1), \nonumber
	\\[-1.5ex] \label{forecorrect}  \\[-1.5ex]
	C_{\rm b}(k) &=& \hat{C}(k)\hat{\mathcal W}_{2n}(k) = C(k) {\mathcal W}_{2n}(k) T^{-1}(k+1). \nonumber
\end{eqnarray}
Since $\hat{B}(k)=T(k+1) B(k)$ and $\hat{D}(k)=D(k)$, we do not need to change them. Thus, $T(k+1)$
maps the modified quadruple $(A(k){\mathcal W}_{2n}(k),B(k), C(k){\mathcal W}_{2n}(k),D(k))$ to 
${\mathcal P}_{\rm b}(k)=({A}_{\rm b}(k),\hat{B}(k),{C}_{\rm b}(k),\hat{D}(k))$. It is a time-varying 
similarity transformation and if ${\mathcal W}_{2n}(k)=I_n$, we then have 
${\mathcal P}_{\rm f}(k) \sim {\mathcal P}(k)$.

The block-columns of ${\mathcal R}_{2n}(k-1)$ start with $B(k-1)$ and for $2 \leq j \leq 2n$, they are given by 
\begin{eqnarray}\label{denklem}
	\Phi(k,k-j+1)B(k-j) &=&  A(k-1) \,\cdots \, A(k-j+1)B(k-j).	
\end{eqnarray}	
We need $B(k-2n)$, and $B(k-j)$, $A(k-j)$, $0 \leq j < 2n$ to calculate ${\mathcal W}_{2n}(k)$. They are equal to 
$B(k_i)$ or $A(k_i)$ for all $k \in [k_i+2n\;\;k_{i+1}-1]$ if $\delta_i(\chi)\geq 2n+1$ and  $0<i<i^*$. Then, 
${\mathcal R}_{2n}(k)={\mathcal R}_{2n}(k-1)$ and 
${\mathcal W}_{2n}(k)={\mathcal R}_{2n}(k-1){\mathcal R}_{2n}^\dag (k)=I_n$. Hence, ${\mathcal P}_{\rm b}(k) 
\sim {\mathcal P}(k_i)$ if $k_i$ and $k_{i+1}$ are known.

Consider $0<i<i^*$ case first as before and set $B_i=B(k_i)$, $A_i=A(k_i)$, $\check{B}=B(k-2n)$. 
Assume $k \in [k_i+2n-1\;\;k_{i+1})$. Then,
\begin{eqnarray}
	{\mathcal R}_{2n}(k)& =& [B(k)\;\;A(k)B(k-1)\;\; \cdots \;\;A(k)\,\cdots \,A(k-2n+2)
	B(k-2n+1)] \nonumber \\
	&=& [B_i\;\;A_iB_i\;\; \cdots \;\;A_i^{2n-1}B_i], \nonumber 
	\\[-1ex] \label{alicengiz2} \\[-1ex]
	{\mathcal R}_{2n}(k-1) &=& [B(k-1)\;\;A(k-1)B(k-2)\;\; \cdots \;\;A(k-1)\,\cdots\, 
	A(k-2n+1)B(k-2n)] \nonumber \\
	&=& [B_i\;\;A_iB_i\;\; \cdots \;\;A_i^{2n-1}\check{B}]. \nonumber
\end{eqnarray}
From (\ref{alicengiz2}),
\begin{eqnarray}
	{\mathcal R}_{2n}(k-1){\mathcal R}_{2n}^T (k) &=&{\mathcal R}_{2n}(k)
	{\mathcal R}_{2n}^T(k)+A_i^{2n-1}(\check{B}-B_i)B_i^T  (A_i^{2n-1})^T \nonumber 
	\\[-.5ex]   \\[-.5ex]
	&=& \left(I_n +A_i^{2n-1}(\check{B}-B_i) B_i^T (A_i^{2n-1})^T {\mathcal G}_{\rm c}^{-1}(k+1;2n)\right)
	{\mathcal G}_{\rm c}(k+1;2n). \nonumber 
\end{eqnarray}
Hence,
\begin{eqnarray}
	\hat{\mathcal W}_{2n}(k) &=&  T(k){\mathcal R}_{2n}(k-1){\mathcal R}_{2n}^\dag (k) T^{-1}(k+1) \nonumber 
	\\[-1.5ex] \label{haci2} \\[-1.5ex] 
	&=& T(k)\left(I_n +A_i^{2n-1}(\check{B}-B_i) B_i^T (A_i^{2n-1})^T {\mathcal G}_{\rm c}^{-1}(k+1;2n)
	\right)T^{-1}(k+1). \nonumber
\end{eqnarray}
Since ${\mathcal M}$ is invariant to similarity transformations, 
\begin{equation}\label{armutcu}
{\mathcal M}\left( \hat{\mathcal W}_{2n}(k)\right) = {\mathcal M}\left(I_n +A_i^{2n-1}(\check{B}-B_i) B_i^T 
(A_i^{2n-1})^T{\mathcal G}_{\rm c}^{-1}(k+1;2n)\right). 
\end{equation}
Since $k \geq k_i+2n \Rightarrow \check{B}=B_i$, from (\ref{armutcu}) we have ${\mathcal M}(\hat{\mathcal W}_{2n}(k))=n$. 
Conversely suppose ${\mathcal M}(\hat{\mathcal W}_{2n}(k^\sharp)) \neq n$ implies that $k^\sharp=k_i+2n-1$. It is the smallest 
$k$ satisfying ${\mathcal M}(\hat{\mathcal W}_{2n}(k)) \neq n$. Then, $\check{B}=B(k_i-1)=B(k_{i-1})$. Note that $k_i-1$ is a 
switch in the reverse direction. Next, we choose an initial value $\tilde{k}$ for $k$. From Lemma~\ref{hoppala1}, the 
distance of $S_i=[\alpha_i\;\;\beta_i]$ to $k_i$ satisfies $0 \leq d(S_i,k_i) \leq 2n$. If we choose $\tilde{k}=\alpha_i$,
then $\alpha_i=k_i$ is a possible value, yet the assumption $\tilde{k} \geq k_i+2n-1$ is violated. Therefore,
we choose $\tilde{k}=\alpha_i+2n-1$. Starting from $\tilde{k}$, we reach to $k^\sharp$ in $\tilde{k}-k^\sharp=\alpha_i-k_i$ 
steps. Upper and lower bounds on $\tilde{k}-k^\sharp$ are derived as follows
\begin{equation*}
	0 \leq \tilde{k}-k^\sharp=\alpha_i-k_i \leq 2n.
\end{equation*}
This case extends to $i=i^*$ without modification. For $i=0$, we simply let $\varphi(k)=\varphi(\alpha_0)$ for all 
$k \in [k^{\prime}\;\;\alpha_0)$ since $k^{\prime}$ is not a switch. The switch detectability conditions are stated
as follows.

\begin{assumption}\label{assmp4}
	The MIMO--SLS model (\ref{ssx})--(\ref{varphit}) satisfies the following conditions
	\begin{enumerate}
		\item[(a)] ${\mathcal M}\left(I_n+{\mathcal G}_{\rm o}^{-1}(k_{i+1}-2n-1;{2n+1})[A^T(k_i)]^{2n}
		C^T(k_i)(C(k_{i+1})-C(k_i) ) [A(k_i)]^{2n}\right) \neq n$ for all $0\leq i<i^*$ with the convention
		$A(k_0)=A(k^\prime)$ and $C(k_0)=C(k^\prime)$
		\item[(b)] ${\mathcal M}\left(I_n +[A(k_i)]^{2n-1}(B(k_{i-1})-B(k_i)) B^T(k_i) 
		[A^T(k_i)]^{2n-1}{\mathcal G}_{\rm c}^{-1}(k_i+2n;2n)\right) \neq n$, $0<i \leq i^*$.
	\end{enumerate}
\end{assumption}

If Assumption~\ref{assmp4} holds, the two conditions mentioned in the forward and backward iterations are satisfied. A 
different feature than ${\mathcal M}$ to detect switches is $\|D(k+1)-D(k)\|$ where $\| \cdot\|$ is a matrix norm if 
$D(\varphi(k)) = D(\varphi(l))$ iff $\varphi(k) =\varphi(l)$. Let us examine Conditions (a)--(b) in Assumption~\ref{assmp4} 
through simple examples.

\begin{example}
	Let $X$ be defined by
	$$
	X=\left[\begin{array}{cc} 0 & \lambda \\ 0 & 0 \end{array} \right], \qquad \lambda \in \mathbb{R}.
	$$ 
	Then, ${\mathcal M}(I_2+X)=2$ for all $\lambda \in \mathbb{R}$, yet $I_2+X \neq I_2$ except $\lambda =0$. 
\end{example}
If the second term in the argument of ${\mathcal M}(\cdot)$ is a strictly upper or lower triangular matrix, then 
$k_{i+1}$ or $k_i$ cannot be detected by the proposed algorithm. We present two 
more examples.

\begin{example}
	$A(k_i)$ is idempotent, {\em i.e.}, $[A(k_i)]^{2n}=0$ and/or $[A(k_i)]^{2n-1}=0$, yet $A(k_i) \neq 0$. Either 
	or both of the conditions in Assumption~\ref{assmp4} fail since ${\mathcal M}(\hat{V}_{2n+1}(k_{i+1}-2n-1))=n$ 
	and/or ${\mathcal M}(\hat{W}_{2n}(k_i+2n-1))=n$. 
\end{example}

\begin{example}
	If $C(k_i) \perp C(k_{i+1})-C(k_i)$, ${\mathcal M}(\hat{V}_{2n+1}(k_{i+1}-2n-1))=n$ and $k_{i+1}$ cannot be 
	detected by the algorithm to be proposed. If $B(k_i) \perp B(k_{i-1})-B(k_i)$, ${\mathcal M}(\hat{V}_{2n+1} 
	(k_{i+1}-2n-1))=n$ and $k_{i-1}$ cannot be detected.
\end{example}
Despite these and many other examples which can be constructed, Assumption~\ref{assmp4} holds for almost
all switch pairs and model sets. The above derivations are summarized as Algorithm~3 in Appendix. This algorithm
detects a switch or switch pair subject to the dwell time constraints in Lemma~\ref{hoppala1}. In particular, if 
$N_S>5n$, then all switches in $\chi$ are detected. When a dwell time constraint fails,  alternative
procedures presented in the following sections may be resorted. The main result derived in this section for the noiseless
data case is summarized in the following.

\begin{theorem}\label{main23}
	Consider Algorithm~3 driven by the noise-free Markov parameters of (\ref{ssx})--(\ref{varphit}) 
	and the outputs of Algorithm~2. Suppose that $\chi$ and $\mathcal{P}$ satisfy Assumptions~\ref{sysasmp}---\ref{varphiassmp}, 
	\ref{unimodality}, 
	and \ref{assmp4}. Then, Algorithm~3 recovers $\varphi$ on 
	the segments with dwell times satisfying $\delta_i(\chi) \geq 4n+2$ for $0<i<i^*$, $\delta_0(\chi)\geq 6n+2$, and 
	$\delta_{i^*}(\chi) >8n$.
\end{theorem}

If $N_S>5n$, each segment of $\chi$ satisfies a dwell time condition in Theorem~\ref{main23} and Algorithm~3 detects
all switches in $\chi$. Recall that Algorithm~2 recovers the discrete states in the segments satisfying one inequality from
$\delta_0(\chi) > 11n$, $\delta_i(\chi) >9n$, $0<i\leq i^*$, and $\delta_{i^*}(\chi) \geq 13n$. Algorithm~3 aplied to 
these segments, detects the switches at the end points of these segments. The lower bounds on the dwell times  in 
Theorem~\ref{main23} are required for advancements in both directions. If advancements are made only in one direction, they
can be reduced significantly. It should also be noticed that Algorithm~3 does not have to start from very long segments.
It suffices to start with a correct discrete state and satisfy one of the dwell conditions in Theorem~\ref{main23}.
If the second part of Assumption~\ref{sysasmp2} does not hold, short segments appearing in rows may form large sets 
which are difficult to discern from the long segments. This pathological case is, however, unlikely to occur if a 
mixing condition on $\varphi$ is satisfied. 

\subsection{Time complexity of Algorithm~3}\label{sebsecalg3}

For each $k$, time complexities of $\hat{\mathcal V}_{2n+1}(k)$ and $\hat{\mathcal W}_{2n}(k)$ are calculated similarly 
to $\hat{A}(k)$ as in Section~\ref{complexity1} and we find them $O(poly(n))$ for a third order polynomial in $n$. Finding 
eigenvalues of an $n$ by $n$ matrix has complexity $O(n^3)$. Since the number of steps to detect a switch is 
bounded above by $2n+1$, detection of one switch has a worst-case complexity $O(poly(n))$ for a fourth order poynomial in $n$. 
The worst-case time complexity of Algorithm~3 is $O(N poly(n))$ for a third order polynomial in $n$ since there are at most 
$O(Nn^{-1})$ switches when Assumption~\ref{varphiassmp} holds.

\subsection{Robustness of Algorithm~3}
From Theorem~\ref{Alg1robustness}, we have ${\mathcal P}^\varepsilon(k) \rightarrow \hat{\mathcal P}(k)$ uniformly in $k$
as $\varepsilon \rightarrow 0$. This means $\hat{\mathcal V}_{2n+1}^\varepsilon(k) \rightarrow \hat{\mathcal V}_{2n+1}(k)$ 
and ${\mathcal W}_{2n}^\varepsilon(k)\rightarrow \hat{\mathcal W}_{2n}(k)$ uniformly in $k$ as $\varepsilon \rightarrow 0$.
Since eigenvalues of a matrix are continuous functions of matrix entries, 
${\mathcal M}(\hat{\mathcal V}_{2n+1}^\varepsilon(k)) \rightarrow {\mathcal M}(\hat{\mathcal V}_{2n+1}(k))$ 
and ${\mathcal M}({\mathcal W}_{2n}^\varepsilon(k))\rightarrow {\mathcal M}(\hat{\mathcal W}_{2n}(k))$ uniformly in 
$k$ as $\varepsilon \rightarrow 0$. Recall that ${\mathcal M}(\cdot)$ is invariant to similarity transformations.
Plugging the perturbed matrices $\tilde{\mathcal V}_{2n+1}(k)$ and $\tilde{\mathcal W}_{2n}(k)$ on the left
hand-sides, we see that the threshold criteria in Algorithm~3 never fail as $\varepsilon \rightarrow 0$  under 
Assumptions~\ref{sysasmp}, \ref{unimodality}--\ref{sysasmp2}, and \ref{assmp4}. The following extends Theorem~\ref{main23} 
to the noisy Markov parameters case.

\begin{theorem}\label{main23r}
	Consider Algorithm~3 driven by the noisy Markov parameters in (\ref{corruptMarkov}) and the outputs of Algorithm~2. 
	Suppose that $\chi$ and $\mathcal{P}$ satisfy Assumptions~\ref{sysasmp}---\ref{varphiassmp}, \ref{unimodality}, and
	\ref{assmp4}. Then, Algorithm~3 recovers $\varphi$ on the segments with dwell times satisfying 
	$\delta_i(\chi) \geq 4n+2$ for $0<i<i^*$, $\delta_0(\chi)\geq 6n+2$, and $\delta_{i^*}(\chi) >8n$ as  
	$\varepsilon \rightarrow 0$.
\end{theorem}

\section{Switch detection from Markov parameters}\label{Markovmatch}\label{switch-estII}

Let us reconsider the intervals $S_i=[\alpha_i,\beta_i]$ with the midpoints $\gamma_i=(\alpha_i+\beta_i)/2$, 
$0\leq i \leq i^*$ in Lemma~\ref{hoppala1}. We showed that $\hat{\mathcal P}(\gamma_i) =(\hat{A}(\gamma_i),\hat{B}(\gamma_i),\hat{C}(\gamma_i),\hat{D}(\gamma_i))$ is similar to the true model 
${\mathcal P}(\gamma_i)=(A(\gamma_i),B(\gamma_i),C(\gamma_i),D(\gamma_i)) $. This means that 
$A(\gamma_i)=T_i^{-1}\hat{A}(\gamma_i)T_i$, $C(\gamma_i)=\hat{C}(\gamma_i) T_i$, $B(\gamma_i)= T_i^{-1}\hat{B}(\gamma_i)$,
and $\hat{D}(\gamma_i)=D(\gamma_i)$ for a non-singular matrix $T_i$. Moreover, $0 \leq \alpha_i-k_i \leq 2n$.

Let $\ell >\beta_i$ and suppose we have shown that $\varphi(k)=\varphi(\gamma_i)$ for all 
$\ell-2n \leq k \leq \ell-1$. Plugging $\ell=\beta_i+1$, from $\ell-2n \geq \alpha_i$  we see that 
$\beta_i-\alpha_i \geq 2n-1$ holds. Then, from (\ref{defMarkov}) 
for $0\leq j \leq 2n$,
\begin{equation}\label{hk0i}
	h(\ell,\ell-j) = \left\{ \begin{array}{ll} C(\ell) \, A(\gamma_i)^{j-1}B(\gamma_i), & j \neq 0 \\ 
		D(\ell), & j=0. 
	\end{array} \right. 
\end{equation} 
We want to cluster $\varphi(\ell)$ to either $\varphi(\gamma_i)$ or $\varphi(\gamma_{i+1})$. Plug the 
similarity equations for $\hat{\mathcal P}(\gamma_i)$ in (\ref{hk0i}) to get
\begin{equation}\label{hk0i1}
	h(\ell,\ell-j) = \left\{ \begin{array}{ll} \check{C} \, [\hat{A}(\gamma_i)]^{j-1}\hat{B}(\gamma_i), 
		& j \neq 0 \\ \check{D}, & j=0 \end{array} \right. 
\end{equation} 
where $\check{C}= C(\ell) T_i^{-1}$ and $\check{D}=D(\ell)$. Treating $\check{C}$ and $\check{D}$ as 
unknowns, the right-hand side of (\ref{hk0i1}) becomes a linear estimator of $h(\ell,\ell-j)$ denoted 
by $\hat{h}(\ell,\ell-j)$. We find $\check{C}$ and $\check{D}$ by solving the linear least-squares 
minimization problem
\begin{equation}\label{minCD}
	\min_{\check{C} \in \mathbb{R}^{p \times n},\, \check{D}\in \mathbb{R}^{p \times m}} f(\check{C},\check{D})
\end{equation}
where
\begin{equation}
	f({\check{C},\check{D}})=\sum_{j=0}^{2n} \|\hat{h}(\ell,\ell-j)-h(\ell,\ell-j)\|_F^2.
\end{equation}
The minimization problem (\ref{minCD}) is separable in $\check{C}$ and $\check{D}$. Thus, 
$\check{D}=D(\ell)$. Suppose a switch detectability condition $D(\ell) \neq D(k) 
\Longleftrightarrow \varphi(\ell) \neq \varphi(k)$ is in force.  Then, $\check{D}=\hat{D}(\ell)$ and 
$\check{C}$ and $\varphi(\ell)$ are determined unambiguously. When this switch detectability condition 
does not hold, we determine $\check{C}$ by solving the normal equations ${\mathcal H}_{1,2n}(\ell) 
= \check{C} \hat{\mathcal R}_{2n}(\ell-1)$ where $\hat{\mathcal R}_{2n}(\ell-1)$ is formed from the 
controllability pair $(\hat{A}(\gamma_i),\hat{B}(\gamma_i))$. Hence, the unique minimizer is found as
\begin{equation}\label{checkC2} 
	\check{C} = {\mathcal H}_{1,2n}(\ell)\hat{\mathcal R}_{2n}^\dag(\ell-1) = 
	C(\ell){\mathcal R}_{2n}(\ell-1) \hat{\mathcal R}_{2n}^\dag(\ell-1) = C(\ell) T_i^{-1}.	
\end{equation} 
Not surprisingly $\hat{h}(\ell,\ell-j)=h(\ell,\ell-j)$ for all $0\leq j \leq 2n$. On substituting $\check{C}$ 
from (\ref{checkC2}) into (\ref{hk0i1}) and comparing it with (\ref{hk0i}), we see this. To carry on,
we impose  a switch detectability condition.

\begin{assumption}\label{assmp5}
	The MIMO--SLS model (\ref{ssx})--(\ref{varphit}) satisfies
	\begin{eqnarray}
		[C(k) \;\; D(k)] &\neq& [C(l) \;\; D(l)] 
		\Longleftrightarrow \varphi(k) \neq \varphi(l), \label{sd1} \\
		\left[\begin{array}{c} B(k) \\ D(k) \end{array}\right] &\neq& 
		\left[\begin{array}{c}B(l) \\ D(l)\end{array}\right] 
		\Longleftrightarrow \varphi(k) \neq \varphi(l). \label{sd2}
	\end{eqnarray}
\end{assumption}

If $\varphi(\ell)=\varphi(\gamma_i)$, $C(\ell)=C(\gamma_i)$, $D(\ell)=D(\gamma_i)$ and from (\ref{checkC2})
$\check{C}=\hat{C}(\gamma_i)$, $\check{D}=\hat{D}(\gamma_i)$. Conversely, suppose $\check{C}=\hat{C}(\gamma_i)$ 
and $\check{D}=\hat{D}(\gamma_i)$, yet $\varphi(\ell) \neq \varphi(\gamma_i)$, {\em that is}, 
$\varphi(\ell)=\varphi(\gamma_{i+1})$. Then, $C(l)=C(\gamma_{i+1})$ 
and $D(l)=D(\gamma_{i+1})$. Hence, from (\ref{checkC2}) $\check{C}=C(\gamma_{i+1}) T_i^{-1}=
\hat{C}(\gamma_{i+1}) T_{i+1}T_i^{-1}$ where $T_{i+1}$ is the similarity transformation in 
$\hat{\mathcal P}(\gamma_{i+1}) \sim {\mathcal P}(\gamma_{i+1})$. Moreover, $\check{D}=D(\ell)=D(\gamma_{i+1})$. 
Since by assumption $\check{D}=\hat{D}(\gamma_i)$, from $\hat{D}(\gamma_i)=D(\gamma_i)$ we then get 
$D(\gamma_i)=D(\gamma_{i+1})$. Furthermore, we derive $C(\gamma_i)=C(\gamma_{i+1})$ from the chain of 
equalities $C(\gamma_{i})=\hat{C}(\gamma_i) T_i=\check{C} T_i= \hat{C}(\gamma_{i+1}) T_{i+1} =C(\gamma_{i+1})$. Therefore, 
from (\ref{sd1}) in Assumption~\ref{assmp5} $\varphi(\gamma_i) =\varphi(\gamma_{i+1})$. We reach a contradiction. 
With $\check{D}=\hat{D}(\ell)$ and $\check{C}={\mathcal H}_{1,2n}(\ell) \hat{\mathcal R}_{2n}^\dag(\ell-1)$, we conclude  
$[\check{C} \;\;\check{D}] = [\hat{C}(\gamma_i)\;\;\hat{D}(\gamma_i)]\Longleftrightarrow \varphi(\ell)=\varphi(\gamma_i)$  
when Assumption~\ref{assmp5} holds. Until $k_{i+1}$ is detected, $\ell$ is increased. 

We now proceed along the negative axis. Let $\ell <\alpha_i$ and suppose for all $\ell+1 \leq k \leq \ell+2n$, we showed 
that $\varphi(k)=\varphi(\gamma_i)$, which requires $\beta_i-\alpha_i \geq 2n-1$. This inequality follows from 
$\ell+n \leq \beta_i$ on putting $\ell=\alpha_i-1$. We want to cluster $\varphi(\ell)$ to either 
$\varphi(\gamma_{i})$ or $\varphi(\gamma_{i-1})$. From (\ref{defMarkov}) for $0\leq j \leq 2n$,
\begin{equation}\label{hk0i2}
	h(\ell+j,\ell) = \left\{ \begin{array}{ll} C(\gamma_i) \, A(\gamma_i)^{j-1}B(\ell), & j \neq 0 \\ 
		D(\ell), & j=0. 
	\end{array} \right. 
\end{equation} 
Plugging the similarity equations for $\hat{\mathcal P}(\gamma_i)$ in, we get
\begin{equation}\label{hk0i3}
	h(\ell+j,\ell) = \left\{ \begin{array}{ll} \hat{C}(\gamma_i) \, [\hat{A}(\gamma_i)]^{j-1}\check{B}, 
		& j \neq 0 \\ \check{D}, & j=0 \end{array} \right. 
\end{equation} 
where $\check{B}= T_i B(\ell)$ and $\check{D}=D(\ell)$. Treating $\check{B}$ and $\check{D}$ as unknowns, 
the right-hand side of (\ref{hk0i3}) becomes a linear estimator of $h(\ell+j,\ell)$ denoted by 
$\hat{h}(\ell+j,\ell)$. We find $\check{B}$ and $\check{D}$ by solving the linear least-squares minimization 
problem
\begin{equation}\label{minCD2}
	\min_{\check{B} \in \mathbb{R}^{n \times m}, \, \check{D} \in \mathbb{R}^{p \times m}} g(\check{B},\check{D})
\end{equation}
where
\begin{equation}
	g({\check{B},\check{D}})=\sum_{j=0}^{2n} \|\hat{h}(\ell+j,\ell)-h(\ell+j,\ell)\|_F^2.
\end{equation}

The normal equations for $\check{B}$ are ${\mathcal H}_{2n,1}(\ell+1) = \hat{\mathcal O}_{2n}(\ell+1) 
\check{B}$ with $\hat{O}_{2n}(\ell+1)$ formed from the pair $(\hat{C}(\gamma_{i}),\hat{A}(\gamma_{i}))$. 
Furthermore, $\check{D}=D(l)$. 
Using the similarity relations, we derive
\begin{equation}\label{checkB2}
	\check{B} = \hat{\mathcal O}_{2n}^\dag(\ell+1) {\mathcal H}_{2n,1}(\ell+1)=
	\hat{\mathcal O}_{2n}^\dag(\ell+1)^\dag {\mathcal O}_{2n}(\ell+1) B(\ell) =T_{i} B(\ell). 
\end{equation}
Note for all $0\leq j \leq 2n$ that $\hat{h}(\ell+j,\ell)=h(\ell+j,\ell)$. 

If $\varphi(\ell)=\varphi(\gamma_i)$, $B(\ell)=B(\gamma_i)$, $D(\ell)=D(\gamma_i)$. Then, from (\ref{checkB2})
we derive $\check{B}=T_{i} B(\ell)=T_{i}B(\gamma_i)=\hat{B}(\gamma_i)$, and $\check{D}=D(\ell)=D(\gamma_i)=\hat{D}(\gamma_i)$. 
Conversely suppose $\check{B}=\hat{B}(\gamma_i)$ and $\check{D}=\hat{D}(\gamma_i)$, yet $\varphi(\ell) \neq 
\varphi(\gamma_i)$, {\em that is}, $\varphi(\ell)=\varphi(\gamma_{i-1})$. Then, $B(l)=B(\gamma_{i-1})$ and 
$D(l)=D(\gamma_{i-1})$. Hence, from (\ref{checkB2}) we derive $\check{B}=T_i B(\gamma_{i-1})=T_i T_{i-1}^{-1} 
\hat{B}(\gamma_{i-1})$ where $T_{i-1}$ is the similarity transformation in $\hat{\mathcal P}(\gamma_{i-1}) \sim {\mathcal P}
(\gamma_{i-1})$. Thus, $B(\gamma_i)=T_i^{-1}\hat{B}(\gamma_{i})=T_{i}^{-1}\check{B}=T_{i-1}^{-1} \hat{B}(\gamma_{i-1}) 
=B(\gamma_{i-1})$. Furthermore, $\check{D}=D(l)={D}(\gamma_{i-1})$ and $\check{D}=\hat{D}(\gamma_i)=D(\gamma_i)$
imply $\hat{D}(\gamma_i)=\hat{D}(\gamma_{i-1})$. Hence, $\varphi(\gamma_i) =\varphi(\gamma_{i-1})$ from (\ref{sd2}) 
in Assumption~\ref{assmp4}. We arrive a contradiction. We conclude that with $\check{D}=\hat{D}(\ell)$ and 
$\check{B}=\hat{\mathcal O}_{2n}^\dag(\ell+1) {\mathcal H}_{2n,1}(\ell+1)$,
$[\check{B}^T \;\check{D}^T]^T = [\hat{B}^T(\gamma_i)\;\hat{D}^T(\gamma_i)]^T 
\Longleftrightarrow \varphi(\ell)=\varphi(\gamma_i)$ when Assumption~\ref{assmp4} holds. Until $k_{i}-1$ 
is reached, $\ell$ is decreased. Since $\varphi(k_i-1) \neq \varphi(k_i)=\varphi(k_i+1)$, it is necessary 
to pass $k_i$ left to detect a switch at $k_i$.

Starting from $\beta_i+1$ the scheme presented above detects $k_{i+1}$ in $k_{i+1}-\beta_i-1$ steps, 
i.e., between $0$ and $2n+1$ steps, and starting from $\alpha_i-1$ it reaches to $k_i-1$ in 
$\alpha_i-k_i$ steps, i.e., between $0$ and $2n$ steps. The numbers of the required steps are the 
same both in this scheme and Algorithm~3.  The differences between the schemes appear to be the 
requirements on $\beta_i-\alpha_i$ and the switch detectability conditions. While 
$\delta_i(\chi) \geq 4n+2$ is stipulated on Algorithm~3, this scheme, called Algorithm~3$^\prime$ 
replaces it with $\beta_i-\alpha_i \geq 2n-1$. If $\delta_0(\chi) \geq (6+\nu)n+1$,
$\delta_*(\chi) \geq (4+\nu)n+1$, and $\delta_{i^*}(\chi) \geq (8+\nu)n$, each segment of $\chi$
contains an interval $S_i$ for some $0\leq i \leq i^*$ with at least $\nu n$ points. Set $\nu\geq 2$.
Then, the requirement $\beta_i-\alpha_i \geq 2n-1$ is satisfied. Note that Assumption~\ref{sysasmp2} 
already requires $\nu>5$. The steps of Algorithm~3$^\prime$ are outlined in Appendix. The results derived 
above are captured in the following.

\begin{theorem}\label{main232nl}
	Consider Algorithms~3$^\prime$ with the noiseless Markov parameters. Suppose that $\chi$ and $\mathcal{P}$ 
	satisfy Assumptions~\ref{sysasmp}--\ref{varphiassmp}, \ref{unimodality}, and \ref{assmp5}. Then,
	Algorithm~3$^\prime$ recovers $\varphi$ on segments with dwell times $\delta_0(\chi) \geq 8n+1$, 
	$\delta_i(\chi) \geq 6n+1$ for $0<i<i^*$, and 
	$\delta_{i^*}(\chi) \geq 10n$.
\end{theorem}

\subsection{Time complexity of Algorithm~3$^\prime$}
The most time-consuming operations in Algorithm~3$^\prime$ are the calculations of $\check{B}$ and $\check{C}$. 
Similarly to Section~\ref{sebsecalg3}, we find their time complexities $O(n^3)$. The number of the steps to hit a switch is 
at most $O(n)$. Hence, the detection of one switch or all will have worst-case time complexities $O(n^4)$ or $O(Nn^3)$,
respectively.  The latter may be observed when the switches are uniformly distributed in $\chi$.

\subsection{Robustness of  Algorithm~3$^\prime$}
First, note that the Lyapunov transformation $S(k)$ in Theorem~\ref{Alg1robustness} is absorbed in the
similarity transformations $T_i$ in the formulas for $\check{C}$ and $\check{B}$. Therefore, $\mathcal{P}^\varepsilon(k)$
may replace the perturbed realization $\tilde{\mathcal P}(k)$ in the calculations. From Theorem~\ref{Alg1robustness},
$\mathcal{P}^\varepsilon(k) \rightarrow \hat{\mathcal P}^\varepsilon(k)$ as $\varepsilon \rightarrow 0$. 
As a result the threshold criteria in Algorithm~3$^\prime$ never fail as $\varepsilon \rightarrow 0$ under Assumptions~\ref{sysasmp}, \ref{unimodality}--\ref{sysasmp2}, and \ref{assmp5}. The following result extends
Theorem~\ref{main232nl} to the noisy Markov parameters case.

\begin{theorem}\label{main232robust}
	Consider Algorithms~3$^\prime$ with the noisy Markov parameters in (\ref{corruptMarkov}). Suppose that $\chi$ 
	and $\mathcal{P}$ satisfy Assumptions~\ref{sysasmp}--\ref{varphiassmp}, \ref{unimodality}, and \ref{assmp5}. 
	Then, Algorithm~3$^\prime$ recovers $\varphi$ on segments with dwell times $\delta_0(\chi) \geq 8n+1$, 
	$\delta_i(\chi) \geq 6n+1$ for $0<i<i^*$, and $\delta_{i^*}(\chi) \geq 10n$ as $\varepsilon \rightarrow 0$.
\end{theorem}

\subsection{Estimation of $\varphi(k)$ on very short segments}\label{switch-estIII}

Algorithms~2 and 3 or 3$^\prime$ combined estimates $\varphi$ on segments with length at least $4n+2$, 
but leaves it undetermined on shorter segments. Let $[k_i,k_{i+1})$ be such a segment and suppose 
that $\hat{\mathcal P}(\gamma_{i-1})$ and $\varphi(k)$ were estimated on $[k_{i-1},k_i)$, but 
$\hat{\mathcal P}(\gamma_{i})$ and $k_{i+1}$ need to be estimated. Here, we are considering a situation
in which the discrete state set is completely identified up to a similarity transformation, but we don't know 
which discrete state is active in $[k_i,k_{i+1})$ because none of the dwell time constraints in either 
Theorem~\ref{main23} or Theorem~\ref{main232nl} apply to $\delta_i(\chi)$. Set $k=k_i+n$, $q=n+1$, 
and $r=n$ in (\ref{gHankel}) if $k_i+2n \leq  k_{i+1}-1$, i.e., $\delta_i(\chi) >2n$. 
The resulting Hankel matrix can be expressed in terms of the state-space matrices of
${\mathcal P}(\gamma_i)$ and $\hat{\mathcal P}(\gamma_i)$ as follows
\begin{eqnarray}
	{\mathcal H}_{n+1,n}(k_i+n) &=& \left[\begin{array}{ccc} C(\gamma_i) B(\gamma_i) &  & C(\gamma_i) [A(\gamma_i)]^{n-1} 
	B(\gamma_i) \\  & \ddots & \vdots \\ C(\gamma_i) [A(\gamma_i)]^n B(\gamma_i) & \cdots & C(\gamma_i) A^{2n-1}(\gamma_i)
    B(\gamma_i) \end{array} \right] \nonumber \\
	&=& \left[\begin{array}{ccc} \hat{C}(\gamma_i) \hat{B}(\gamma_i) &  & \\  & \ddots & \vdots 
	\\   & \cdots & \hat{C}(\gamma_i) [\hat{A}(\gamma_i)]^{2n-1}\hat{B}(\gamma_i) \end{array} \right] \\
	&=& \hat{\mathcal H}(\hat{\mathcal P}(\gamma_i)). \hspace{5cm} \label{hoppaladedik}
\end{eqnarray}
Note the factorization ${\mathcal H}_{n+1,n}(k_i+n)={\mathcal O}_{n+1} {\mathcal O}_n$ which determines 
${\mathcal P}(\gamma_{i})$ uniquely up to a similarity transformation. This observation suggests that 
${\mathcal P}(\gamma_{i})$ can be identified easily as
\begin{equation}\label{oflenbe1}
	\hat{\mathcal P}(\gamma_i) =\arg \min_{j \in \mathbb{S}} 
	\|{\mathcal H}_{n+1,n}(k_i+n)-\hat{\mathcal H}(\hat{\mathcal P}(\gamma_j))\|_F.
\end{equation}
Then, we declare $\varphi(k)=\varphi(\gamma_i)$ on the interval $[k_i,k_i+2n]$. Since the length of 
this interval is $2n+1$, by Algorithm~3$^\prime$ we can extend $\hat{\mathcal P}(\gamma_i)$ to $(k_i+2n,k_{i+1})$
until we hit $k_{i+1}$. The entire process starts again at $k_{i+1}$ and continues until left endpoint of a large 
segment is encountered. For the segment $[k_{i^*}\;\;k^{\prime\prime}]$, we only need to estimate $\hat{P}(\gamma_{i^*})$
since $k^{\prime\prime}$ is not a switch. Substituting $k^{\prime\prime}=k_{i+1}-1$ in the above calculations,
we derive the recovery condition $k^{\prime\prime}-k_{i^*} \geq 2n$. From $k^{\prime\prime}=N-4n$ and 
$\delta_{i^*}(\chi)=N-k_{i^*}$, we get $\delta_{i^*}(\chi) \geq 6n$. For the segment $[k^\prime\;\;k_1]$, if $k_1$
is not known this procedure is applied to estimate both ${\mathcal P}(\gamma_0)$ and $k_1$ by substituting $k_0=k^\prime$. 
The recovery condition is then, $k_1-k^\prime >2n$. From $k^\prime=2n+1$ and $\delta_0(\chi)=k_1-1$, we get 
$\delta_0(\chi) > 4n$.

In the second case, we assume that $\hat{\mathcal P}(\gamma_{i+1})$ and $\varphi(k)$ have been estimated on 
$[k_{i+1},k_{i+2})$ and we want to determine $\hat{\mathcal P}(\gamma_{i})$, $k_i$, and $\varphi(k)$ on 
$[k_i,k_{i+1})$. This is the previous case if one notes that $\hat{\mathcal H}(\hat{\mathcal P}(\gamma_i))$
does not change when ${\mathcal H}_{n+1,n}(k_i+n)$ is replaced with ${\mathcal H}_{n+1,n}(k_{i+1}-n-1)$.  
We then determine $\hat{\mathcal P}(\gamma_i)$ from 
\begin{equation}\label{oflenbe2}
	\hat{\mathcal P}(\gamma_i) =\arg \min_{j \in \mathbb{S}} 
	\|{\mathcal H}_{n+1,n}(k_{i+1}-n-1)-\hat{\mathcal H}(\hat{\mathcal P}(\gamma_j))\|_F.
\end{equation}
The condition on $\delta_i(\chi)$ is found $\delta_i(\chi) > 2n$ as before from the inequalities 
$k=k_{i+1}-n-1$ and $k-n \geq k_i$.  For the segment $[k^\prime\;\;k_1]$, if $k_1$
is not known this procedure is applied to estimate both ${\mathcal P}(\gamma_0)$ and $k_1$ by substituting $k_0=k^\prime$. 
The recovery condition is then, $k_1-k^\prime >2n$. From $k^\prime=2n+1$ and $\delta_0(\chi)=k_1-1$, we get 
$\delta_0(\chi) > 4n$. For the segment $[k_{i^*}\;\;k^{\prime\prime}]$, if $k_{i^*}$
is not known this procedure is applied to estimate both ${\mathcal P}(\gamma_{i^*})$ and $k_{i^*}$ by substituting 
$k_{i^*+1}=k^{\prime\prime}+1$. We summarize the above calculations in Algorithm~3$^{\prime\prime}$
detailed in Appendix. The result derived above is captured in

\begin{theorem}\label{main250}
	Consider Algorithms~3$^{\prime\prime}$ with the noise-free Markov parameters. Suppose that $\chi$ and $\mathcal{P}$ 
	satisfy Assumptions~\ref{sysasmp}--\ref{varphiassmp}, and \ref{unimodality}. Then, 
	Algorithm~3$^{\prime\prime}$ recovers $\varphi$ on segments with dwell times satisfying 
	$\delta_0(\chi) > 4n$, $\delta_i(\chi) \geq 2n+1$ for $0<i<i^*$, and 
	$\delta_{i^*}(\chi) \geq 6n$.
\end{theorem}

\subsection{Time complexity of Algorithm~3$^{\prime\prime}$} 
    Computationally two expensive processes are the calculations of $\hat{\mathcal P}(\gamma_j)$ in (\ref{oflenbe1})
    and (\ref{oflenbe2}) for $j=1,\cdots,\sigma$, but they are done only once. Moreover, it suffices to consider
    only the calculation of $\hat{C}(\gamma_i)[\hat{A}(\gamma_i)]^{2n-1}\hat{B}(\gamma_i)$. Recall that eigen decomposition
    has time complexity $O(n^3)$. Then, $[\hat{A}(\gamma_i)]^{\ell}$ for $\ell=1,\cdots,2n-1$  are efficiently
    calculated from this decomposition. Products of three matrices forming Markov parameters have time complexity $O(n^3)$.
    Thus, $\hat{\mathcal P}(\gamma_j)$ has time complexity $O(n^3)$ for each $j$. There are $\sigma$ discrete states, hence
    the overall time complexity of (\ref{oflenbe1}) or (\ref{oflenbe2}) becomes $O(\sigma n^3)$. Now, (\ref{oflenbe1}) or 
    (\ref{oflenbe2}) are followed by Algorithm~3$^\prime$ which has time complexity $O(n^4)$. It follows that 
    Algorithm~3$^{\prime\prime}$ has complexity $O(d_1 \sigma n^3+d_2 n^4)$ for some $d_1,d_2>0$ to detect one switch.
    If the switches are evenly distributed in $\chi$, this complexity grows to $O(d_1 \sigma Nn^2+d_2 Nn^3)$. 
    
\subsection{Robustness of Algorithm~3$^{\prime\prime}$}      
Since Markov parameters of topologically equivalent systems are identical, from Theorem~\ref{Alg1robustness} 
the perturbed Hankel matrix in (\ref{hoppaladedik}) satisfies $\tilde{\mathcal H}({\mathcal P}(\gamma_i)) 
\rightarrow {\mathcal H}({\mathcal P}(\gamma_i))$ as $\varepsilon \rightarrow 0$. Then, Algorithm~3$^{\prime\prime}$
uniquely identify the active discrete state. The rest follows from the robustness of Algorithm~3$^\prime$ to amplitude
bounded noise. We summarize this result in the following.

\begin{theorem}\label{main2500}
	Consider Algorithms~3$^{\prime\prime}$ with the noisy Markov parameters in (\ref{corruptMarkov}). Suppose 
	$\chi$ and $\mathcal{P}$ satisfy Assumptions~\ref{sysasmp}--\ref{varphiassmp}, and \ref{unimodality}. Then, 
	Algorithm~3$^{\prime\prime}$ recovers $\varphi$ on segments with dwell times satisfying $\delta_0(\chi) > 4n$, 
	$\delta_i(\chi) \geq 2n+1$ for $0<i<i^*$, and 
	$\delta_{i^*}(\chi) \geq 6n$ as $\varepsilon \rightarrow 0$.
\end{theorem}

In passing, among the recovery conditions the least stringent is the one in Theorem~\ref{main250}.
The next one is stated in Theorem~\ref{main23}. When Assumption~\ref{sysasmp2} holds, any one of 
Algorithms~3--3$^{\prime\prime}$ may be combined with Algorithm~2 to form a meta-algorithm since
all switch detection conditions will automatically be satisfied by Algorithm~2. Therefore, a meta 
algorithm may include them all. In Section~\ref{912}, this will be demonstrated in a numerical example.

\section{Basis transformation problem for discrete state estimates}~\label{basistransform}

The choice of a basis for the discrete state estimates requires attention if the estimated SLS model 
will be used for predicting outputs to prescribed input sequences. To understand how this issue arises, 
re-write the Markov parameters in (\ref{defMarkov})
\begin{equation}\label{X1}
h(k,l)=C(k)A(k-1) \cdots A(l+1)B(l), \;\;k>l+1,
\end{equation}
$h(k,k-1)=C(k)B(k-1)$, and $h(k,k)=D(k)$. Recall that the discrete state estimates in $\hat{\mathcal P}$ are
similar to the true ones in ${\mathcal P}$, {\em that is}, for every ${\mathcal P}_j \in {\mathcal P}$,
there exists a $\hat{\mathcal P}_j \in \hat{\mathcal P}$ and a $T_j \in \mathbb{R}^{n \times n}$
satisfying $A_j=T_j^{-1} \hat{A}_j T_j$, $C_j=\hat{C}_j T_j$, $B_j=T_j^{-1} \hat{B}_j$, and $D_j=\hat{D}_j$ where
${\mathcal P}_j=(A_j,B_j,C_j,D_j)$ and $\hat{\mathcal P}_j=(\hat{A}_j,\hat{B}_j,\hat{C}_j,\hat{D}_j)$.
In (\ref{X1}), plug $k=k_i+\xi$, $0 \leq \xi \leq n$ and $l=k_i-\eta$, $1 \leq \eta \leq n$ in. Suppose 
${\mathcal P}_{j_1}$ and ${\mathcal P}_{j_2}$ are the active discrete states on $[k_{i-1},k_i)$ and $[k_i,k_{i+1})$,
respectively. Assuming $\delta_*(\chi) \geq 2n$ and using the similarity relations for ${\mathcal P}_{j_1}$ and 
${\mathcal P}_{j_2}$, we derive
\begin{equation}\label{X11}
h(k_i+\xi,k_i-\eta)=\hat{C}_{j_2} \hat{A}_{j_2}^\xi T_{j_2} T_{j_1}^{-1}\hat{A}_{j_1}^{\eta-1}\hat{B}_{j_1}.
\end{equation}
This equation covers all possibilities for $T_{j_2} T_{j_1}^{-1}$ since it is at the leftmost position 
when $\xi=0$ and at the rightmost position when $\eta=1$. Note that $T_{j_2} T_{j_1}^{-1}$ appears only once 
in the string
of state-space matrices since $\xi+\eta \leq 2n$. 

Fix $\eta \geq 1$, evaluate (\ref{X11}) for $\xi$ between $1$ to $n$, and concatenate the resulting 
equations to derive
\begin{equation}\label{esek1}
Z_{1,\eta} \stackrel{\Delta}{=} \left[\begin{array}{c} h(k_i+1,k_i-\eta)  \\ \vdots \\ 
	h(k_i+n,k_i-\eta) \end{array} \right] = \hat{\mathcal O}_n(j_2)  X_\eta 
\end{equation}
where $\hat{\mathcal O}_n(j_2)$ is the extended observability matrix constructed from the pair 
$\hat{C}_{j_2}$ and $\hat{A}_{j_2}$ and
\begin{equation}\label{X2} 
X_\eta=\hat{A}_{j_2} T_{j_2} T_{j_1}^{-1} \hat{A}^{\eta-1}_{j_1}\hat{B}_{j_1}.
\end{equation}
Then, from (\ref{esek1}) 
\begin{equation}\label{esek2}
	X_\eta=\hat{\mathcal O}_n^\dag (j_2) Z_{1,\eta}, \qquad \eta=1,\cdots,n.
\end{equation}
Evaluate (\ref{X2}) for $\eta=1,\cdots,n$ and concatenate them to get
\begin{equation}\label{esek3}
	[X_1 \;\cdots\;X_n]=\hat{A}_{j_2} T_{j_2} T_{j_1}^{-1}\hat{\mathcal C}_n(j_1)
\end{equation}
where $\hat{\mathcal C}_n(j_1)$ is the extended controllability matrix constructed from the pair 
$\hat{A}_{j_1}$ and $\hat{B}_{j_1}$.
Then,
\begin{eqnarray}
	\hat{A}_{j_2} T_{j_2} T_{j_1}^{-1} &=& [X_1 \;\cdots\;X_n]\hat{\mathcal C}_n^\dag(j_1) \nonumber
	\\[-1.5ex] \label{esek4} \\[-1.5ex] 
	&\stackrel{\Delta}{=}& Y_{i,i-1}. \nonumber
\end{eqnarray}
Assuming $A_{j_2}$ has no eigenvalues at the origin, we get $T_{j_2} T_{j_1}^{-1}=\hat{A}^{-1}_{j_2} 
Y_{i,i-1}$. We derive $T_{j_1}T_{j_2}^{-1} =Y_{i,i-1}^{-1}\hat{A}_{j_2}$ on inverting this equation. 
Suppose we want to estimate $T_{j_3} T_{j_1}^{-1}$. Write it as a product 
$T_{j_3} T^{-1}_{j_1}=(T_{j_3} T_{j_2}^{-1})(T_{j_2} T_{j_1}^{-1})$ which is readily calculated if we know 
the products. At our disposition, we have Assumption~\ref{sysasmp2} which tells us that every discrete state 
can be reached from another by crossing a finite number of the switches. Since there are only $\sigma$ 
discrete states, we start from $\hat{P}_\sigma$, estimate $T_\sigma T_{\sigma-1}^{-1}$, move to 
$T_{\sigma-1} T_{\sigma-2}^{-1}$, and so on. This procedure is repeated $\sigma-1$ times. Some steps may not 
be implemented directly. In this case, a path is chosen and the inversion and the product 
formulas are applied. 

We define a basis transform by letting $\check{A}_j=\Pi_j^{-1} \hat{A}_j\Pi_j$, 
$\check{B}_j=\Pi_j^{-1} \hat{B}_j$, $\check{C}_j=\hat{C}_j\Pi_j$, $\check{D}_j=\hat{D}_j$ 
where $\Pi_j=T_j T_1^{-1}$ and setting $\check{\mathcal P}_j=(\check{A}_j,\check{B}_j,\check{C}_j,\check{D}_j)$.
Rewrite (\ref{X11}) with the state-space matrices of $\check{\mathcal P}_{j_1}$ and $\check{\mathcal P}_{j_2}$:
\begin{eqnarray*}
	h(k_\ell+\xi,k_\ell-\eta)&=&\check{C}_{j_2} \check{A}^\xi_{j_2} \Pi_{j_2}^{-1} 
	T_{j_2} T_{j_1}^{-1} \Pi_{j_1}\check{A}^{\eta-1}_{j_1}\check{B}_{j_1} \\
	&=& \check{C}_{j_2} \check{A}^\xi_{j_2} T_1 T_{j_2}^{-1} T_{j_2} T_{j_1}^{-1} T_{j_1}T_1^{-1}
	\check{A}^{\eta-1}_{j_1}\check{B}_{j_1} \\
	&=& \check{C}_{j_2} \check{A}^\xi_{j_2} \check{A}^{\eta-1}_{j_1}\check{B}_{j_1}.
\end{eqnarray*}
But,
\begin{eqnarray*}
	{\check C_{{j_2}}} &=& {{\hat C}_{{j_2}}}{\Pi _{{j_2}}} = {C_{{j_2}}}T_{{j_2}}^{ - 1}{T_{{j_2}}}
	T_{{1}}^{ - 1} = {C_{{j_2}}} T_{{1}}^{ - 1},\\
	{\check A_{{j_2}}} &=& \Pi _{{j_2}}^{ - 1}{{\hat A}_{{j_2}}}{\Pi _{{j_2}}} = {T_{{1}}} 
	T_{{j_2}}^{ - 1}{T_{{j_2}}}{A_{{j_2}}}T_{{j_2}}^{ - 1}{T_{{j_2}}}T_{{1}}^{ - 1} = {T_{{1}}}A_{{j_2}}
	T_{{1}}^{ - 1},\\
	{\check A_{{j_1}}} &=& \Pi _{{j_1}}^{ - 1}{{\hat A}_{{j_1}}}{\Pi _{{j_1}}} = {T_{{1}}}T_{{j_1}}^{ - 1}{T_{{j_1}}}{A_{{j_1}}}T_{{j_1}}^{ - 1}{T_{{j_1}}}T_{{1}}^{ - 1} = {T_{{1}}}{A_{{j_1}}}T_{{1}}^{ - 1},\\
	{\check B_{{j_1}}} &=& \Pi _{{j_1}}^{ - 1}{{\hat B}_{{j_1}}} ={T_{{1}}} T_{{j_1}}^{ - 1}{T_{{j_1}}}{B_{{j_1}}} = {T_{{1}}}			{B_{{j_1}}}.
\end{eqnarray*}

Replacing back leads to $h({k_\ell } + \xi ,{k_\ell } - \eta ) = {C_{{j_2}}}A_{{j_2}}^\xi A_{{j_1}}^{\eta  - 1}{B_{{j_1}}}$. 
Hence, the Markov parameters of (\ref{ssx})--(\ref{varphit}) are matched as desired. Since ${\Pi _1}=I_n$, there is no need 
to operate on $\hat{\mathcal P}_{j_1}$. Clearly, this matching is applicable to any pair of the discrete states and the 
Markov parameters with large lags exhibiting several switches. The transformed discrete states and the switching signal may 
be used to calculate the response of the system to arbitrary inputs. The steps of this transformation is formalized as 
Algorithm~4 in Appendix. Time complexity of $\Pi_j$ is $O(poly(n))$ for each $j$. The worst-case time complexity 
of Algorithm~4 is then $O(N poly(n))$ and it happens when the switches are uniformly distributed in $\chi$ and $j$ assumes 
values from a set with cardinality $O(Nn^{-1})$. Before summarizing the result derived in this section, we state the sole 
requirement as follows.

\begin{assumption}\label{asmmpbasis}
	The discrete states of the MIMO--SLS model (\ref{ssx})--(\ref{varphit}) have no poles at zero.
\end{assumption}

\begin{theorem}
	Suppose Assumption~\ref{asmmpbasis} holds and $\delta_*(\chi) \ge 2n$. Then, Algorithm~4 solves the
	basis transformation problem  for the MIMO--SLS model (\ref{ssx})--(\ref{varphit}).
\end{theorem}
	
\section{From Markov parameter estimates to MIMO-SLS models}\label{hyperalg}
\begin{figure}[hbt!]
	\centering
	\includegraphics[width=1\textwidth]{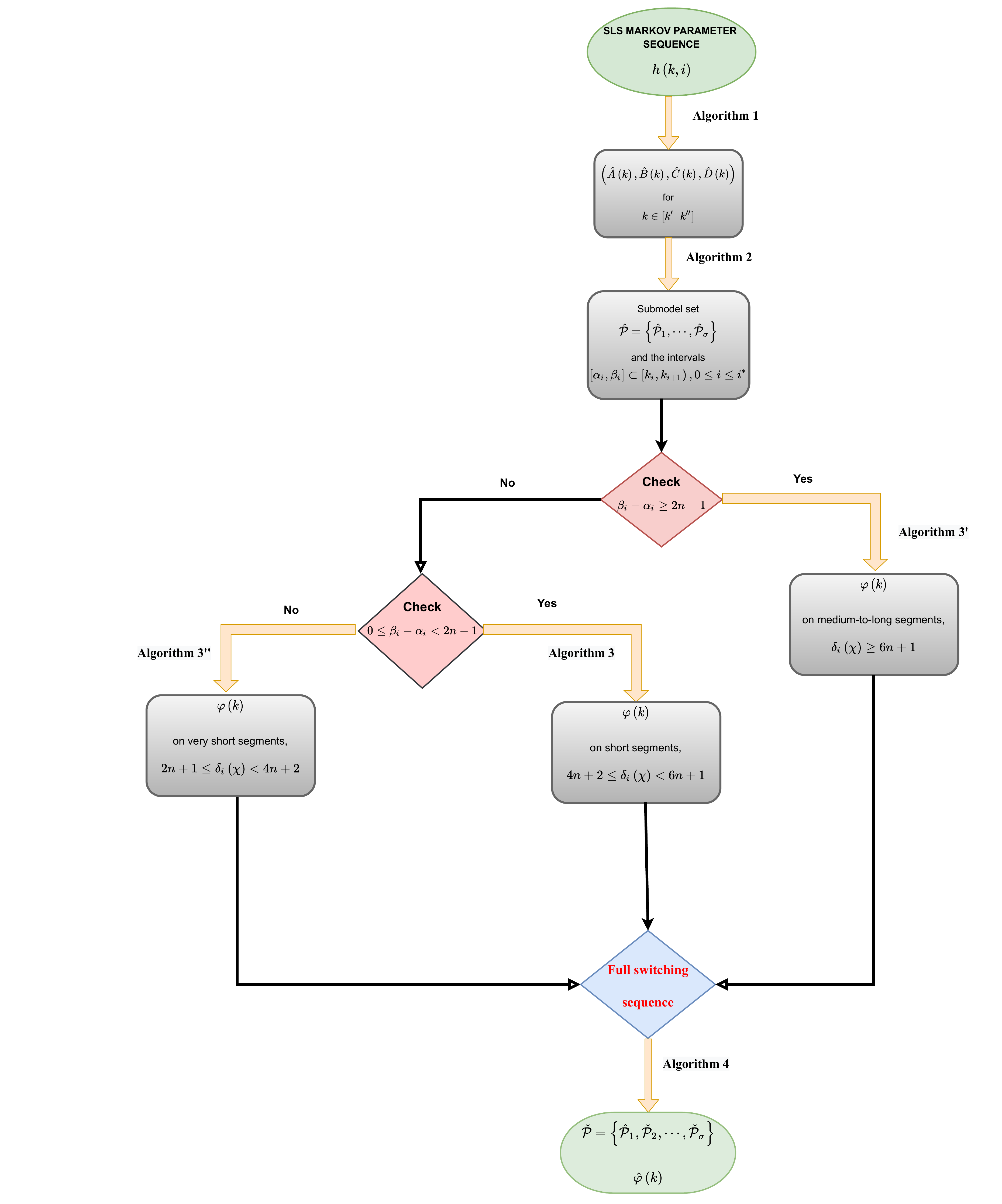}
	\caption{A meta-algorithm for the SLS realization from Markov parameters sequence.} 
	\label{zamazingo}
\end{figure}

In this section, we combine the results derived in Sections~\ref{probform}--\ref{basistransform} to form 
a meta-algorithm to identify the MIMO-SLS model (\ref{ssx})--(\ref{varphit}) from 
its noisy Markov parameters. The steps of this meta-algorithm are illustrated in Figure~\ref{zamazingo} 
as a flowchart. We collect the results derived in Sections~\ref{LTVhankel}--\ref{basistransform} in the following main result.

\begin{theorem}
	Consider (\ref{ssx})--(\ref{varphit}) with the noisy Markov parameters in (\ref{corruptMarkov}). Assume that $\chi$, and 
	$\mathcal{P}$ satisfy Assumptions~\ref{sysasmp}--\ref{varphiassmp}, \ref{unimodality}, \ref{assmp4}, \ref{assmp5}, and \ref{asmmpbasis}. 
	If Assumption~\ref{sysasmp2} holds, then the meta-algorithm recovers ${\mathcal P}$ and $\chi$ as $\varepsilon \rightarrow 0$. 
	Suppose that the first part of Assumption~\ref{sysasmp2} holds and $\varphi$ satisfies a mixing condition. If each segment in 
	$\chi$ satisfies a dwell requirement in one of Theorems~\ref{main23}, \ref{main232nl}, and \ref{main250}, then the meta-algorithm 
	recovers ${\mathcal P}$ and $\chi$ as $\varepsilon \rightarrow 0$.
\end{theorem}	 

In all algorithms we developed so far, observability and controllability concepts have played a 
foundational role. We have assumed that the discrete states have a common Macmillan degree. When this 
assumption is dropped, identifiability, controllability, and observability for the MIMO-SLS models 
become much harder to analyze. For example, identifiability of an SLS does not imply identifiability 
of its submodels \cite{Petreczky&Bako&Schuppen:2010}. Pivotal role played by observability in the 
estimation/realization of the SLSs has been noted early in hybrid system literature  
\cite{Vidal&Chiuso&Soatto:2002}. The observability part of Lemma~\ref{lemreal} appeared as 
Theorem~2 in \cite{Bakoetal:2009a}. Extension of the Ho-Kalman realization theory for the LPV state-space 
models was reported in \cite{Petreczkyetal:2017}. 

This paper did not address the issue of estimating Markov parameters from input-output data. The reasons are twofold. 
First, the realization problem for SLSs is interesting in its own \cite{Petreczky:2011,Petreczkyetal:2013,Petreczky&Schuppen:2010,Hossain&Trenn:2021}. See also the references therein.
An equally important subject in SLSs is model order reduction. We refer to recent works \cite{Birouche&Mourllion&Basset:2012,Petreczky&Wisniewski&Leth:2013,Goseaetal:2018}. Second, we decoupled identification
and realization problems and treated the latter in great detail in this paper though both can be treated in a unified manner.
This approach has several merits. Identification of an  SLS is normally conducted in two stages: estimation of the 
discrete states and detection of the switches, with the order being interchangeable. The realization problem solved in this paper
recovers the switching sequence in addition to the discrete states. The switching sequence is an arbitrary time-varying
signal subject to mild restrictions in contrast to existing works in hybrid systems literature. By lifting the issue of
estimating the switching sequence to the realization stage, attention is directed to estimating the Markov parameters
in parsimonious models. 

One approach for estimating Markov parameters is to use multiple trajectories when it is possible \cite{Majjietal:2010,Verhaegen&Yu:95,Liu:1997}. But, this approach has limited applications to system identification. 
Markov parameter estimation problem is not tractable for generic LTV systems due to the curse of dimensionality, 
unless further system assumptions such as slowly varying or smooth, mode switching with long waiting times, {\em etc.} 
are made. In \cite{Ohlsson&Ljung:2013}, switched ARX models were identified by a kernel based estimation method. This 
method encompasses the parameter estimation and the switch detection stages by solving an optimization problem. Recent 
work \cite{Bencherki&Turkay&Akcay:2020} used deadbeat observers to transform state-space identification problem
into an SARX identification problem. This transformation generates parsimonious models by packing infinite strings of
Markov parameters into finite sequences. In principle, any of these methods can be used to estimate Markov parameters.
Work is under progress for direct estimation of the Markov parameters from input-output data.

\section{Numerical example}\label{example}
Consider the following MIMO--SLS state-space model adapted from \cite{Bakoetal:2013} 
\begin{eqnarray*}
	{A_1} &=& \left[ {\begin{array}{*{20}{c}}{0.15}&{0.40}&{ - 0.65}\\
			{ - 0.75}&{0.1}&{ - 0.35}\\{0.20}&{0.70}&{0.20}\end{array}} \right],\;\;
	{B_1} = \left[ {\begin{array}{*{20}{c}}{ - 0.20}&{0.45}\\{ - 0.06}&0\\{0.22}&0\end{array}} \right]\\
	{C_1} &=& \left[ {\begin{array}{*{20}{c}}0&{0.40}&{0.45}\\{ - 1}&{ - 0.60}&{0.90}\end{array}} \right],\;\;
	{D_1} = \left[ {\begin{array}{*{20}{c}}0&{ - 0.35}\\{ - 1.70}&{ - 0.25}\end{array}} \right]
\end{eqnarray*}

\begin{eqnarray*}
	{A_2} &=& \left[ {\begin{array}{*{20}{c}}{0.27}&{0.24}&{ - 0.55}\\
			{0.24}&{0.65}&{0.30}\\{ - 0.55}&{0.30}&{0.27}\end{array}} \right],\;\;
	{B_2} = \left[ {\begin{array}{*{20}{c}}{ - 0.55}&0\\{ - 1.40}&1\\{0.05}&{ - 0.72} \end{array}} \right]\\
	{C_2} &=& \left[{\begin{array}{*{20}{c}}{0.70}&1&{ - 0.27}\\{ - 0.35}&0&{ - 1.10} \end{array}} \right],\;\;
	{D_2} = \left[ {\begin{array}{*{20}{c}}{2.15}&{0.25}\\0&{ - 0.36}\end{array}} \right]
\end{eqnarray*}

\begin{eqnarray*}
	{A_3} &=& \left[ {\begin{array}{*{20}{c}}{0.45}&{0.02}&{0.42}\\{-0.17}&{0.53}&{0.20}\\
			{0.38}&{0.26}&0\end{array}}\right],
	{B_3} = \left[{\begin{array}{*{20}{c}}0&{0.15}\\{0.27}&{ - 0.46}\\ {0.07}&{0.54}\end{array}} \right],\;\;\\
	{C_3} &=& \left[ {\begin{array}{*{20}{c}}0&{0.60}&{0.28}\\0&{0.86}&{0.45}\end{array}} \right],\;\;
	{D_3} = \left[ {\begin{array}{*{20}{c}}0&{ - 0.90}\\0&{0.85}\end{array}} \right]
\end{eqnarray*}
where the problem of identifying a state-space SLS model for the case when the continuous state is known was treated. 
In the sequel, we address the problem of discrete state estimation given an exact sequence of Markov parameters.

\subsection{SLS estimation in a noiseless setup}

We first deal with the estimation of the discrete-states.

\subsubsection{Discrete state estimation}\label{911}

\begin{figure}[hbt!]
	\centering
	\includegraphics[width=0.5\textwidth]{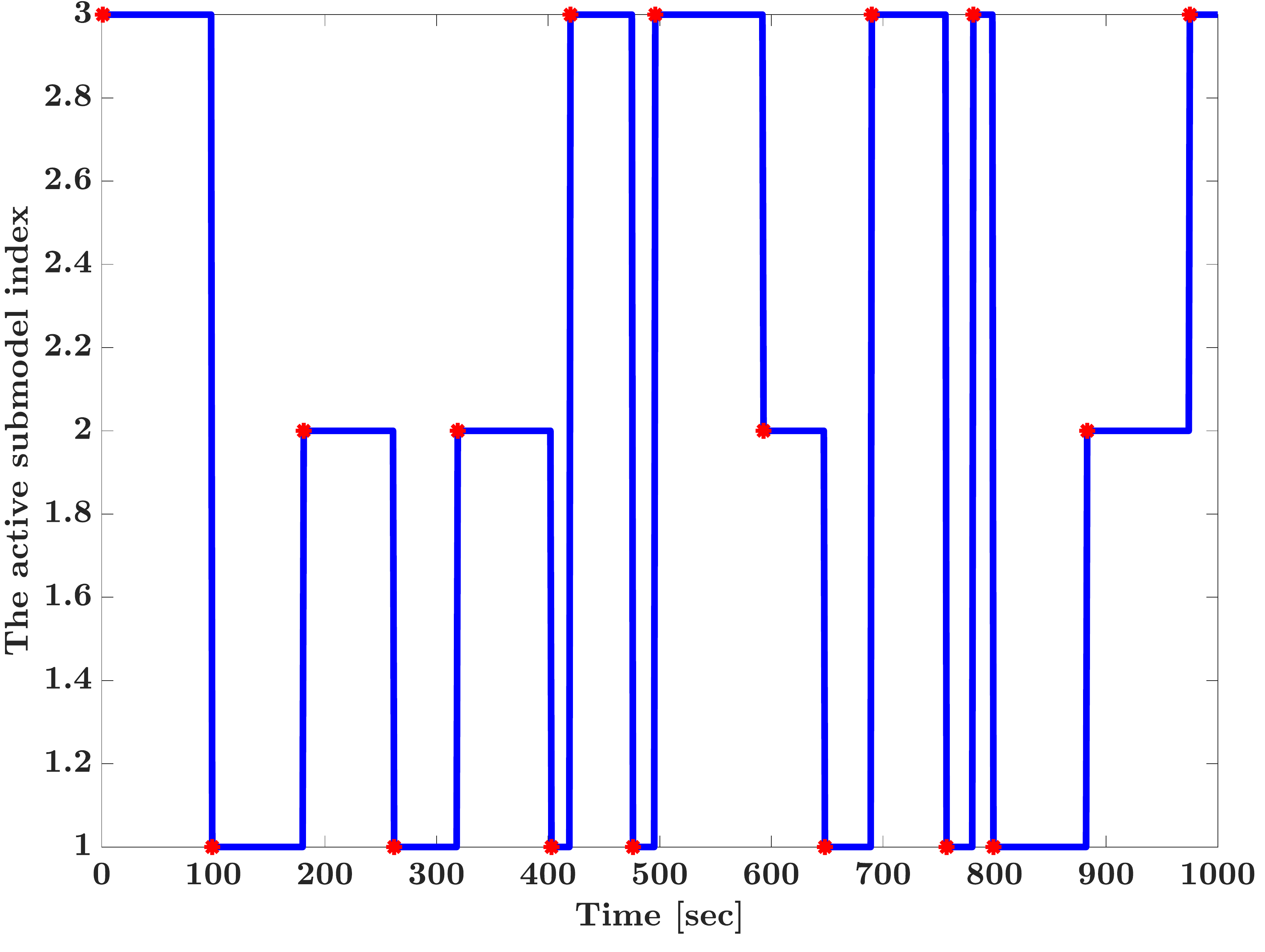}
	\caption{The switching sequence in the example.}
	\label{fig1}
\end{figure}

\begin{figure}[hbt!]
	\centering
	\begin{subfigure}{.45\linewidth}
		\centering
		\includegraphics[width=1\textwidth]{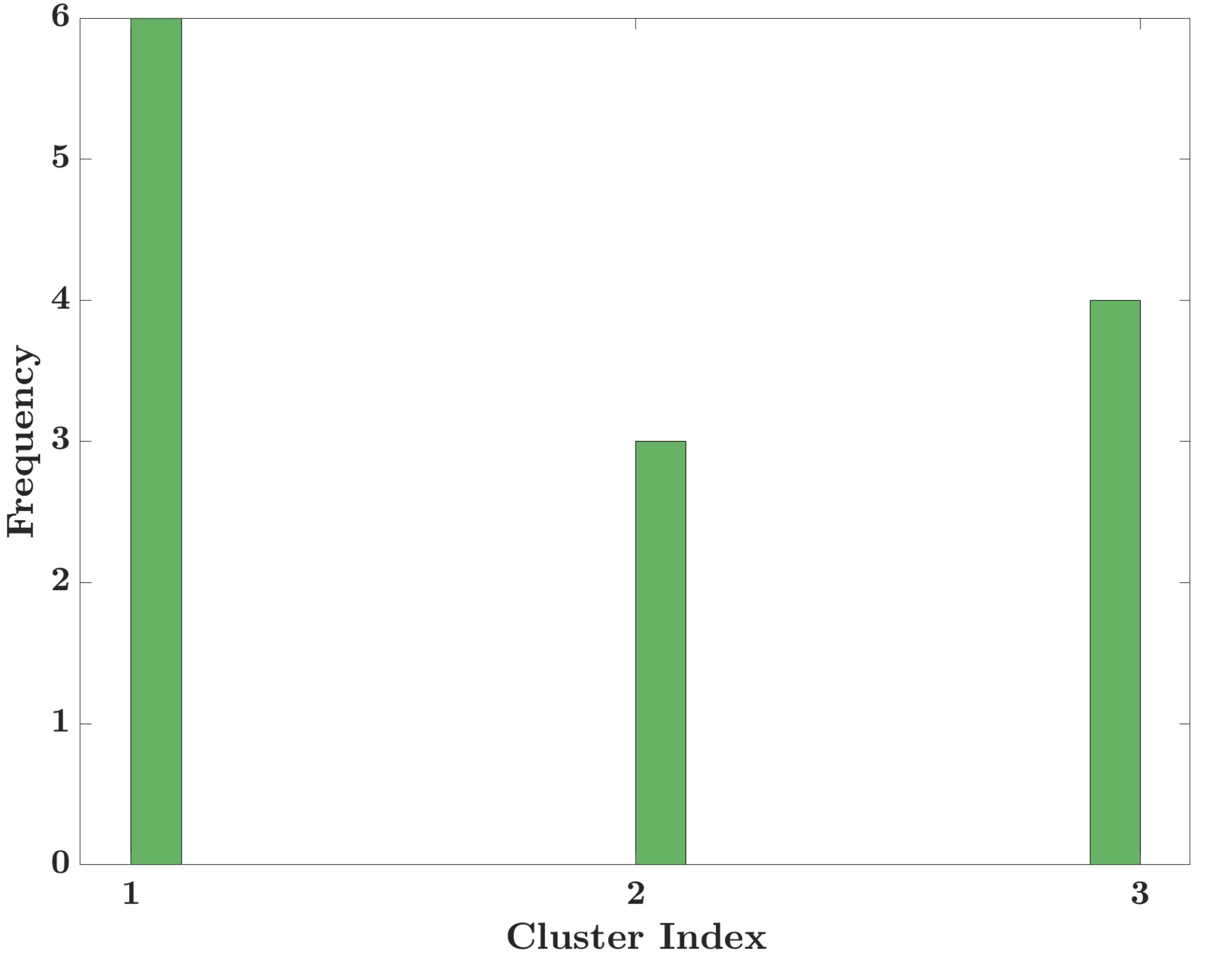}
		\caption{The histogram of clusters over the $1$-d feature space $\left\{ {{\cal M}
				(\hat A({\gamma _i})),i \in \mathcal I} \right\}$.}
		\label{fig2}
	\end{subfigure}
	\begin{subfigure}{.45\linewidth}
		\centering
		\includegraphics[width=1\textwidth]{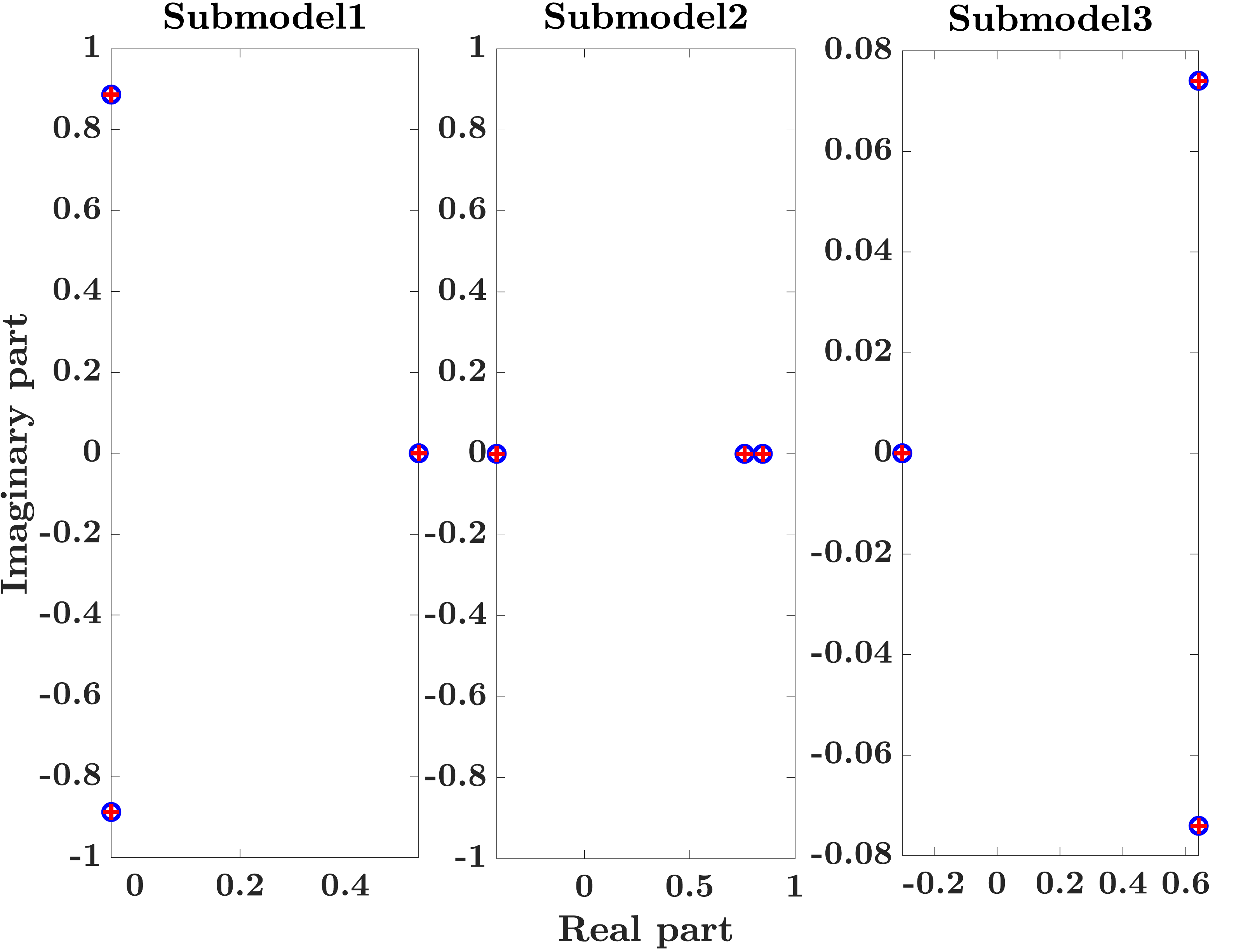}
		\caption{The estimated '$\circ$' and the true '{\bf +}' submodel eigenvalues in a noiseless setup. }
		\label{fig3}
	\end{subfigure}
	\caption{The discrete state estimation via Algorithm~2 (see Appendix).}
\end{figure}

A switching sequence that conforms with Assumption~4.2 in all but few segments was generated via sampling from a random 
uniform distribution. It is shown in Figure~\ref{fig1} where the red dots represent the switches. The discrete state estimation 
was accomplished via Algorithm~2. As inputs to the algorithm, we selected $\epsilon_{\mathbb{Z}}=10^{-4}$, $\nu=6$, and estimated 
the state-space quadruples $\hat{\mathcal P}_k$ over the time range $k' \le k \le k''$ from Algorithm~1. The {\tt dbscan} 
command in MATLAB was implemented by selecting {\it epsilon=$10^{-5}$} as a threshold value for the neighborhood search 
radius  and {\it minpts=1} for the minimum number of neighbors. As an entry to {\tt dbscan}, we supplied
${\mathcal M}(\hat{A}(\gamma_i))$ as the clustering feature where $\gamma_i$ is the  midpoint of the targeted 
intervals which satisfy $\beta_i-\alpha_i \ge 6n$. Let $\mathcal{I} = \left\{ {i:\;{\beta _i}- {\alpha _i} \ge vn} \right\}$
denote the set of indices satisfying this constraint. Figure~\ref{fig2} shows the clustering result over 
${\mathcal M}(\hat{A}(\gamma_i))$, $i \in \mathcal I$. As anticipated, $\sigma$ is correctly estimated, i.e., 
$\hat \sigma  = 3$. Using clusters, we now determine the set of submodel estimates $\hat{\cal P}_j$, $j \in \mathbb{S}$.
Notice that not all segments satisfy $\beta_i-\alpha_i \geq \nu n$, yet recovery is perfect.

To assess the fitting capability of the estimates, the submodel estimates' eigenvalues were compared to the true ones 
as shown in Figure~\ref{fig3}. Observe the exact matches to the true eigenvalues.

\vspace{3mm}

\subsubsection{Switching sequence estimation}\label{912}

\begin{figure}[hbt!]
	\centering
	\begin{subfigure}{.45\linewidth}
		\centering
		\includegraphics[width=1\textwidth]{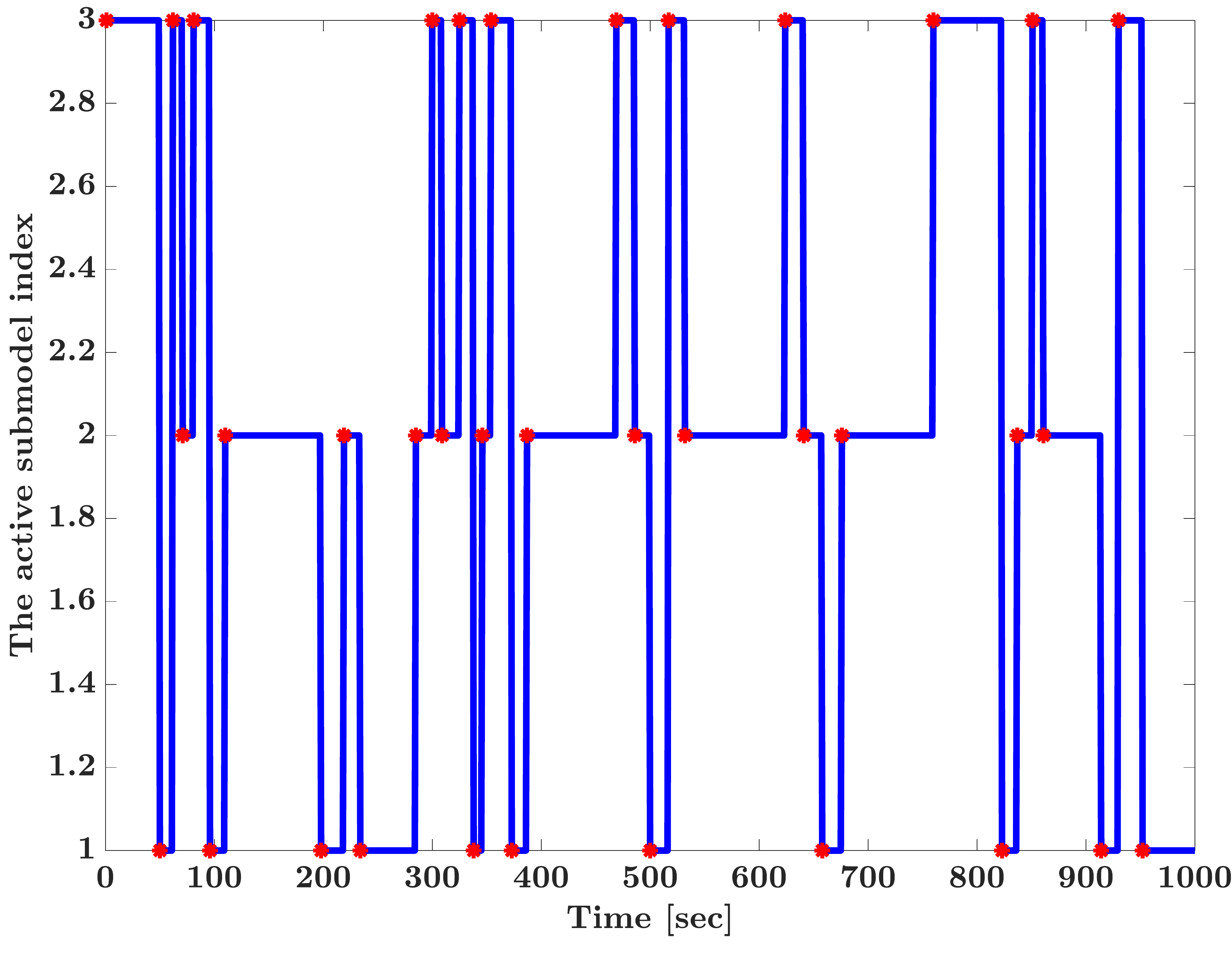}
		\caption{The switching sequence.}
		\label{fig4}
	\end{subfigure}
	\begin{subfigure}{.45\linewidth}
		\centering
		\includegraphics[width=1\textwidth]{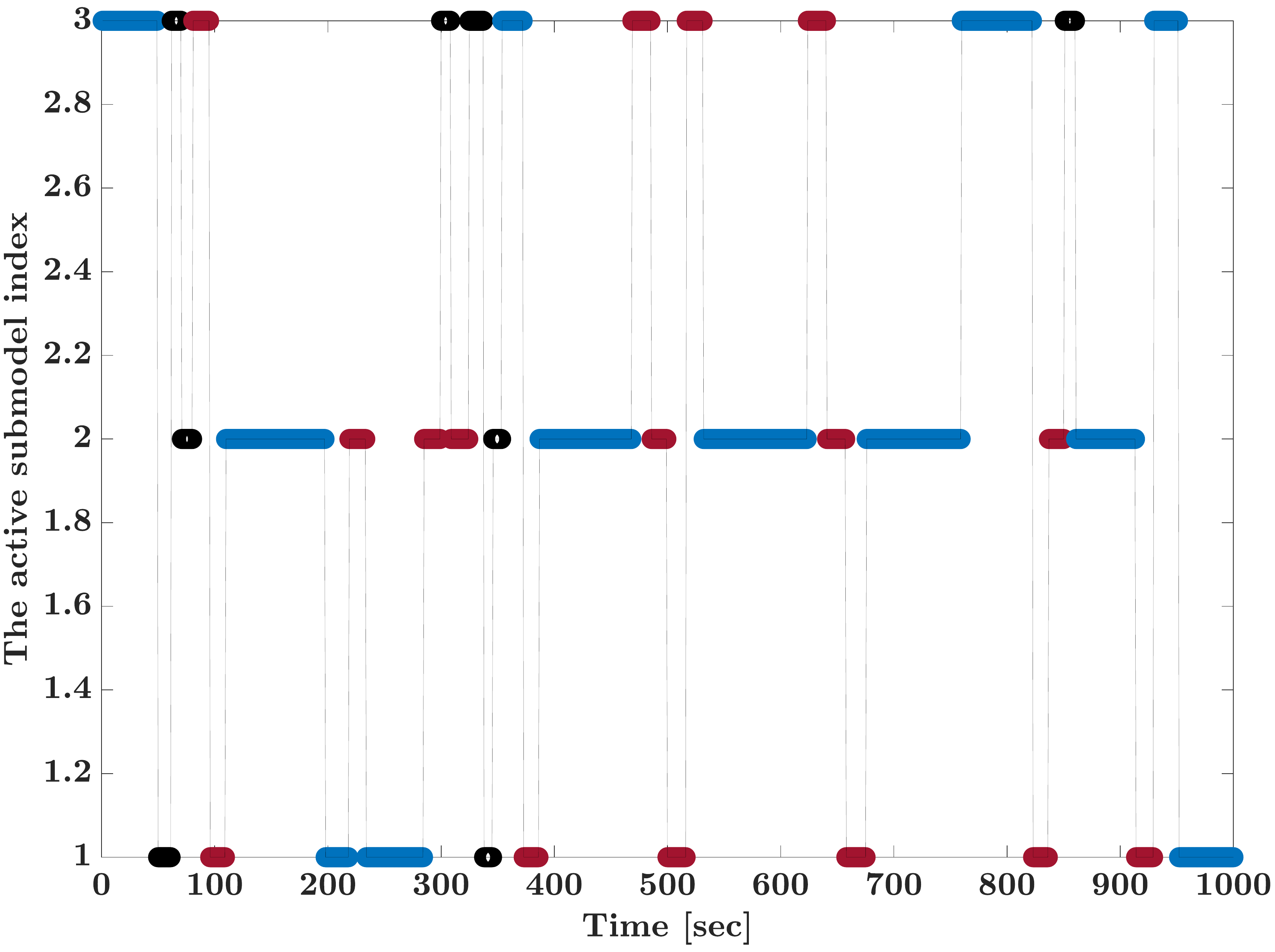}
		\caption{The medium-to-long 'blue', the short 'brown', and the very short 'black' segments in the 
			switching sequence of Figure~\ref{fig4}.}
		\label{fig5}
	\end{subfigure}
	\caption{The switching sequence generated in Section~\ref{912}.}
\end{figure}

Now we address the issue of switching sequence estimation assuming that the submodel set was estimated in the
previous stage. This could be achieved via Algorithm~$3'$ solely if our SLS has a minimum dwell time of at least 
$6n+1$, i.e, ${\delta _*}\left( \chi  \right) \ge 6n +1$. On the other hand, Algorithm~3 can be used if 
$\delta _*\left( \chi  \right) \ge 4n+2$. Lastly, if there exist very short segments with a minimum dwell time of at 
least ${\delta _*}\left( \chi  \right) \ge 2n + 1$, we use Algorithm~$3''$. We call a segment $[k_i,k_{i+1})$ 
satisfying $\delta_i(\chi)\geq 6n+1$ 'medium-to-long', a segment $[k_i,k_{i+1})$ with $6n+1>\delta_i(\chi) \geq 4n+2$ 
'short', and lastly a segment $[k_i,k_{i+1})$ satisfying $4n+2>\delta_i(\chi) \geq 2n+1$ 'very short'. 
Algorithms~$(3',3,3'')$ estimate the medium-to-long, the short, and the very short segments, respectively. 
To show their working mechanisms, we generated a switching sequence containing segments of the three types. 
Different type segments were marked with distinctive colors for better discernibility: medium-to-long 'blue', short 
'brown', and very short 'black'. The switching sequence generated is shown in Figure~\ref{fig4}. 
In Figure~\ref{fig5}, the segments in $\chi$ are classified according to their types.

\begin{figure}[hbt!]
	\centering
	\begin{subfigure}{1\linewidth}
		\centering
		\includegraphics[width=0.5\textwidth]{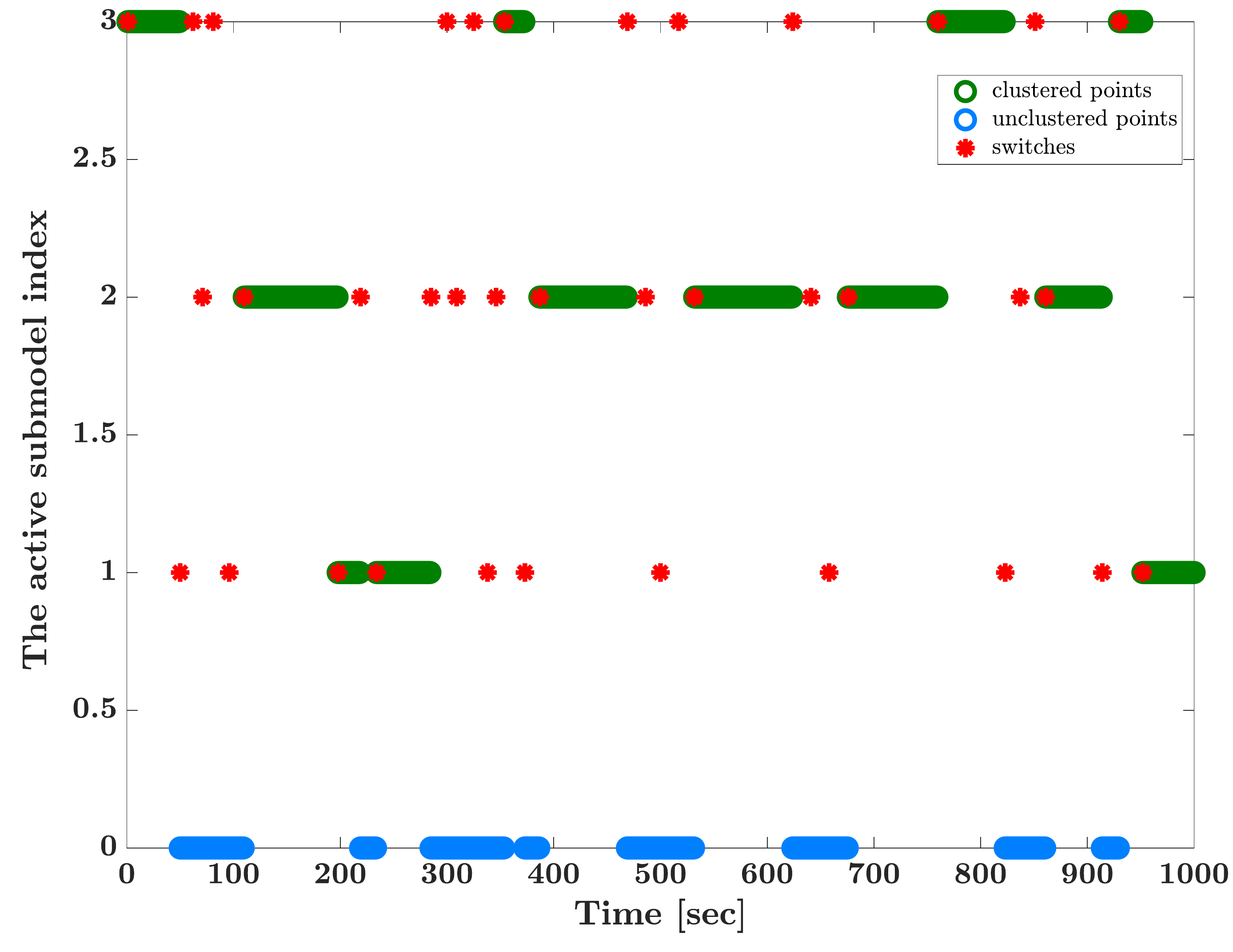}
		\caption{The medium-to-long segment estimation via Algorithm~$3'$.}
		\label{fig6}
	\end{subfigure} 
	\begin{subfigure}{1\linewidth}
		\centering
		\includegraphics[width=0.5\textwidth]{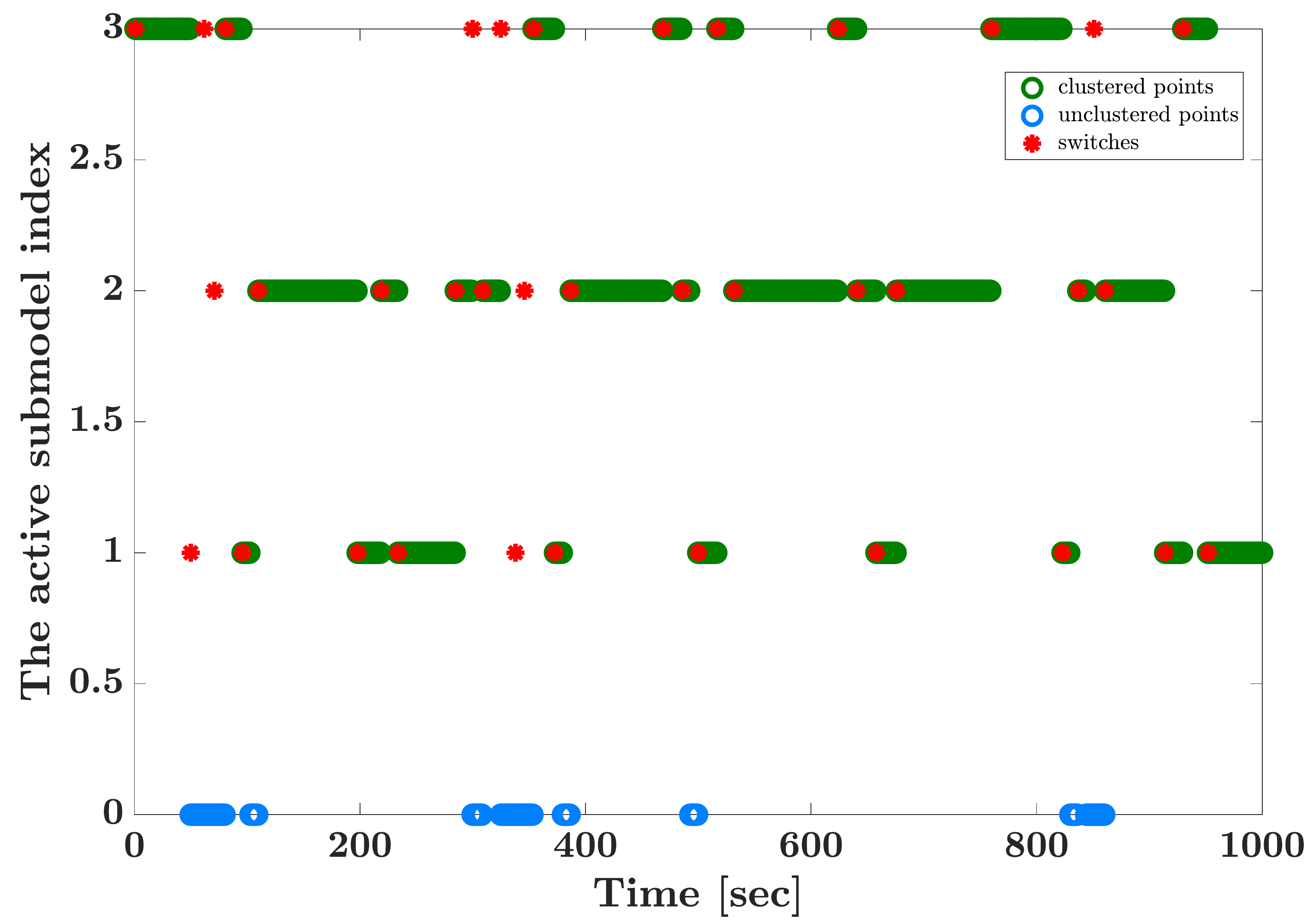}
		\caption{The short segment estimation via Algorithm~3.}
		\label{fig7}
	\end{subfigure} 
	\begin{subfigure}{1\linewidth}
		\centering
		\includegraphics[width=0.5\textwidth]{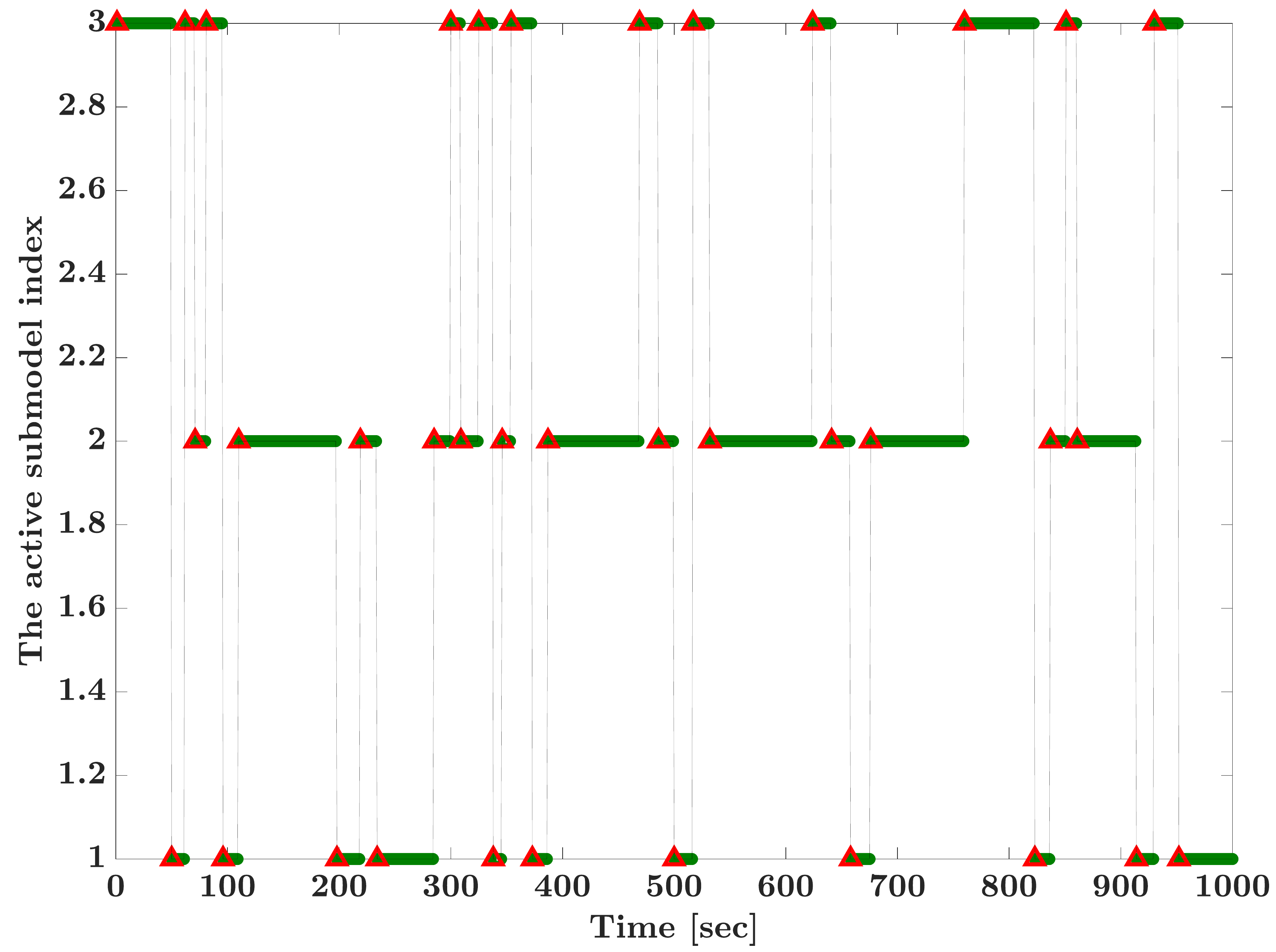}
		\caption{The final switching sequence estimate '$\circ$', the true switching sequence '{\bf -}', 
			and the switches '$\triangle$'.}
		\label{fig8}
	\end{subfigure}
	\caption{The switching sequence estimation via Algorithms $(3',3,3'')$ in order.}
\end{figure}

We start by estimating the submodels, and this is accomplished via Algorithm 2 with $\nu=6$. Next, we sequentially run 
the switch detection algorithms. We first try Algorithm~$3'$ over 
the medium-to-long segments, but it fails to estimate the switching sequence at some points in the short and 
the very short segments. This explains why a large set of points are not yet attributed to any of the submodels 
as shown in Figure~\ref{fig6}. These points were put in Cluster 0 to distinguish them from the points in the 
segments which have already been attributed a correct submodel index. The red dots are the switches as before. 
Next, we try Algorithm~$3$ to estimate the switching sequence over the short segments, 
but it fails at some points in the very short segments. The switching sequence estimate is shown in Figure~\ref{fig7}. 
Finally, we run Algorithm $3''$ to estimate the rest of the switching sequence, i.e., over the very short segments. 
The final switching sequence estimate and its zoomed image are shown in Figure~\ref{fig8} and Figure~\ref{fig9}. 
It is an exact match to the true one. As a final remark, note that Algorithm~3' operates on the intervals satisfying
$\beta_i-\alpha_i \ge \nu n$. This condition was used in Algorithm~2 to estimate the discrete states.
	
\begin{figure}[hbt!]
	\centering
	\includegraphics[width=0.6\textwidth]{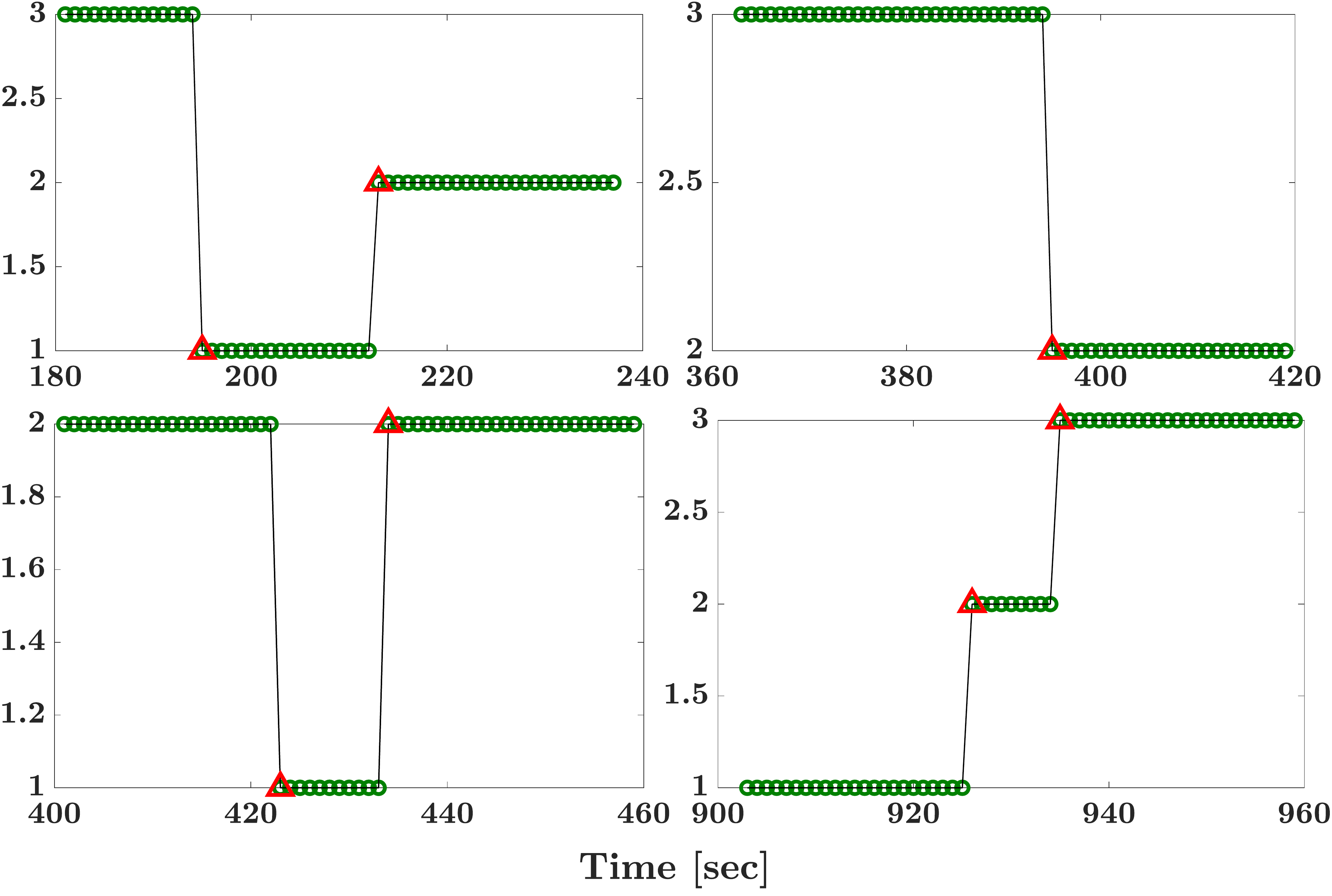}
	\caption{The true '$\circ$' and the estimated '{\bf -}' switching sequences zoomed from Figure~\ref{fig8}.}
	\label{flowchartfig}
\end{figure}

\subsubsection{Basis transform for the discrete states}\label{913}

This subsubsection is concerned with performing the necessary discrete state transformations to render the SLS 
model usable for predicting output given an input signal. To perform that, one needs the estimated sub-models 
from Algorithm~2 and the estimated switching sequence from Algorithms~$(3,3',3'')$. Having that, we employ 
Algorithm~4 to obtain ${\Pi _j} = {T_j}T_1^{ - 1}$ for $j=2,3$. They were calculated as follows
\begin{eqnarray*}
	{\Pi _2} = \left[ {\begin{array}{*{20}{c}}
			{{\rm{ - 0}}.{\rm{4549}}}&{{\rm{0}}.{\rm{0311}}}&{{\rm{0}}.{\rm{1299}}}\\
			{{\rm{ - 0}}.{\rm{0140}}}&{{\rm{0}}.{\rm{5774}}}&{{\rm{ - 0}}.{\rm{2281}}}\\
			{{\rm{ - 0}}.{\rm{4472}}}&{{\rm{ - 0}}.{\rm{3224}}}&{{\rm{ - 0}}.{\rm{8538}}}
	\end{array}} \right],\\
	{\Pi _3} = \left[ {\begin{array}{*{20}{c}}
			{{\rm{ - 0}}.{\rm{4595}}}&{{\rm{0}}.{\rm{1516}}}&{{\rm{0}}.{\rm{9003}}}\\
			{{\rm{ - 0}}.{\rm{0115}}}&{{\rm{ - 0}}.{\rm{1930}}}&{{\rm{0}}.{\rm{2244}}}\\
			{{\rm{ - 0}}.{\rm{2481}}}&{{\rm{ - 0}}.{\rm{1360}}}&{{\rm{ - 0}}.{\rm{0860}}}
	\end{array}} \right].
\end{eqnarray*}

\begin{figure}[hbt!]
	\centering
	\begin{subfigure}{1\linewidth}
		\centering
		\includegraphics[width=0.45\textwidth]{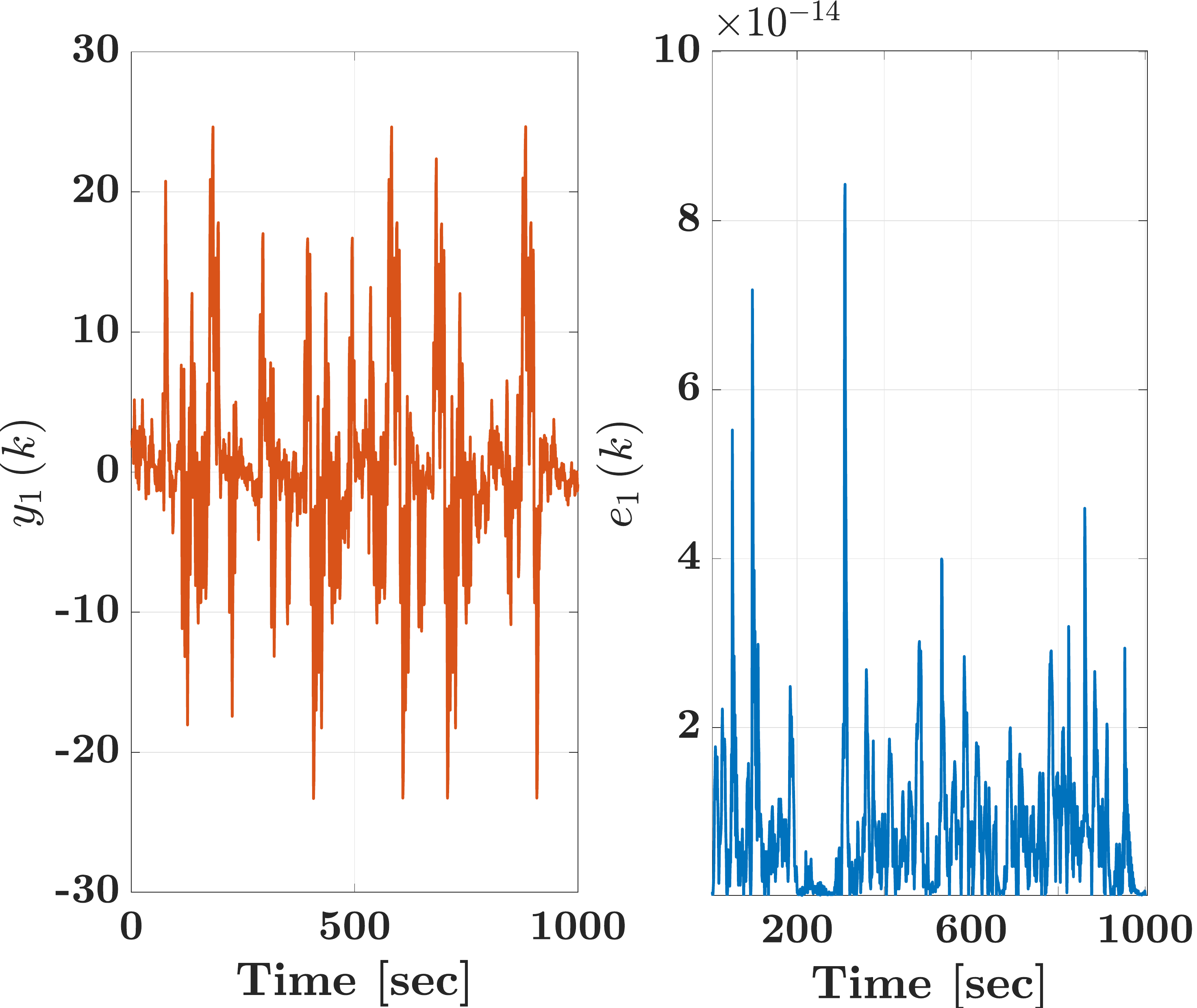}
		\caption{True Output~1 (left) and the estimation error (right).}
	\end{subfigure} 
	\begin{subfigure}{1\linewidth}
		\centering
		\includegraphics[width=0.45\textwidth]{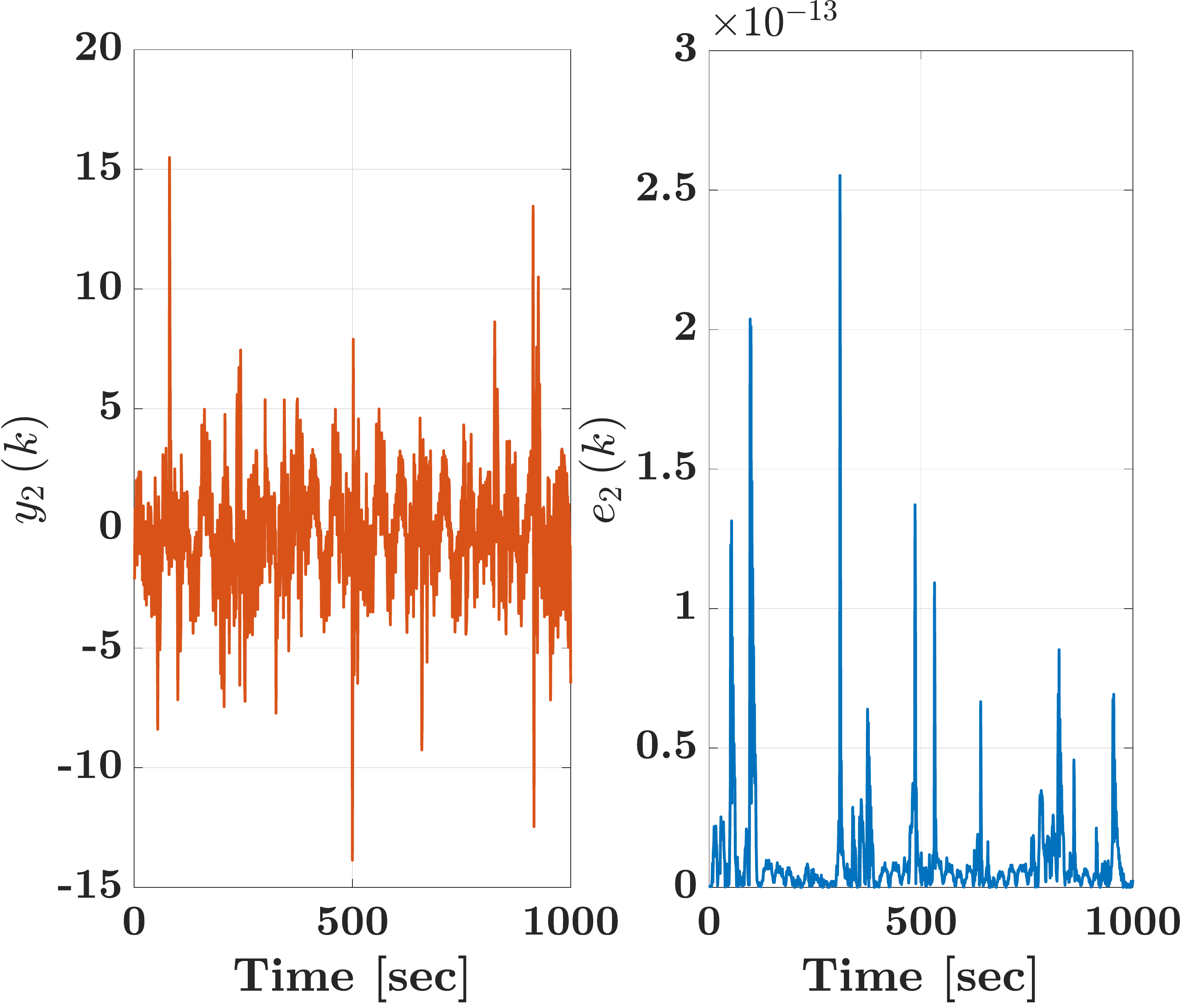}
		\caption{True Output~2 (left) and the estimation error (right).}
	\end{subfigure}     
	\caption{The outputs and the estimation errors.}
	\label{fig9}
\end{figure}

After that, we apply ${\Pi _j}$ to $\mathcal {\hat P}$ to put all the submodels in a common state basis. 
In particular, we apply ${\Pi _2}$ to ${{\hat {\cal P}}_2}$ to render it in a common state basis as of that 
corresponding ${{\hat {\cal P}}_1}$. In a similar manner, ${\Pi _3}$ is applied to ${{\hat {\cal P}}_3}$ 
which transforms its basis to that of ${{\hat {\cal P}}_1}$. To demonstrate the effectiveness of this approach, 
we generated an input signal as a sum of harmonics, and we simulated its output through the SLS model. We did that 
both for the original SLS and the estimated one, and we inspected the match between the two. In generating the 
output signal we assumed that there were no switches occurring before $k'$ and after $k''$, hence we extended 
$\hat \varphi \left( k \right)$ to $1$ and $N$. Both the true and the modeled SLSs were started at rest, i.e., 
the initial state was taken to be $x_0=[0\;\;0\;\;0]^T$. Figure~\ref{fig9} displays the true output signals and the 
estimation errors side by side.

To assess the fidelity of these estimates, we introduce the variance-accounted-for criterion  
defined by
\begin{equation}\label{pezzo}
	VAF = \left( {1 - \frac{{{\rm{var}}\left( {y\left( k \right) - \hat y\left( k \right)} \right)}}
		{{{\rm{var}}\left( {y\left( k \right)} \right)}}} \right) \times 100\% .
\end{equation}
We calculated (\ref{pezzo}) for both outputs. For Output~1, we got  VAF=99.99$\%$ and for Output~2, VAF=99.99$\%$
indicating perfect match between the two. This perfect match resulted from the agreement between the pairs true switching sequence-Markov parameter set and their estimates since the I/O map explicitly depends on them. 

\subsection{SLS recovery from noisy Markov parameters measurements}\label{SLS-noisy}

We now study SLS recovery in a noisy setup. The same SLS model as in the noiseless case was adopted here. 
Table~I lists the smallest nonzero singular values in the noiseless Hankel matrices of the submodels, which 
according to \cite{Oymak&Ozay:2019} determine how robustly a realization algorithm could retrieve a particular 
discrete state. It is clear from the table that Submodels~1 and 2 can be learned better than Submodel 3 
when the exact Markov parameters pertaining to these submodels are perturbed  with the same noise level. 

\begin{table}[ht]
	\caption{The smallest nonzero singular values in the Hankel matrices of the individual submodels.}
	\centering 
	\begin{tabular}{c c c c} 
		\hline\hline 
		Submodel & $\mathcal P_1$ & $\mathcal P_2$ & $\mathcal P_3$ \\ [0.5ex] 
		\hline % inserts single horizontal line
		${\sigma _{\min }}$ & 0.4063 & 0.3560 & 0.0180  \\ [1ex] % 
		\hline 
	\end{tabular}
	\label{table:1} 
\end{table}

\begin{figure}[hbt!]
	\centering
	\begin{subfigure}{1\linewidth}
		\centering
		\includegraphics[width=0.5\textwidth]{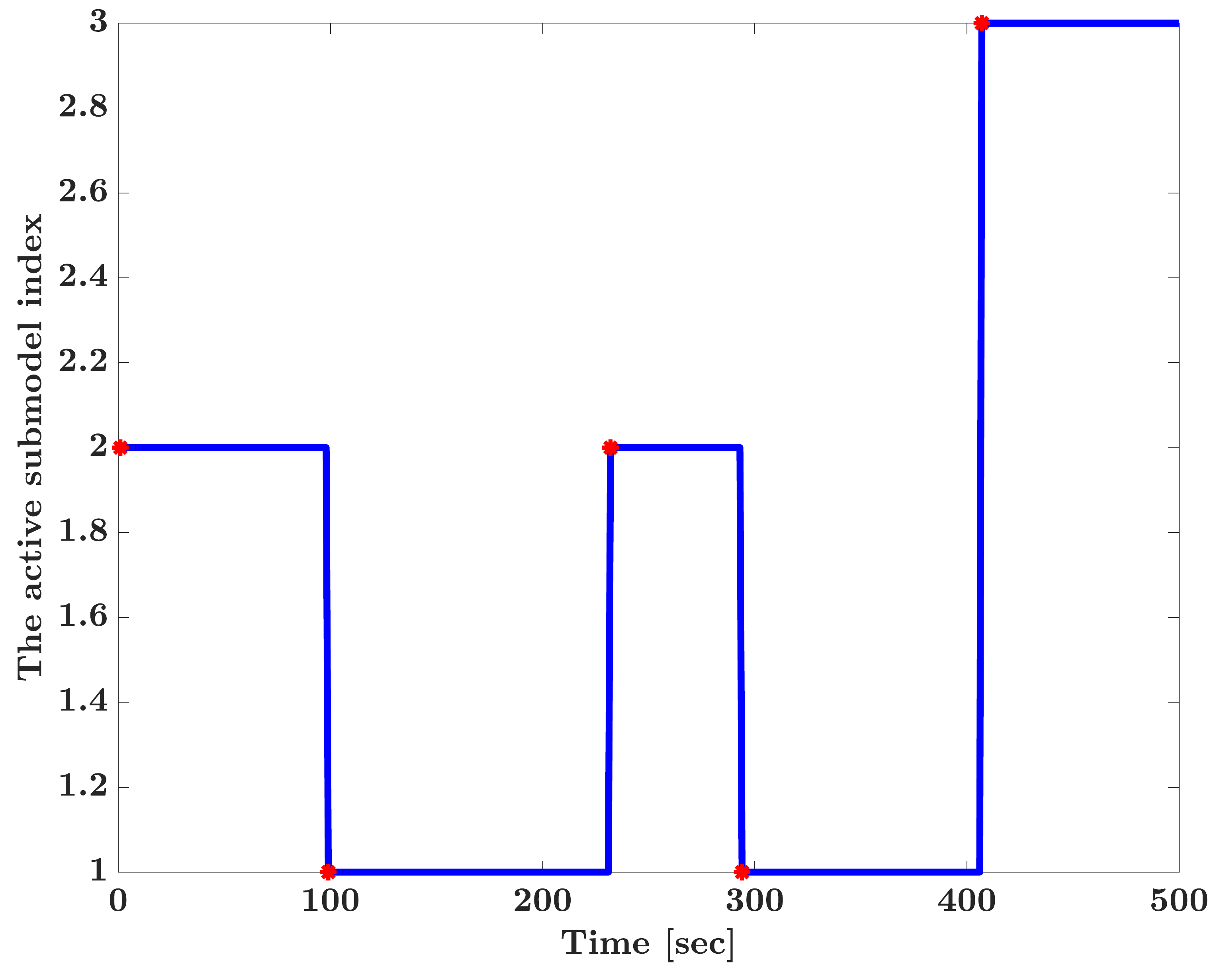}
		\caption{The switching sequence in Section~\ref{SLS-noisy}.}
	\end{subfigure} 
	\begin{subfigure}{1\linewidth}
		\centering
		\includegraphics[width=0.5\textwidth]{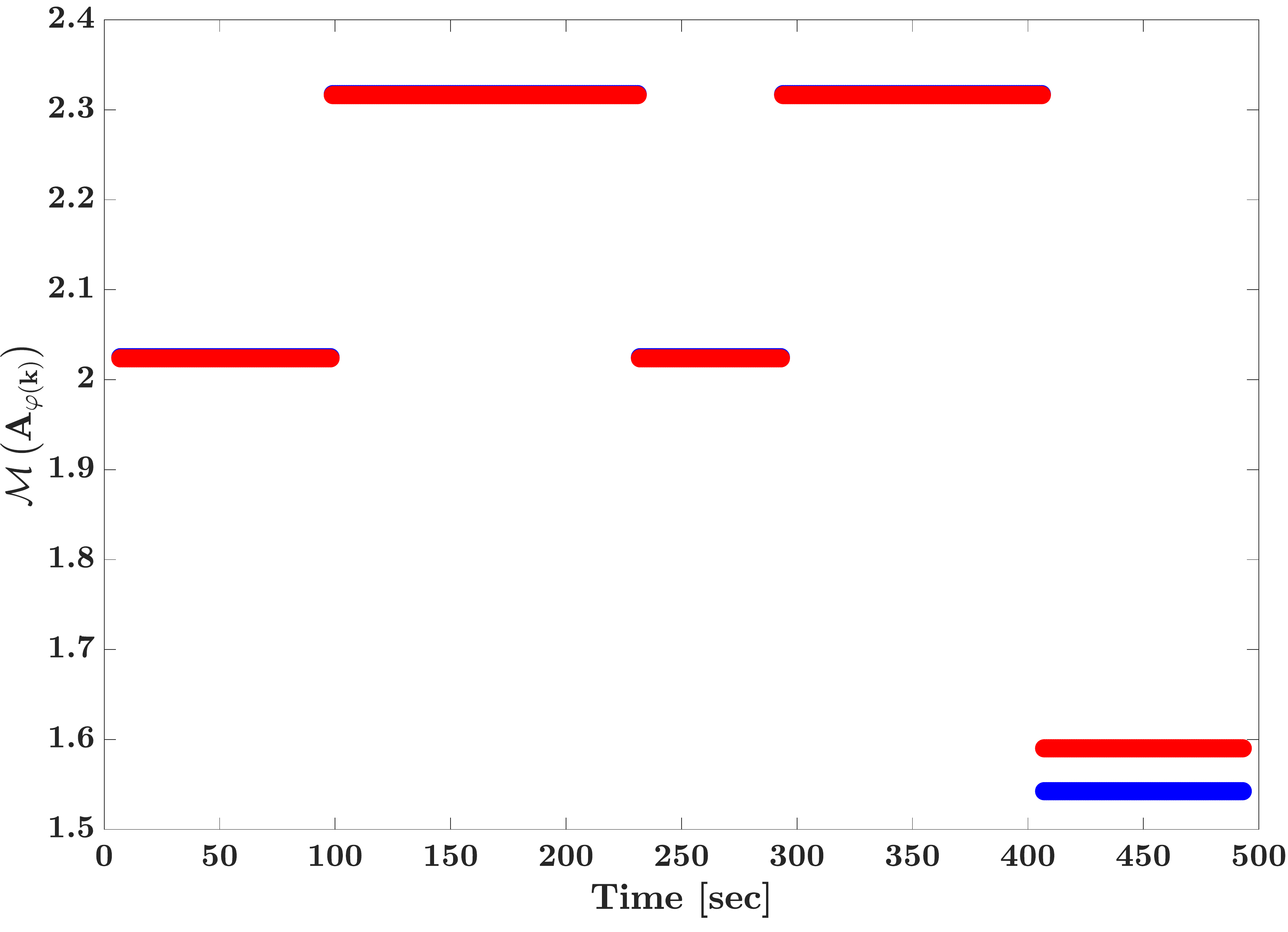}
		\caption{${\cal M}({\hat A}(k))$ (blue) and ${\mathcal M}(A(k))$ (red) for one noise realization at SNR=40dB.}
	\end{subfigure}     
	\caption{The switching sequence and the feature used for clustering in (\ref{Meigdef}).}
	\label{fig10}
\end{figure}

\begin{figure}[hbt!]
	\centering
	\includegraphics[width=0.5\textwidth]{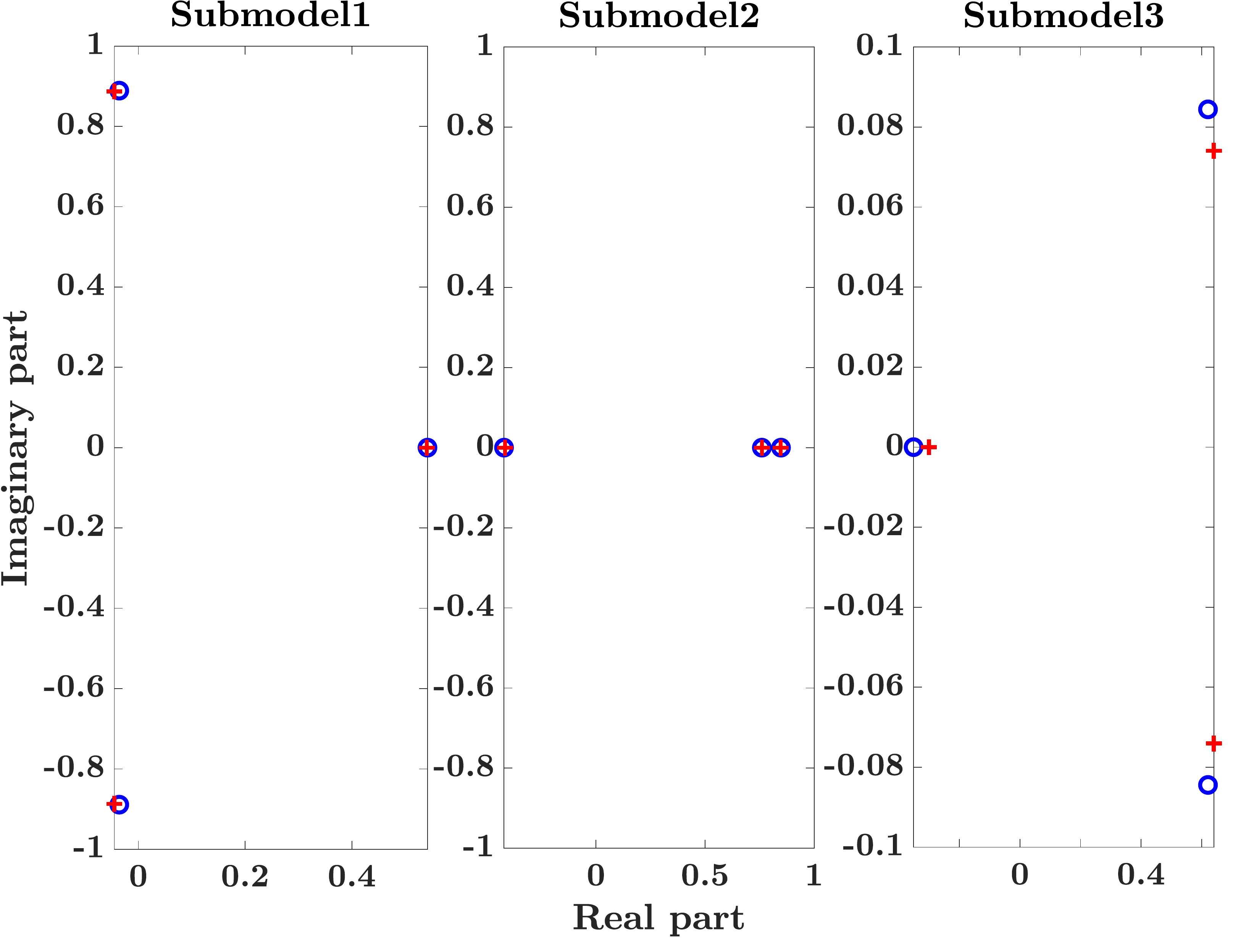}
	\caption{The eigenvalue estimates for one noise realization at SNR=40dB: the true '{\bf +}'and 
		the estimates '$\circ$'.}
	\label{fig11}
\end{figure}

We generated a switching sequence by sampling from a uniform distribution. Instead of using the exact 
Markov parameters, we contaminated them via an additive white noise. From the noisy doubly indexed Markov 
parameters, the SLS identification was conducted similarly to the previous case. Algorithm~2 recovered the 
submodels. These submodels were exploited to retrieve the switching sequence by running Algorithms~$(3,3',3'')$ 
depending on the type of segment considered. Algorithm~4 transformed the discrete states estimated in a 
precedent stage to a common state basis so that the output prediction could be conducted harmlessly. 
This could be as well inspected by checking that the true Markov parameters and the estimated ones are 
identical through the entire time interval.  The individual steps are summarized in the flowchart of 
Figure~\ref{zamazingo}. Figure~\ref{fig10} plots the feature used for clustering defined in (\ref{Meigdef}) 
together with the switching sequence for one noise realization with signal-to-noise ratio (SNR) 40dB. 
Note the gross error in the last segment in agreement with Table~I, that is, the realization algorithm 
performs poorly when it tries to learn Submodel~3. The estimated SLS 
eigenvalues shown in Figure~\ref{fig11} highly mismatch the eigenvalues of Submodel~3.

\begin{figure}[hbt!]
	\centering
	\begin{subfigure}{1\linewidth}
		\centering
		\includegraphics[width=0.45\textwidth]{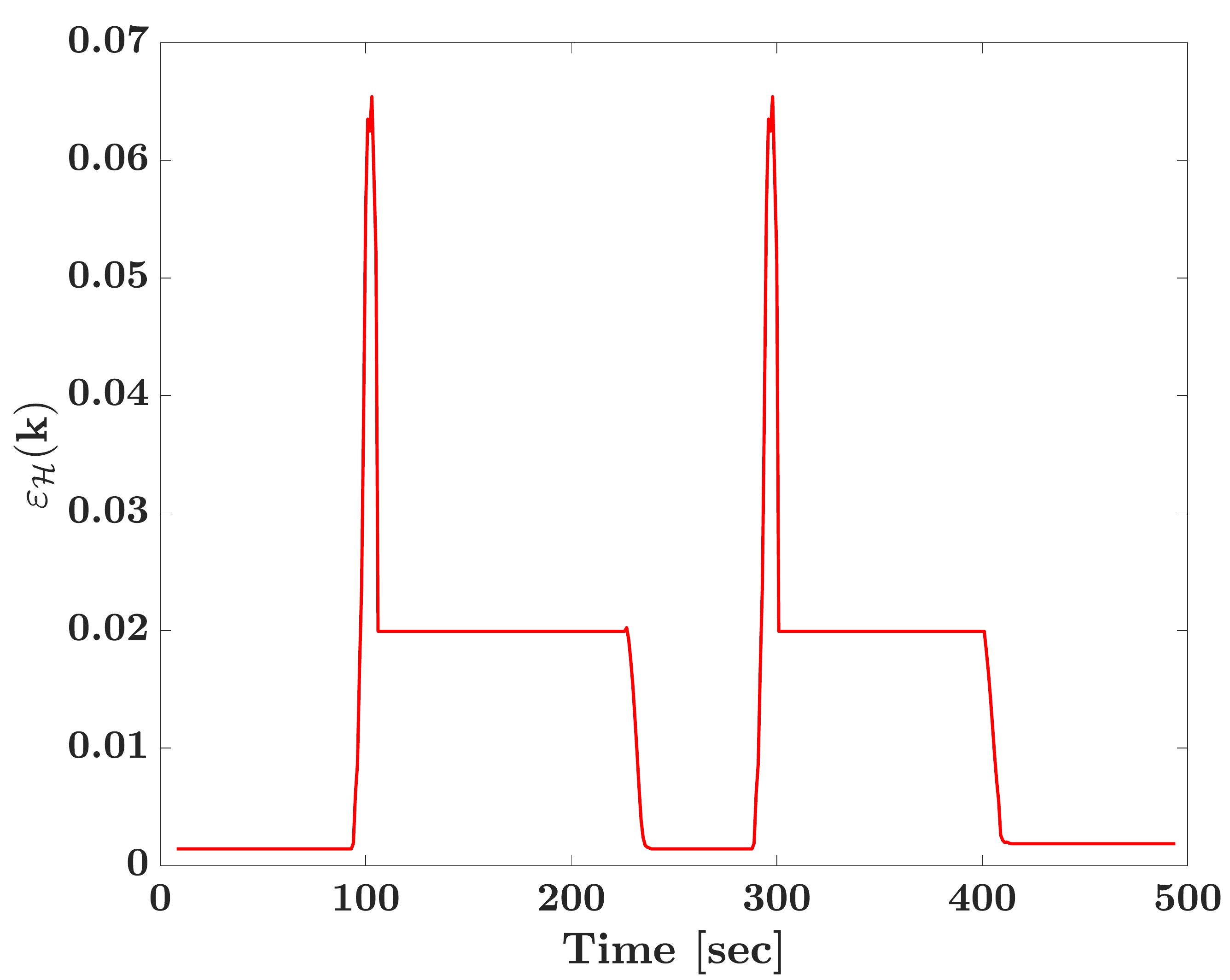}
		\caption{SNR=45dB.}
	\end{subfigure} 
	\begin{subfigure}{1\linewidth}
		\centering
		\includegraphics[width=0.45\textwidth]{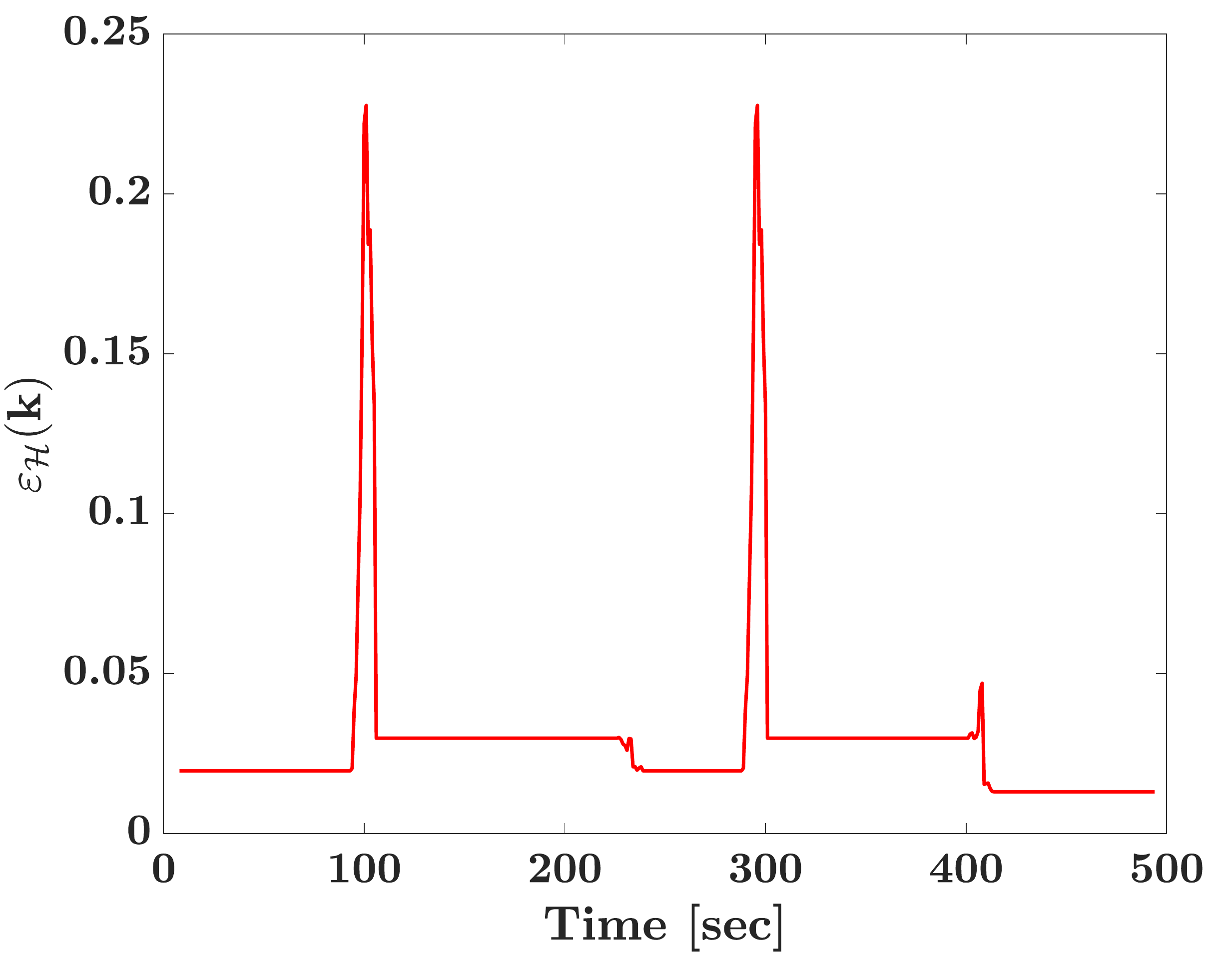}
		\caption{SNR=40dB.}
	\end{subfigure}
	\begin{subfigure}{1\linewidth}
		\centering
		\includegraphics[width=0.45\textwidth]{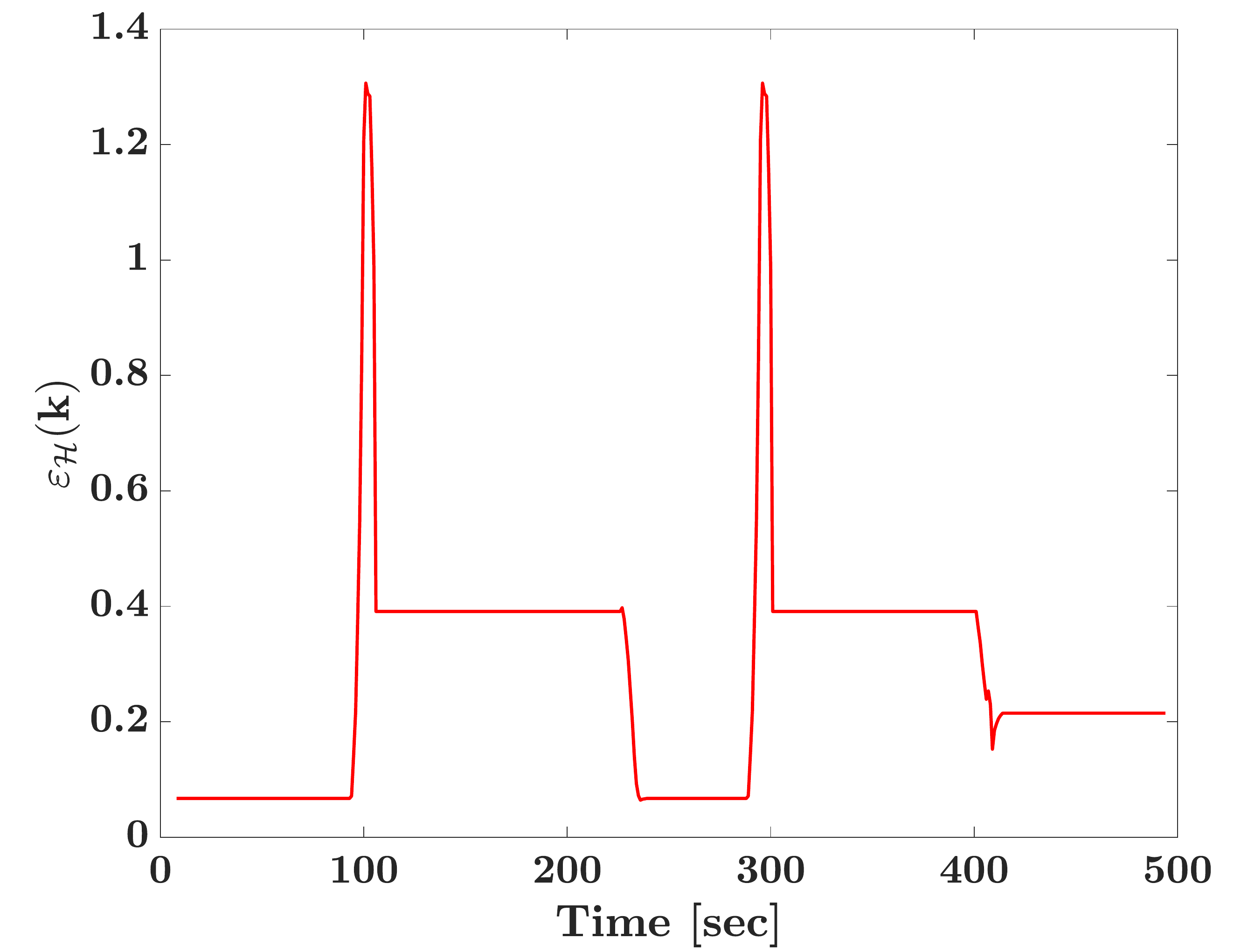}
		\caption{SNR=35dB.}
	\end{subfigure}   
	\caption{The mismatch error (\ref{error}) at different SNRs.}
	\label{fig12}
\end{figure}

To study robustness of the meta-algorithm to noise, we calculated the mismatch error in estimating the true 
Hankel matrix in (\ref{gHankel}) defined pointwise in time by
\begin{equation}\label{error}
	{\varepsilon _\mathcal H}\left( k \right) = {\left\| {{\mathcal H_{2n + 1,2n}}\left( k \right) - {{\hat {\mathcal H}}_{2n + 1,2n}}\left( k \right)} \right\|_F},\quad \forall k \in [k'\;\;k'']
\end{equation}
where ${{{\hat {\mathcal H}}_{2n + 1,2n}}\left( k \right)}$ is the Hankel matrix estimate. Since the SLS realization problem 
involves both $\varphi$ and $\mathcal{P}$, we selected a criterion reflecting both through Markov parameters. Alternatively,
we could have conducted output predictions to prescribed inputs. Figure~\ref{fig12} displays the mismatch error (\ref{error}) 
as a function of time for a range of SNRs. The mismatch error is flat over the segments with the exception of the switches 
where the changes in the discrete states are abrupt. As SNR is decreased, the mismatch error increases rapidly.

\subsection{Realization from input-output data}
In this subsection we demonstrate how our meta-algorithm could be incorporated in a complete system identification package 
that estimates an SLS starting from the input-output measurements rather than the Markov parameters. To achieve this goal, 
we append our meta-algorithm to an identification algorithm proposed in \cite{Bencherki&Turkay&Akcay:2020} to form a 
two-stage algorithm. The algorithm in \cite{Bencherki&Turkay&Akcay:2020} is a standalone algorithm which can be used to 
fully estimate an SLS in the state-space form from the input-output data. It can also be used to estimate the Markov parameters 
by solving a sparse optimization problem. We emphasize that this algorithm successfully retrieves the Markov parameters 
from a single trajectory, unlike the earlier works \cite{Verhaegen&Yu:95,Liu:1997,Majjietal:2010} where multiple
trajectories are assumed to be available.

Solution of sparse optimization problem essentially grants us a finite set of observer Markov parameters which encodes 
entire knowledge of the system dynamics. From observer parameters, one can retrieve infinite strings of the Markov 
parameters. Without going into technical details, it suffices to say that conditions on the
input-output data, the dwell times, persistence of excitation on the inputs, and some identifiability criteria 
permit successful recovery of the Markov parameters. The meta-algorithm of this paper complements this 
scheme. Let us consider the SLS model 

\begin{eqnarray*}
	{A_1} &=& \left[{\begin{array}{*{20}{c}}0&{0.8}\\{ - 0.8}&{0.5}\end{array}} \right],\;\;
		{B_1} = \left[ {\begin{array}{*{20}{r}}{1.4}\\{ - 2}\end{array}} \right],\;\;
		C_1 =  - \left[{1\;\;1.5} \right],\;\;
		{D_1} = 1\\
	{A_2} &=& \left[{\begin{array}{*{20}{c}}{ - 0.9}&{ - 1.25}\\ {1.8}&{1.5}\end{array}} \right],\;\;
	{B_2} = \left[{\begin{array}{*{20}{r}}{ - 1}\\{0.5}\end{array}} \right],\;\;
		C_2 = \left[{ - 1\;\;1} \right],\;\;{D_2} =  - 1.5;\\
	{A_3} &=& \left[ {\begin{array}{*{20}{c}}{0.8}&0\\0&{ - 0.3}\end{array}} \right],\;\;
	{B_3} = \left[ {\begin{array}{*{20}{r}} 1\\ 2\end{array}} \right],\;\;
	C_3 = \left[{2\;\;3} \right],\;\;
	{D_3} = 2.
\end{eqnarray*}
We generated a switching sequence of length $N=1,000$ by sampling from a uniform distribution.  The SLS model was 
excited with a multi-sine input starting from the initial condition $x(0)=(0\;\;0)^T$. Running Algorithm~6 in \cite{Bencherki&Turkay&Akcay:2020} by using the {\tt CVX package} \cite{Grant&Boyd:2014} with the observer order 
$\tau=2n$, we get the observer Markov parameter estimates $h_{\rm o}(k,k-\ell)$ for $2n < k \le N - 2n$ and 
$0 \le \ell < 2n$. Let us call Algorithm~1 in \cite{Bencherki&Turkay&Akcay:2020} Algorithm~1$^\prime$ to avoid 
confusion with Algorithm~1 in this paper. Algorithm~1$^\prime$ driven by the observer Markov parameters returns 
$h(k,k-\ell)$ for $0 \leq \ell \leq k$ and $2n < k \le N - 2n$. Algorithm~1 relies only on the first $4n+1$ Markov 
parameters. Thus, running Algorithm~1$^\prime$ we get as many Markov parameter estimates as we want. Having estimated 
the Markov parameters, we run the meta-algorithm. We will perform a Monte-Carlo simulation study for $100$ noise 
realizations. 

\begin{figure}[hbt!]
	\centering
	\includegraphics[width=0.5\textwidth]{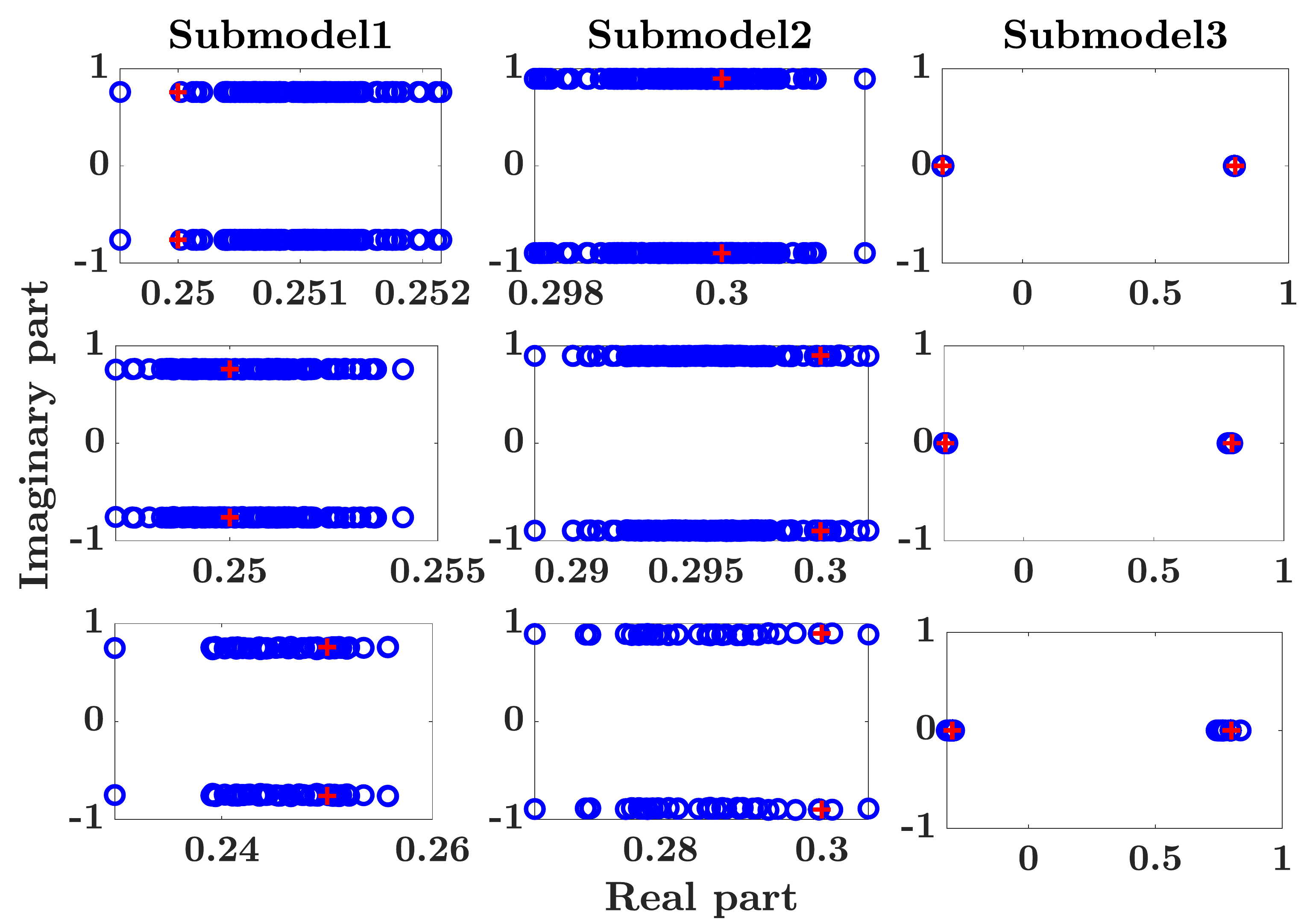}
	\caption{The true '{\bf +}' and the estimated '$\circ$' eigenvalues for $100$ noise realizations at $20,30$, and 
		$40$dB SNRs (top-to-bottom).}
	\label{fig17}
\end{figure}

The eigenvalues estimated by this two-stage scheme are plotted in Figure~\ref{fig17}. The average case performance will be assessed  by first computing the relative errors $\|M_j-\hat{M}_j\|_F/\|M_j\|_F$, $j\in \mathbb S$ where 
${\hat {\cal P}_j} = ({{\hat A}_j },{{\hat B}_j },{{\hat C}_j},{{\hat D}_j})$ and
\begin{equation*}
	{{\hat M}_j } = \left[ {\begin{array}{*{20}{c}}
			{{{\hat A}_j }}&{{{\hat B}_j }}\\
			{{{\hat C}_j }}&{{{\hat D}_j }}
	\end{array}} \right],
\end{equation*}
summing over $\mathbb{S}$, and averaging over $100$ realizations. The result is denoted by $\delta_{\mathcal P}$. 
In Table~\ref{table101}, $\delta_{\cal P}$ is displayed for $20,30$, and $40$dB SNRs. In the same table we compared the two 
stage scheme reported here to the algorithm in \cite{Bakoetal:2013}, which assumes the availability of state measurements 
and proceeds by solving a sparse SARX regression optimization problem. However, the availability of the state measurements 
is a stringent assumption in practice.

\begin{table}[ht]
	\caption{$\delta_{\cal P}$ calculated over the 100 noise realizations.}
	\begin{center}
		\begin{tabular}{||c|c|c|c||} 
			\hline\hline
			\textbf{Schemes}  & \multicolumn{3}{c|}{$\delta_{\cal P}$}  \\
			\hline 
			SNR (dB) & 40dB & 30dB & 20dB  \\ % [0.5ex] 
			\hline
			\texttt{Proposed}  & 0.0169 & 0.0297 &  0.0839  \\ 
			\hline
			\cite{Bakoetal:2013}  & 0.0104 &  0.0404 &  0.1581 \\
			\hline\hline
		\end{tabular}
		\label{table101}
	\end{center}
\end{table}

\begin{figure}[hbt!]
	\centering
	\includegraphics[width=0.6\textwidth]{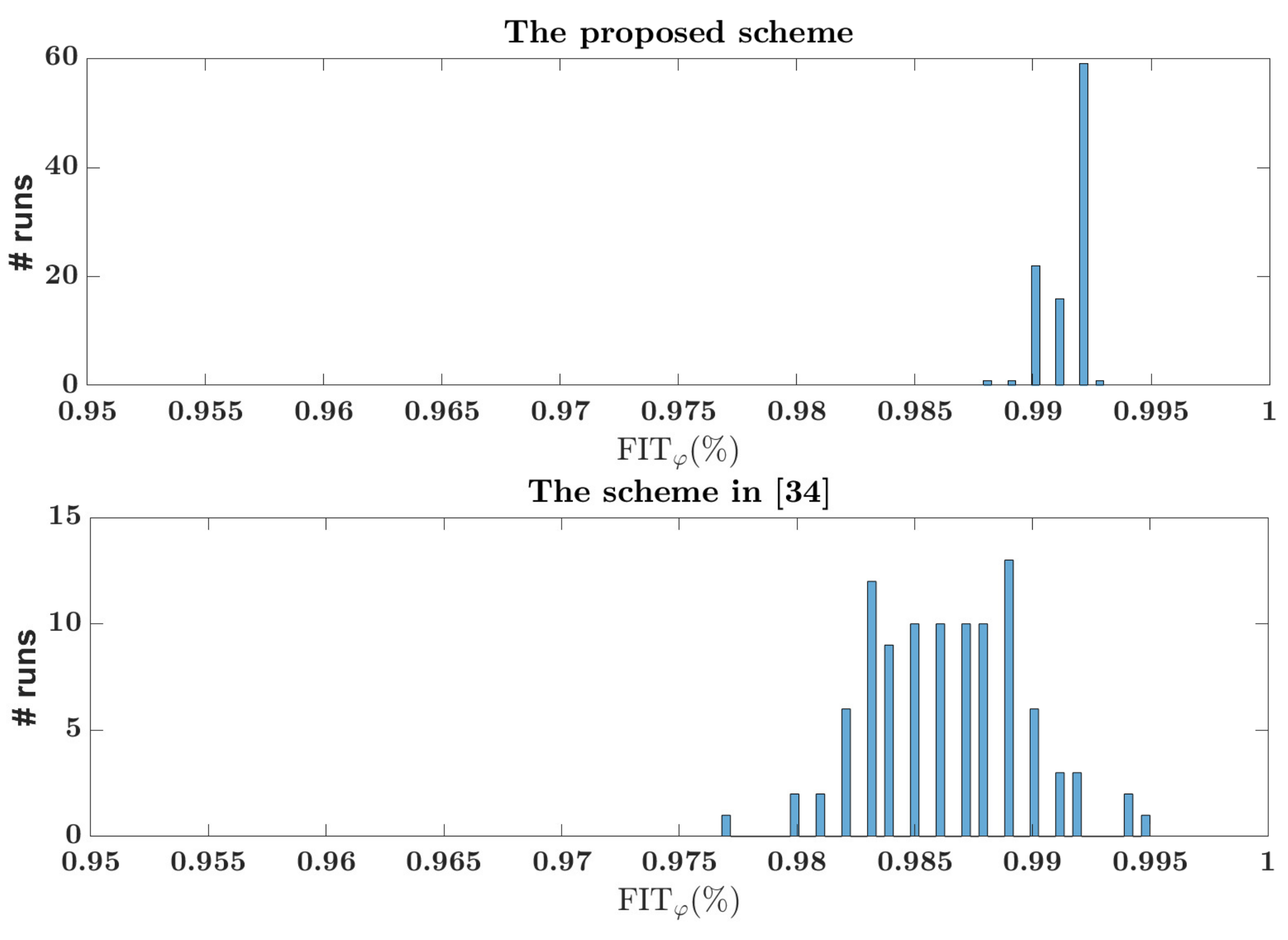}
	\caption{The histogram of ${\rm{FI}}{{\rm{T}}_\varphi } $ over $100$ noise realizations at $SNR=30$dB.}
	\label{fig18}
\end{figure}

A measure of fit for the switching sequence estimates is the percentage of correctly classified points. It is called 
FIT$_\varphi$ and calculated by the formula 
\begin{equation}\label{fit-phi}
{\rm FI}{\rm T}_\varphi = \left(1 - \frac{1}{N} \sum\limits_{k=1}^N 
sign\left|\hat{\varphi}(k) - \varphi( k)\right| \right) \times 100\% .
\end{equation}
Figure~\ref{fig18} plots the histogram of FIT$_\varphi$ over $100$ noise realizations at SNR=$30$dB for both our approach and the scheme in \cite{Bakoetal:2013}. From Table~\ref{table101} and Figure~\ref{fig18} we see that the two-stage scheme outperforms the algorithm in \cite{Bakoetal:2013} and have an edge in the sense that it operates under mild assumptions.

\subsection{Realization of randomly generated SLSs}
\begin{figure}[hbt!]
	\centering
	\includegraphics[width=0.6\textwidth]{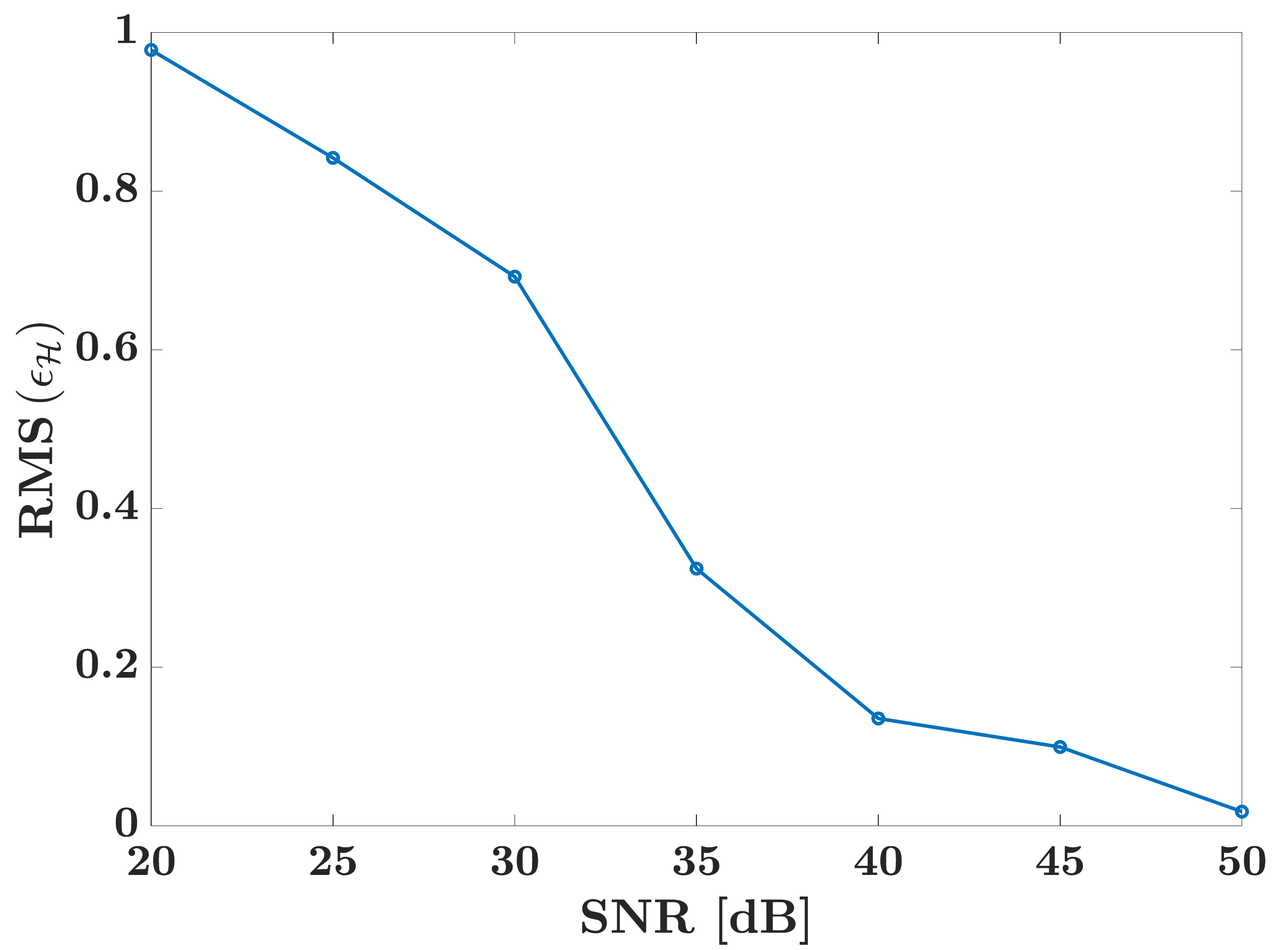}
	\caption{The average {\tt RMS} of the error in (\ref{error}) calculated over $50$ runs at different SNRs.}
	\label{fig19}
\end{figure}

In this subsection, we examine the performance of the meta-algorithm in estimating randomly generated SLSs. The discrete states 
were sampled from a normal distribution. The switching sequences were generated by sampling from a uniform distribution 
similarly to the previous subsections. The SLSs and the switching sequences satisfy Assumptions~\ref{sysasmp}--\ref{varphiassmp}, \ref{unimodality}, \ref{assmp4}, \ref{assmp5}, and \ref{asmmpbasis}. Randomly generated SLSs were corrupted by noise to 
achieve a certain SNR prior to carrying out identification. We then generated the Markov parameters to drive 
the meta-algorithm. In this study, we considered SISO-SLSs with $n=2$ and $\sigma=3$. We picked $N=650$ for Monte-Carlo simulations and assessed the performance of our algorithm using first $\delta_{\cal P}$ and 
FIT$_\varphi$. The results in Table~\ref{table3} were obtained by first estimating an SLS and $\varphi$ for 
each run and averaging the sum of the relative errors and ${\rm FIT}_\varphi$ over 50 runs. For each run, 
we picked a random SLS and a random noise realization. We repeat calculations for a range of SNRs to fill
the table. As another measure of performance, for each run next we calculated the RMS value of (\ref{error})
over $k$ and averaged it over $50$ runs. The results are plotted in Figure~\ref{fig19} as a function of SNR 
levels.

\begin{table}[hbt!]
	\begin{center}
	\caption{The average performance of the meta-algorithm in estimating randomly generated SLSs.}
		\begin{tabular}{|c|c|c|c|c|c|c|c|c|} 
			\hline
			\textbf{Errors}  & \multicolumn{4}{c|}{$\delta_{\cal P}$} & \multicolumn{4}{c|}{\textbf{FIT$_\varphi$} ($\%$)} \\
			\hline 
			SNR (dB) & 50dB & 40dB & 30dB & 20dB & 50dB & 40dB & 30dB & 20dB \\ % [0.5ex] 
			\hline
			 Average Values & 0.0110 & 0.0423 &  0.1818 & 0.7651 & 100 & 99.98 & 98.20 & 94.54\\ 
			\hline
			\hline
		\end{tabular}
		\label{table3}
	\end{center}
\end{table}

\section{Conclusions}\label{conclus}

This paper laid forward a four-stage algorithm for the realization of MIMO-SLSs from Markov parameters. 
Its first stage recovers an LTV realization that is topologically equivalent to the SLS under mild
system and dwell time assumptions. Since topological equivalence does not guarantee similarity to
the discrete states on any segment, it was necessary to apply forward/backward corrections point-wise 
in time to the LTV realization, to reveal the segments. This process established a time-varying
similarity to the discrete states. Next, the stationary point set on which the LTV realization has an LTI pulse
response were sought. This search was effected by a clustering algorithm using a feaure space that is
invariant to similarity transformations. Combination of these schemes formed Stage~2 of the realization
algorithm. In this stage, the discrete state estimates similar to the true ones were extracted
from the LTV realization under some dwell time and identifiability conditions. In the third stage,
the switching sequence was estimated by three schemes, complementing each other. The first scheme 
operates on the forward/backward corrections and estimates the switching sequence over short segments. 
The second scheme is based on matching the estimated and the true Markov parameters for LTV systems and 
it can be applied to medium-to-long segments. The third scheme is also based on matching the Markov parameters, 
but for LTI systems only via Hankel matrix factorization. It can be applied to segments of even shorter length 
which cannot be handled by the first two schemes. The schemes are applied in the order second-to-first-to-third. 
The issue of discrete states residing in different state bases were resolved by applying a basis correction in 
Stage~4; this equips the retrieved SLS with the potential to conduct output predictions.

The success of the various schemes in the realization algorithm put us face to face with the compulsion 
to impose several restrictions on the discrete state set and their dwell times; nonetheless, all that can 
be said is that they are mild in nature. The involvement of the observability and controllability matrices 
in the schemes presented so far was quite noticeable, as noted in several other related works in the 
literature on SLS identification and realization. Future work will focus on developing an identification 
algorithm to estimate the Markov parameters from input-output measurements of a single trajectory.

\section*{Acknowledgment}

This research did not receive any specific grant from funding agencies in the public, commercial, or 
not-for-profit sectors. The authors would like to thank professor Laurent Bako for providing the codes 
for his algorithm in \cite{Bakoetal:2013}.


\newpage

\appendix

\section*{Appendix: Pseudo codes for the algorithms}

\begin{table}[h!]	
	\small
	\begin{center}
		\begin{tabular}{l}
			\hline 
			{\bf Algorithm~1. LTV realization from Markov parameters}  \\
			\hline
			Set $q=2n+1$ and $r=2n$. \\
			{\bf Inputs:} Markov parameters $h(k+i,k-j)$ for $0\leq i <q$, \\ 
			$1 \leq j \leq r$, $k \in [k^\prime\;\;k^{\prime\prime}]$  \\
			{\bf while} $k \in [k^\prime\;\;k^{\prime\prime}]$  \\
			\hspace{4mm} 1: Compute ${\mathcal H}_{q,r}(k)$ and ${\mathcal H}_{q,r}(k+1)$ from 
			(\ref{gHankel12})--(\ref{gHankel2}) \\
			\hspace{4mm} 2: Compute the SVDs in (\ref{svdHqr})  \\
			\hspace{4mm} 3: Compute the extended observability/controllability \\
			\hspace{8mm}   matrices estimates in (\ref{Oqr}-\ref{Obqrp}) \\
			\hspace{4mm} 4: Estimate $\hat{A}(k)$, $\hat{C}(k)$, $\hat{B}(k)$ from (\ref{Aes}) and (\ref{CBes}),  \\
			\hspace{8mm} and $\hat{D}(k)=h(k,k)$\\
			\hspace{4mm} 5: $k \leftarrow k + 1$\\
			{\bf end while} \\ 
			{\bf Outputs:} $\hat{\mathcal P}(k)=(\hat{A}(k),\hat{B}(k),\hat{C}(k),\hat{D}(k))$, 
			$\forall k \in \mathcal{K}$.\\
			\hline
		\end{tabular}
	\end{center}
\end{table}

\begin{table}[h!]	
	\small
	\begin{center}
		\begin{tabular}{l}
			\hline 
			{\bf Algorithm~2. Discrete state estimation}  \\
			\hline
			{\bf Inputs:} Markov parameters $h(k+i,k-j)$ for $0\leq i <2n+1$, \\ 
			$1 \leq j \leq 2n$, $k \in [k^\prime \;\;k^{\prime\prime}]$, a small 
			$\varepsilon_{\mathbb{Z}}>0$, and $\nu>5$. \\
			1: Compute $\delta_{\mathcal H}(k)$, for all $k \in [k^\prime\;\;k^{\prime\prime}]$ from (\ref{firstdiff}) \\
			2: Initialize $\mathbb{Z}_{{\mathcal H},\varepsilon_{\mathbb{Z}}}=\phi$ \\ \textcolor{blue}{}
			3: {\bf while} $k \in [k^\prime\;\;k^{\prime\prime}]$  \\
		         \hspace{8mm} \textbf{if} $\left\|\delta_{\mathcal H}(k)\right\|_F 
				\leq \varepsilon_{\mathbb{Z}} $\\
			        \hspace{12mm} $\mathbb{Z}_{{\mathcal H},\varepsilon_{\mathbb{Z}}}=
				\mathbb{Z}_{{\mathcal H},\varepsilon_{\mathbb{Z}}} \cup \{k\}$\\
			    \hspace{8mm} \textbf{end if}\\
			\hspace{8mm} $k \leftarrow k + 1$\\
			 \hspace{3mm} {\bf end while}  \\ 
			4: Identify maximal disjoint intervals $[\alpha_i,\beta_i] \subset \mathbb{Z}_{{\mathcal H},
					\varepsilon_{\mathbb{Z}}}$, $\beta_i-\alpha_i \geq \nu n $ \\
			5: Retrieve $\hat{A}(\gamma_i)$, $\gamma_i=(\alpha_i+\beta_i)/2$ from the output Algorithm~1 . \\	
			6: Compute ${\mathcal M}(\hat{A}(\gamma))$  \\
			7: Estimate $\mathcal P$ by running the clustering algorithm in \cite{Esteretal:96} 
			over $\gamma$\\	
			\hspace{3mm} and re-clustering if necessary \\	
			{\bf Outputs:} $\hat{\cal P}_j$, $j \in \mathbb{S}$ and the intervals in Step~4 containing them.\\
			\hline
		\end{tabular}
	\end{center}
\end{table}

\begin{table}[hbt!]	
	\small
	\begin{center}
		\begin{tabular}{p{9cm}}\small
			\hrule
			\vspace{0.1cm}
			{\bf Algorithm~3. Switch detection based on corrections} 
			\vspace{0.1cm}
			\hrule
			\vspace{0.1cm}
			{\bf Inputs:} $S_i=[\alpha_i,\beta_i]$, $0\leq i\leq i^*$; $h(k+i,k-j)$, $0\leq i <2n+1$, \\
			$1 \leq j \leq 2n$, $k \in [k^\prime \;\;k^{\prime\prime}]$ \\
			Run the following loop:\\		
			{\bf for} $0 \leq i \leq i^*$\\ 
			\hspace{4mm} {\bf if} ($i==0$ and $\delta_0(\chi) \geq 6n+2$) or ($i \neq 0$, $i\neq i^*$, and $\delta_i(\chi) \geq 4n+2$)\\ 
			\hspace{8mm} Initialize $k=\beta_i-2n$ \\ 		
			\hspace{8mm} {\bf while} $|{\mathcal M}(\hat{\mathcal V}_{2n+1}(k))-n|<\varepsilon$\\		
			\hspace{12mm} k=k+1 \\ 		
			\hspace{8mm} {\bf end while}\\
			\hspace{8mm} $k_{i+1}=k+2n+1$ \\
			\hspace{4mm} {\bf end if} \\
			\hspace{4mm}  {\bf if} ($i== i^*$ and $\delta_{i^*}(\chi) > 8n$) or ($i\neq 0$, $i \neq i^*$, and $\delta_i(\chi) \geq 4n+2$)\\
			\hspace{8mm} Initialize $k=\alpha_i+2n-1$ \\
			\hspace{8mm} {\bf while} $|{\mathcal M}(\hat{\mathcal W}_{2n}(k))-n|<\varepsilon$ \\		
			\hspace{12mm} k=k-1 \\ 		
			\hspace{8mm} {\bf end while}\\
			\hspace{8mm} $k_i=k-2n+1$ \\
			\hspace{4mm} {\bf end if} \\ 
			{\bf end for}\\		
			{\bf Outputs:} $\varphi(k)$ on segments satisfying  ${\delta _i}\left( \chi  \right) \ge 4n + 2$ 
			\vspace{0.01cm}	
			\hrule
		\end{tabular}
	\end{center}
\end{table}

\begin{table}[hbt!]
	\small
	\begin{center}
		\begin{tabular}{p{9cm}}\small
			\hrule
			\vspace{0.1cm}
			{\bf Algorithm~3$^\prime$. Switch detection from Markov parameters}
			\vspace{0.1cm}
			\hrule
			\vspace{0.1cm}
			{\bf Inputs:} $\hat{\mathcal P}(\gamma_i)$, $\gamma_i=(\alpha_i+\beta_i)/2$, $S_i=[\alpha_i,\beta_i]$, $0\leq i\leq i^*$; \\ $h(k+i,k-j)$, $0\leq i \leq 2n$, 
			$1 \leq j \leq 2n$, $k \in [k^\prime \;\;k^{\prime\prime}]$, $\varepsilon>0$ 	\\
			Run the following loop:\\
			{\bf for} $0\leq i \leq i^*$\\ 
			\hspace{4mm} {\bf if} ($i==0$ and $\delta_0(\chi) \geq 8n+1$) or ($i \neq 0$, $i\neq i^*$, and $\delta_i \geq 6n+1$)\\	
			\hspace{8mm} Initialize $\ell=\beta_i-1$, $\check{C}=\hat{C}(\gamma_i)$, $\check{D}=\hat{D}(\gamma_i)$ \\			
			\hspace{8mm} {\bf while} $\|\check{C}-\hat{C}(\gamma_i)\|_F+\|\check{D}-\hat{D}(\gamma_i)\|_F<\varepsilon$ \\
			\hspace{12mm} $\ell=\ell+1$ \\
			\hspace{12mm} $\check{C} = {\mathcal H}_{1,2n}(\ell)\hat{\mathcal R}_{2n}^\dag(\ell-1)$,
			$\check{D} = \hat{D}(\ell)$ \\		
			\hspace{8mm} {\bf end while}\\							
			\hspace{8mm} Set $k_{i+1}=\ell$ \\	
			\hspace{4mm} {\bf end if}\\
			\hspace{4mm} {\bf if} ($i==i^*$ and $\delta_0(\chi) \geq 10n$) or ($i \neq 0$, $i\neq i^*$, and $\delta_i \geq 6n+1$)\\	
			\hspace{8mm} Initialize $\ell=\alpha_i+1$, $\check{B}=\hat{B}(\gamma_i)$, $\check{D}=\hat{D}(\gamma_i)$ \\			
			\hspace{8mm} {\bf while} $\|\check{B}-\hat{B}(\gamma_i)\|_F+\|\check{D}-\hat{D}(\gamma_i)\|_F<\varepsilon$ \\
			\hspace{12mm} $\ell=\ell-1$ \\
			\hspace{12mm} $\check{B} = \hat{\mathcal O}_{2n}^\dag(\ell+1){\mathcal H}_{2n,1}(\ell+1)$, 
			$\check{D} = \hat{D}(\ell)$ \\	
			\hspace{8mm} {\bf end while}\\							
			\hspace{8mm} Set $k_{i-1}=\ell+1$ \\
			\hspace{4mm} {\bf end if}\\
			{\bf end for}\\	
			{\bf Outputs:} $\varphi(k)$ on segments satisfying  ${\delta _i}\left( \chi  \right) \ge 6n + 1$
			\vspace{0.1cm}	
			\hrule
		\end{tabular}
	\end{center}
\end{table}

\begin{table}[hbt!]	
	\small
	\begin{center}
		\begin{tabular}{p{10cm}}\small
			\hrule
			\vspace{0.1cm}
			{\bf Algorithm~3$^{\prime\prime}$. Switch detection on short intervals} 
			\vspace{0.1cm}
			\hrule
			\vspace{0.1cm}
			\textbf{Inputs:} $\hat{\mathcal P}$, one switch from each segment, Markov parameters, and $\varepsilon>0$\\		
			Run the following loop:\\
			{\bf for $1 \leq i < i^*$}\\
			\hspace{4mm}{\bf while} $\delta_i(\chi) > 2n$ and $k_i$ is known 'forward estimation'\\
			\hspace{8mm} 1: Compute $\hat{\mathcal P}(\gamma_i)$ from (\ref{oflenbe1}) and let
			$\varphi(k)=\varphi(\gamma_i)$ on $[k_i,k_i+2n]$\\  
			\hspace{8mm} 2: Estimate $k_{i+1}$ and extend $\hat{\mathcal P}(\gamma_i)$
			to $(k_i+2n,k_{i+1})$ \\ 
			\hspace{12mm}  by forward updating in Algorithm~3$^\prime$\\
			\hspace{4mm} {\bf end while}\\
			\hspace{4mm} {\bf while} $\delta_{i-1}(\chi) > 2n$ and $k_{i}$ is known 'backward estimation'\\
			\hspace{8mm} 1: Compute $\hat{\mathcal P}(\gamma_{i-1})$ from (\ref{oflenbe2}) 
			and let $\varphi(k)=\varphi(\gamma_{i-1})$ on $[k_i-2n,k_i)$\\
			\hspace{8mm} 3: Estimate $k_{i-1}$ and extend $\hat{\mathcal P}(\gamma_{i-1})$ to 
			$[k_{i-1},k_i-2n)$ \\
			\hspace{12mm} by backward updating in Algorithm~3$^\prime$\\				
			\hspace{4mm} {\bf end while}\\				
			{\bf end for} \\
			{\bf if} $\delta_{i^*}(\chi) \geq 6n$ \\
			\hspace{4mm} `$k_{i^*}$ is known' $\Rightarrow$ 'forward estimation from $k_{i^*}$'\\
			\hspace{4mm} `$k_{i^*}$ is not known' $\Rightarrow$ 'backward estimation from $k^{\prime\prime}+1$'\\
			{\bf end if}\\ 	
			{\bf if} $\delta_0(\chi) > 4n$ \\
			\hspace{4mm} `$k_1$ is known' $\Rightarrow$ 'backward estimation from $k_1$'\\
			\hspace{4mm} `$k_1$ is not known' $\Rightarrow$ 'forward estimation from $k^{\prime}$'\\
			{\bf end if}\\ 	
			{\bf Outputs:} $\varphi(k)$ on segments satisfying  ${\delta _i}\left( \chi  \right) \ge 2n + 1$
			\vspace{0.1cm}	
			\hrule
		\end{tabular}
	\end{center}
\end{table}

%\section{}

\begin{table}[hbt!]	
	\small
	\begin{center}
		\begin{tabular}{p{10cm}}\small
			\hrule
			\vspace{0.1cm}
			{\bf Algorithm~4. Basis transformation for the discrete states} 
			\vspace{0.00cm}
			\hrule
			\vspace{0.1cm}
			\textbf{Inputs:} Markov parameters, $\hat{\mathcal P}$ (ordered), and $\varphi$ 			
			\vspace{0.1cm}	
			
			1: Estimate $\Pi_j=T_j T_1^{-1}$ for $j=2,\cdots,\sigma$ from (\ref{esek1})--(\ref{esek2}) 
			
			\hspace{3mm} and (\ref{esek3})--(\ref{esek4}) with $j$ and $1$ plugged in places of $j_2$ and 
			
			\hspace{3mm} $j_1$  by choosing suitable paths and inversions if needed 
			
			2: Transform $\hat{\mathcal P}_j$ to $\check{\mathcal P}_j$ using $\Pi_j$ as the similarity 
			
			\hspace{3mm} transformations for $j=2,\cdots,\sigma$.         	
			
			\vspace{0.1cm}
			
			\textbf{Outputs:} $\hat{\mathcal P}_1$ and $\check{\mathcal P}_j$, $j=2,\cdots,\sigma$.
			\vspace{0.1cm}	
			\hrule
		\end{tabular}
	\end{center}
\end{table}


\begin{thebibliography}{00}

\bibitem{Ruietal:2016}
	R.~Rui, T.~Ardeshiri, and A.~Bazanella, ``Identification of piecewise affine
	state-space models via expectation maximization,'' in: Proceedings of the {\em IEEE Conference on Computer 
	Aided Control System Design}, Buenos Aires, Argentina, pages~1066--1071, September 2016.
	
\bibitem{Paolettietal:2007}
	S.~Paoletti, A.~L. Juloski, G.~Ferrari-Trecate, and R.~Vidal, ``Identification
	of hybrid systems a tutorial,'' {\em European Journal of Control}, vol.~13,
	no.~2-3, pp.~242--260, 2007.
	
\bibitem{Vidal:2008}
	R.~Vidal, ``Recursive identification of switched ARX systems,'' {\em
	Automatica}, vol.~44, no.~9, pp.~2274--2287, 2008.
	
\bibitem{Ozayetal:2011}
	N.~Ozay, M.~Sznaier, C.~M. Lagoa, and O.~I. Camps, ``A sparsification approach
	to set membership identification of switched affine systems,'' {\em IEEE
	Transactions on Automatic Control}, vol.~57, no.~3, pp.~634--648, 2011.
	
\bibitem{Ozayetal:2015}
	N.~Ozay, C.~Lagoa, and M.~Sznaier, ``Set membership identification of switched
	linear systems with known number of subsystems,'' {\em Automatica}, vol.~51,
	pp.~180--191, 2015.
	
\bibitem{Jin&Huang:2010}
	X.~Jin and B.~Huang, ``Robust identification of piecewise/switching
	autoregressive exogenous process,'' {\em AIChE Journal}, vol.~56, no.~7,
	pp.~1829--1844, 2010.
	
\bibitem{Smith:1998}
	R.~Murray-Smith, ``Modelling human control behaviour with context-dependent
	Markov-switching multiple models,'' {\em IFAC Proceedings Volumes}, vol.~31,
	no.~26, pages~461--466, 1998.
	  
\bibitem{Mestre:2010}
	A.~Mestre, {\em Hybrid subspace identification: an application to HIV
	infection}, PhD thesis, Universidade T{\'e}cnica de Lisboa-Instituto
	Superior T{\'e}cnico, 2010.
	
\bibitem{Bemporadetal:2005}
	A.~Bemporad, A.~Garulli, S.~Paoletti, and A.~Vicino, ``A bounded-error approach
	to piecewise affine system identification,'' {\em IEEE Transactions on
	Automatic Control}, vol.~50, pp.~1567--1580, 2005.
	
\bibitem{Ferrarietal:2003}
	G.~Ferrari-Trecate, M.~Muselli, D.~Liberati, and M.~Morari, ``A clustering
	technique for the identification of piecewise affine systems,'' {\em
	Automatica}, vol.~39, no.~2, pp.~205--217, 2003.
	
\bibitem{Juloskietal:2005}
	A.~L. Juloski, S.~Weiland, and W.~Heemels, ``A Bayesian approach to identification of hybrid 
	systems,'' {\em IEEE Transactions on Automatic Control}, vol.~50, pp.~1520--1533, 2005.
	
\bibitem{Lassoued&Abderrahim:2014}
	Z.~Lassoued and K.~Abderrahim, ``An experimental validation of a novel
	clustering approach to PWARX identification,'' {\em Engineering Applications
	of Artificial Intelligence}, vol.~28, pp.~201--209, 2014.

\bibitem{Vidaletal:2003}
	R.~Vidal, S.~Soatto, Y.~Ma, and S.~Sastry, ``An algebraic geometric approach to
	the identification of a class of linear hybrid systems,'' in: Proceedings of the {\em 42nd IEEE
	International Conference on Decision and Control}, Maui, HI, pages~167--172, December 2003.

\bibitem{Ohlsson&Ljung:2013}
	H.~Ohlsson and L.~Ljung, ``Identification of switched linear regression models using sum-of-norms
	regularization,'' {\em Automatica}, vol.~49, pp.~1045--1050, 2013.
	
\bibitem{Hartmannetal:2015}
	A.~Hartmann, J.~M. Lemos, R.~S. Costa, J.~Xavier, and S.~Vinga, ``Identification of switched ARX 
	models via convex optimization and expectation maximization,'' {\em Journal of Process Control}, 
	vol.~28, pp.~9--16, 2015.
	
\bibitem{Lopesetal:2013}
	R.~V. Lopes, G.~A. Borges, and J.~Y. Ishihara, ``New algorithm for identification of 
	discrete-time switched linear systems,'' in: Proceedings of the {\em 2013 American Control Conference}, 
	Washington, DC, pages~6219--6224, June 2013.

\bibitem{Sefidmazgietal:2016}
    M.~G. Sefidmazgi, M.~M. Kordmahalleh, A.~Homaifar, A.~Karimoddini, and E.~Tunstel, ``A bounded 
    switching approach for identification of switched MIMO systems,'' in: Proceedings of the {\em IEEE International 
    Conference on Systems, Man, Cybernetics},  Budapest, Hungary, pages~4743--4748, October 2016.

\bibitem{Weilandetal:2006}
    S.~Weiland, A.~L. Juloski, and B.~Vet, ``On the equivalence of switched affine models and switched 
    ARX models,'' in: Proceedings of the {\em 45th IEEE Conference on Decision and Control}, San Diego, CA, 
    pages~2614--2618, December 2006.
	
\bibitem{Toth:2010}
    R.~T\'oth, {\em Modeling and Identification of Linear Parameter-Varying Systems}, Lecture Notes in
    Information and Control, vol.~403, Springer: Berlin-Heidelberg, 2010.
    
\bibitem{Lauer&Bloch:2019}
    F.~Lauer and G.~Bloch, ``Hybrid System Identification,'' {\em Hybrid System Identification: 
    Theory and Algorithms for Learning Switching Models},'' pages~77--101, Lecture Notes in
    Information and Control, vol.~478, Springer, 2019.
   
\bibitem{Petreczky:2011}
    M.~Petreczky, ``Realization theory for linear and bilinear switched systems: Formal power series 
    approach--Part I: Realization theory of linear switched systems,'' {\em ESAIM: Control, Optimisation 
    and Calculus of Variations}, vol.~17, pp.~410--445, 2011.  
         
\bibitem{Petreczkyetal:2013}
    M.~Petreczky, L.~Bako, and J.~H. Van~Schuppen, ``Realization theory of discrete-time linear switched 
    systems,'' {\em Automatica}, vol.~49, pp.~3337--3344, 2013.    
    
\bibitem{Ho&Kalman:1965}
    B.~L.~Ho and R.~E.~Kalman, ``Effective construction of linear state-variable models from input/output 
    functions,'' {\em Regelunstechnik}, 1965.      
    
\bibitem{Phanetal:1998}
    M.~Q. Phan, J.-N. Juang, and R.~E. Longman, ``Markov parameters in system identification: old and 
    new concepts,'' in {\em Structronic Systems: Smart Structures, Devices and Systems, Part II: Systems 
    and Control}, pages~263--293, World Scientific, 1998.

\bibitem{Bakoetal:2009a}
    L.~Bako, G.~Merc{\`e}re, R.~Vidal, and S.~Lecoeuche, ``Identification of switched linear state space 
    models without minimum dwell time,'' {\em IFAC Proceedings Volumes}, vol.~42, no.~10, pages~569--574, 
    2009.

\bibitem{Tothetal:2011}
    R.~T{\'o}th, H.~S. Abbas, and H.~Werner, ``On the state-space realization of LPV input-output models: 
    Practical approaches,'' {\em IEEE Transactions on Control Systems Technology}, vol.~20, pp.~139--153, 2011. 

\bibitem{Phan&Longman:1996}
    M.~Phan and R.~Longman, ``Relationship between state-space and input-output models via observer 
    Markov parameters,'' {\em WIT Transactions on the Built Environment}, vol.~19, pp. 185--200, 1996.
    
\bibitem{Majjietal:2010}
    M.~Majji, J.-N. Juang, and J.~L. Junkins, ``Observer/Kalman-filter time-varying system identification,'' 
    {\em Journal of Guidance, Control, and Dynamics}, vol.~33, pp.~887--900, 2010.    
       	
\bibitem{Huangetal:2004}
	K.~Huang, A.~Wagner, and Y.~Ma, ``Identification of hybrid linear time-invariant systems via subspace
	embedding and segmentation (SES),'' in: Proceedings of the {\em 2004 43rd IEEE Conference on Decision and Control},  
	Nassau, Bahamas, pages~3227--3234, December 2004.
	
\bibitem{Vidal&Chiuso&Soatto:2002}
    R.~Vidal, A.~Chiuso, S.~Soatto, ``Observability and identifiability of jump linear systems,''
    in: Proceedings of the {\em 41st IEEE Conf. Decision and Control}, Las Vegas, NV, pages 3614--3619, December 2002.  	

\bibitem{Verdult&Verhaegen:2004}
	V.~Verdult and M.~Verhaegen, ``Subspace identification of piecewise linear systems,'' in: Proceedings of the 
	{\em 43rd IEEE Conference on Decision and Control}, Nassau, Bahamas, pages~3838--3843, December 2004.
	
\bibitem{Borgesetal:2005}
	J.~Borges, V.~Verdult, M.~Verhaegen, and M.~A. Botto, ``A switching detection method based on 
	projected subspace classification,'' in: Proceedings of the {\em 44th IEEE Conf. Decision and Control}, Seville, 
	Spain, pages~344--349,  December 2005.	
	
\bibitem{Pekpeetal:2004}
	K.~M. Pekpe, G.~Mourot, K.~Gasso, and J.~Ragot, ``Identification of switching systems using change 
	detection technique in the subspace framework,'' in: Proceedings of the {\em 43rd IEEE Conference on Decision and 
	Control}, Nassau, Bahamas, pages.~3720--3725, December 2004.
	
\bibitem{Bakoetal:2013}
	L.~Bako, V.~L.~Le, F.~Lauer, and G.~Bloch, ``Identification of MIMO switched state-space models,'' in: Proceedings of 
	the {\em 2013 American Control Conference}, Washington, DC, pages~71--76, June 2013.
	
\bibitem{Bencherki&Turkay&Akcay:2020}
    F.~Bencherki, S.~T\"urkay, and H.~Ak\c{c}ay, ``Observer-based switched-linear system identification, arXiv:2107.14571v2 
    {[eess.SY]}, Preprint, 2021.	
    
\bibitem{Bakoetal:2009b}
    L.~Bako, G.~Merc{\`e}re, and S.~Lecoeuche, ``On-line structured subspace identification with 
    application to switched linear systems,'' {\em International Journal of Control}, vol.~82, 
    pp.~1496--1515, 2009. 
    
\bibitem{Blackmoreetal:2007}
    L.~Blackmore, S.~Gil, S.~Chung, and B.~Williams, ``Model learning for switching linear systems 
    with autonomous mode transitions,'' in: Proceedings of the {\em 46th IEEE Conference on Decision and Control},
    New Orleans, LA,  pages~4648--4655, December 2007.  
    
\bibitem{Gil&Williams:2009}
    S.~Gil and B.~Williams, ``Beyond local optimality: An improved approach to hybrid model learning,'' in: Proceedings
    of the {\em 48h IEEE Conference on Decision and Control and 28th Chinese Control
	Conference}, Shanghai, China, pages~3938--3945, December 2009.         
	
\bibitem{Pekpe&Lecoeuche:2008}
	K.~M. Pekpe and S.~Lecoeuche, ``Online clustering of switching models based on a subspace framework,'' 
	{\em Nonlinear Analysis: Hybrid Systems}, vol.~2, pp.~735--749, 2008.
	
\bibitem{Juang&Pappa:1985}
	J.-N. Juang and R.~S. Pappa, ``An eigensystem realization algorithm for modal parameter identification and 
	model reduction,'' {\em Journal of Guidance, Control, and Dynamics}, vol.~8, pp.~620--627, 1985.
	
\bibitem{Ho&Kalman:66}
    B.~Ho and R.~E.~Kalman, ``Effective construction of linear state-variable models from input/output functions, 
    {\em Regelungstechnik}, vol. 14, pp. 545--548, 1966.	
	
\bibitem{Juangetal:1988}
	J.-N. Juang, J.~Cooper, and J.~Wright, ``An eigensystem realization algorithm using data correlations (ERA/DC) for 
	modal parameter identification,''{\em Journal of Control Theory and Advanced Technology}, vol. 4, pp. 5--14, 1988.
		
\bibitem{Cooper:1997}
	J.~Cooper, ``On-line version of the eigensystem realization algorithm using data correlations,'' 
	{\em Journal of Guidance, Control, and Dynamics}, vol.~20, no.~1, pp.~137--142, 1997.
	
\bibitem{Majjietal:2010b}
    M.~Majji, J.-N. Juang, and J.~L. Junkins, ``Time-varying eigensystem realization algorithm,'' {\em Journal 
	of Guidance, Control, and Dynamics}, vol.~33, no.~1, pp.~13--28, 2010.	
	
\bibitem{Oymak&Ozay:2019}
    S.~Oymak and N.~Ozay, ``Non-asymptotic identification of LTI systems from a single trajectory,''
    in: Proceedings of the {\em American Control Conference}, Philadelphia, PA, pages 5655-5661, July 2019.
    
\bibitem{Sarkaretal:2021}
    T.~Sarkar, A.~Rakhlin, and M.~A.~Dahleh, ``Nonparametric finite time LTI system identification,''
    {\em Journal of Machine Learning Research}, vol. 22, pp. 1--61, 2021.

\bibitem{Tothetal:2011b}
    R.~T{\'o}th, J.~C.~Willems, P.~S.~C.~Heuberger, and P.~M.~J.~Van den Hof, ``The behavioral approach
    to linear parameter-varying systems,'' {\em IEEE Transactions on Automatic Control}, vol. 56, 
    pp. 2499--2514, 2011. 
    
\bibitem{Petreczkyetal:2017} 
    M.~Petreczky, R.~T{\'o}th, and G.~Merc\`ere, ``Realization theory for LPV state-space representations with affine
    dependence,'' {\em IEEE Transactions on Automatic Control}, vol. 62, pp. 4667--4674, 2017. 
    
\bibitem{Shokoohi&Silverman:87}
    S.~Shokoohi and L.~M.~Silverman, ``Identification and model reduction of time-varying discrete-time systems,'' 
    {\em Automatica}, vol. 23, pp. 509--521, 1987.    
    
\bibitem{Costa&Fragoso&Marques:2000}  
    O.~L.~V.~Costa, M.~D.~Fragosa, and R.~P.~Marques, {\em Discrete-Time Markov Jump Linear Systems}, 
    Springer-Verlag: London, 2005.
    
\bibitem{Singh&Sznaier&Ljung:2018}  
    R.~Singh, M.~Sznaier, and L.~Ljung, ``A rank minimization formulation for identification of linear parameter
    varying models,'' {\em IFAC Proceedings Volumes}, vol. 51, no 26, pp. 74--80, 2018.   
      
\bibitem{Ohlsson&Ljung&Boyd:2010}
    H.~Ohlsson, L.~Ljung, and S.~Boyd,  ``Segmentation of ARX-models using sum-of-norms regularization,'' {\em Automatica},
    vol.~46, pp.~1107--1111, 2010.    	
    
\bibitem{Elhamifiar&Petreczky&Vidal:2009}
    E.~Elhamifar, M.~Petreczky, and R.~Vidal, ``Rank tests for the observability of discrete-time jump linear systems 
    with inputs, in: Proceedings of the {\em American Control Conference}, St. Louis, MO, pages 3025--3032, June 2009.
    
\bibitem{Kailath:80}
    T.~Kailath, {\em Linear Systems}, Englewood Cliffs, NJ: Prentice-Hall, 1980.
    
\bibitem{Verhaegen&Yu:95} 
    M.~Verhaegen and X.~Yu, ``A class of subspace model identification algorithms to identify periodically and arbitrarily
    time-varying systems,'' {\em Automatica}, vol. 31, pp. 201--216, 1995.  

\bibitem{Liu:1997}
    K.~Liu, ``Identification of linear time-varying systems,'' {\em Journal of Sound and Vibration}, vol. 206, 
    pp. 487--505, 1997.         
    
\bibitem{Golub&VanLoan:1989}
    G.~H.~Golub and C.~F.~Van~Loan, {\em Matrix Computation}, second edition, The Johns Hopkins University Press: 
    Baltimore, 1989. 
    
\bibitem{McKelvey&Akcay&Ljung:96}
    T.~McKelvey, H.~Ak\c{c}ay, and L.~Ljung, ``Subspace-based multivariable system identification from frequency
    response,'' {\em IEEE Transactions on Automatic Control}, vol.~41, pp. 960--979, 1996.           
    
\bibitem{Esteretal:96}
    M.~Ester, H.~P.~Kriegel, J.~Sander, and X.~Xu, ``A density-based algorithm for
    discovering clusters in large spatial databases with noise,'' in: Proc. {\em $2$nd Int. 
	Conf. Knowledge Discovery and Data Mining}, Portland, OR, pages 226--231, August 1996. 

\bibitem{Arthur&Vassilvitskii:2007}	
    D.~Arthur and S.~Vassilvitskii, ``k-means++: the advantages of careful seeding,'' in: Proc. 
    {\em 18th Annual ACM-SIAM Symposium on Discrete Algorithms}, Philadelphia, PA, pages 1027--1035, 
    January 2007.
    
\bibitem{Hastie&Tibshirani&Friedman:2001}
    T.~Hastie, R.~Tibshirani, and J.~Friedman, {\em The elements of statistical learning: Data mining,
    inference, and prediction}. Springer Series in Statistics, New York, NY: Springer-Verlag, 2001.  

\bibitem{Petreczky&Bako&Schuppen:2010}
    M.~Petreczky, L.~Bako, and J.~H.~Van~Schuppen, ``Identifiability of discrete-time linear 
    switched systems,'' in: Proceedings of the {\em 13th ACM International Conference  on Hybrid Systems: Computation and Control}, Porto, Portugal, pages 141--150, April 2010. 
    
\bibitem{Petreczky&Schuppen:2010} 
    M.~Petreczsky and J.~H.~Van Schuppen, ``Realization theory for linear hybrid systems,'' {\em IEEE Transactions
    on Automatic Control}, vol. 55, no 10, pp. 2282--2297, 2010.  

\bibitem{Hossain&Trenn:2021}
    M.~S.~Hossain and S.~Trenn, ``Minimal realization for linear switched systems with a single switch,'' in: 
    Preoceedings of the  {\em European Control Conference}, Delft, Netherlands, pages 1168--1173, June-July 2021.
    
\bibitem{Birouche&Mourllion&Basset:2012}
    A.~Birouche, B.~Mourllion, and M.~Basset, ``Model order-reduction for discrete-time switched linear systems,''
    International Journal of Systems Science, vol. 43, pp. 1753--1763, 2012.
    
\bibitem{Petreczky&Wisniewski&Leth:2013}
	M.~Petreczky, R.~Wisniewski, and J.~Leth, ``Balanced truncation for linear switched systems,''
    {\em Nonlinear Analysis: Hybrid Systems}, vol. 10, pp. 4--20, 2013.    
       
\bibitem{Goseaetal:2018}
    I.~V.~Gosea, M.~Petreczky, and A.~C.~Antoulas, ``Data-driven model order reduction of linear switched systems 
    in the Loewner framework,'' {\em SIAM Journal on Scientific Computing}, vol. 40, pp. B572--B610, 2018.
    
\bibitem{Grant&Boyd:2014} 
    M.~Grant and S.~Boyd, ``CVX: Matlab software for disciplined convex programming,'' Version~2.1, 2014.
                 
                             
\end{thebibliography}
\end{document}